\documentclass[twocolumn]{aastex7}
\usepackage[utf8]{inputenc}
\usepackage{textgreek}
\usepackage{amsmath}
\usepackage{multirow}
\usepackage{booktabs}
\usepackage{mhchem}

\newcommand{\hi}{H\,{\small I}}
\newcommand{\hii}{H\,{\small II}}
\newcommand{\ha}{H$\alpha$}
\newcommand{\paa}{Pa$\alpha$}
\newcommand{\spr}{$\rho$}
\newcommand{\sighi}{$\Sigma_{HI}$}
\newcommand{\coone}{$^{12}$CO(1-0)}
\newcommand{\cotwo}{CO(2-1)}
\newcommand{\mjysr}{MJy~sr$^{-1}$}
\newcommand{\iffive}{$I_{\rm F560W}$}
\newcommand{\ifseven}{$I_{\rm F770W}$}

\newcommand{\ifeleven}{$I_{\rm F1130W}$}

\begin{document}

\title{JWST Whirlpool Galaxy Treasury: Mid-Infrared Emission in M51 and its Relation to Gas Column and Star Formation}

\author[0000-0002-3472-0490]{Mansi Padave}
\affiliation{Department of Astronomy \& Astrophysics, University of California, San Diego, 9500 Gilman Drive, La Jolla, CA 92093, USA}
\email[show]{mpadave@ucsd.edu}
\author[0000-0002-4378-8534]{Karin M. Sandstrom}
\affiliation{Department of Astronomy \& Astrophysics, University of California, San Diego, 9500 Gilman Drive, La Jolla, CA 92093, USA}
\email{kmsandstrom@ucsd.edu}
\author[0000-0002-5782-9093]{Daniel~A.~Dale}
\affiliation{Department of Physics and Astronomy, University of Wyoming, Laramie, WY 82071, USA}
\email[hide]{ddale@uwyo.edu}

\author[0000-0002-2545-1700]{Adam K. Leroy}
\affiliation{Department of Astronomy, The Ohio State University, Columbus, OH 43210, USA}
\email[hide]{}

\author[0000-0001-9605-780X]{Eric W. Koch}
\affiliation{National Radio Astronomy Observatory, 800 Bradbury SE, Suite 235, Albuquerque, NM 87106, USA}
\email[hide]{}

\author[0009-0005-8923-558X]{Tony D. Weinbeck}
\affiliation{Department of Physics and Astronomy, University of Wyoming, Laramie, WY 82071, USA}
\email[hide]{tonyweinbeck@gmail.com}

\author[0000-0002-8192-8091]{Angela Adamo}
\affiliation{Department of Astronomy, The Oskar Klein Centre, Stockholm University, AlbaNova, SE-10691 Stockholm, Sweden}
\email[hide]{}

\author[0000-0002-9183-8102]{Jessica Sutter}
\affiliation{Whitman College, 345 Boyer Avenue, Walla Walla, WA 99362, USA}
\email{jessica.sutter93@gmail.com}

\author[0009-0005-0750-2956]{Lindsey Hands}
\affiliation{Department of Astronomy \& Astrophysics, University of California, San Diego, 9500 Gilman Drive, La Jolla, CA 92093, USA}
\email{lhands@ucsd.edu}

\author[0000-0002-5666-7782]{Torsten B\"oker}
\affiliation{European Space Agency, c/o STScI, 3700 San Martin Drive, Baltimore, MD 21218, USA}
\email[hide]{}

\author[0000-0003-4850-9589]{Martha L. Boyer}
\affiliation{Space Telescope Science Institute, 3700 San Martin Dr., Baltimore, MD 21218, USA}
\email[hide]{}

\author[0000-0001-8241-7704]{Ryan Chown}
\affiliation{Faculty of Computer Science \& Technology, Algoma University, Sault Ste. Marie, ON
P6A 2G4, Canada}
\email[hide]{ryan.chown@algomau.ca}

\author[0000-0002-0846-936X]{Bruce T. Draine}
\affiliation{Dept. of Astrophysical Sciences, Princeton University, Princeton, NJ 08544, USA}
\email[hide]{}

\author[0000-0002-0846-936X]{Ilse de Looze}
\affiliation{Sterrenkundig Observatorium, Universiteit Gent, Krijgslaan 281-S9, 9000 Gent, Belgium}
\email[hide]{}

\author[0000-0003-4224-6829]{Brandt A. L. Gaches}
\affiliation{Faculty of Physics, University of Duisburg-Essen, Lotharstra{\ss}e 1, 47057 Duisburg, Germany}
\email[hide]{brandt.gaches@uni-due.de}

\author[0000-0001-6708-1317]{Simon C. O. Glover}
\affiliation{Universit\"{a}t Heidelberg, Zentrum f\"{u}r Astronomie, Institut f\"{u}r Theoretische Astrophysik, Albert-Ueberle-Str.\ 2, 69120 Heidelberg, Germany}
\email[hide]{glover@uni-heidelberg.de}

\author[0000-0001-6498-2945]{Dario Colombo}
\affiliation{Argelander-Institut f\"ur Astronomie, University of Bonn, Auf dem H\"ugel 71, 53121 Bonn, Germany}
\email{dcolombo@uni-bonn.de}

\author[0000-0001-5448-1821]{Robert C. Kennicutt, Jr}
\affiliation{Steward Observatory, University of Arizona, 933 N Cherry Avenue, Tucson, AZ 85721, USA}
\affiliation{George P. and Cynthia W. Mitchell Institute for Fundamental Physics \& Astronomy, Texas A\&M University, College Station, TX 77843, USA}
\email[hide]{rck@arizona.edu}

\author[0009-0001-5949-1524]{Hannah Koziol}
\affiliation{Department of Astronomy \& Astrophysics, University of California, San Diego, 9500 Gilman Drive, La Jolla, CA 92093, USA}
\email[hide]{}

\author[0000-0002-0560-3172]{Ralf S. Klessen}
\affiliation{Universit\"{a}t Heidelberg, Zentrum f\"{u}r Astronomie, Institut f\"{u}r Theoretische Astrophysik, Albert-Ueberle-Str.\ 2, 69120 Heidelberg, Germany}
\affiliation{Universit\"{a}t Heidelberg, Interdisziplin\"{a}res Zentrum f\"{u}r Wissenschaftliches Rechnen, Im Neuenheimer Feld 225, 69120 Heidelberg, Germany}
\email[hide]{}

\author[0000-0002-1000-6081]{Sean T. Linden}
\affiliation{Steward Observatory, University of Arizona, 933 N Cherry Avenue, Tucson, AZ 85721, USA}
\email{}

\author[0000-0002-7064-4309]{Desika Narayanan}
\affil{Department of Astronomy, University of Florida, 211 Bryant Space Sciences Center, Gainesville, FL 32611 USA}
\affil{Cosmic Dawn Center at the Niels Bohr Institute, University of Copenhagen and DTU-Space, Technical University of Denmark}
\email[hide]{}

\author[0000-0002-8222-8986]{Alex Pedrini}
\affiliation{Department of Astronomy, The Oskar Klein Centre, Stockholm University, AlbaNova, SE-10691 Stockholm, Sweden}
\email[hide]{}

\author[0000-0001-6326-7069]{Julia Roman-Duval}
\email{duval@stsci.edu}
\affiliation{Space Telescope Science Institute, 3700 San Martin Drive, Baltimore, MD 21218, USA}
\email[hide]{duval@stsci.edu}

\author[0000-0002-3933-7677]{Eva Schinnerer}
\affiliation{Max Planck Institut f\"ur Astronomie, K\"onigstuhl 17, D-69117 Heidelberg, Germany}
\email[]{c}

\author[0000-0003-1545-5078]{J.D.T. Smith}
\affil{Ritter Astrophysical Research Center, University of Toledo, Toledo, OH 43606, USA}
\email{JD.Smith@utoledo.edu}

\author[0000-0002-9333-387X]{Sophia~K.~Stuber}
\affiliation{National Astronomical Observatory of Japan, 2-21-1 Osawa, Mitaka, Tokyo 181-8588, Japan}
\affiliation{Max Planck Institute for Radio Astronomy, Auf dem Hügel 69, 53121 Bonn, Germany}
\email[hide]{astro@sophiastuber.de}

\author[0000-0003-4793-7880]{Fabian Walter}
\affiliation{Max Planck Institut f\"ur Astronomie, K\"onigstuhl 17, D-69117 Heidelberg, Germany}
\affiliation{California Institute of Technology, Pasadena, CA 91125, USA}
\email{walter@mpia.de}


\begin{abstract}
Using JWST/MIRI imaging of M51 in eight broadband filters, we investigate correlations of mid-infrared emission from polycyclic aromatic hydrocarbons (PAHs) and dust continuum with molecular, atomic, and ionized gas traced by \coone, \hi, and \paa, respectively. In molecular gas-dominated regions, PAH-dominated filters (F560W, F770W, F1130W, F1280W) exhibit near-linear correlations with \coone\ at $\sim40$~pc scale, indicating that PAHs are well-mixed with gas and experience relatively constant radiation field intensities. 
The F1500W, F1800W, and F2100W dust continuum-dominated filters show shallower slopes with \coone\, reflecting contributions from star-forming regions with high radiation field intensities. This is reinforced by the near-linear scaling between F2100W and \paa. PAH-dominated bands do not show this linear trend with \paa, likely due to their destruction in ionized regions.
F1000W behaves similarly to PAH bands in its correlations with \coone\ and \paa. Modeling mid-infrared emission with an empirical decomposition into gas- and star-formation–associated components shows that PAH-dominated filters receive comparable contributions from both, while the relative contribution associated with the \paa\ template increases toward longer wavelengths, reaching $\sim$75\% in F2100W. These results demonstrate that mid-infrared simultaneously traces the gas column and star formation, but with a systematic wavelength-dependent shift in what drives the correlations: PAHs being more gas-tracing and dust-continuum reflecting star formation. Lastly, considering both \hi\ and \ce{H2} at $\sim440$~pc resolution, we find a tight, $\sim$linear relation between $\Sigma_{HI+H_2}$ and PAH-dominated filters. Although most of our coverage is in \ce{H2}-dominated regions, we note similar observations with \hi, suggesting that PAHs are also well-mixed with atomic gas.
\end{abstract}


\section{Introduction} 
\label{sec:intro}
Mid-infrared (mid-IR) emission in star-forming galaxies arises primarily from small dust grains. Among these are polycyclic aromatic hydrocarbons \citep[PAHs;][]{puget85, alla89, tielens08}, which produce prominent features at 6.2, 7.7, 8.6, 11.3, 12.7, and 17~\micron\ due to vibrational modes of C–C and C–H bonds in the mid-IR. These undergo stochastic heating by absorbing ultraviolet photons from young main sequence stars and by the diffuse interstellar radiation field \citep[ISRF;][]{draine01, li01, galliano18}. As a result, dust emission in the mid-IR depends on both the dust column, tracing the distribution of interstellar material, and the ambient radiation field. The mid-IR emission thus encodes the radiative coupling of dust with starlight and gas, which governs the structure and heating of the interstellar medium (ISM) \citep{draine11}. 

The ability of dust grains to reprocess stellar radiation and re-radiate it in the IR has enabled its use as a diagnostic of obscured star formation. Empirical relations of mid-IR emission with UV- or \ha-based star formation rates (SFR) have been investigated in several studies \citep{calz07, kenn12, gregg25}. Both the dust continuum and PAH emission have been used in SFR calibrations to account for dust-obscured star formation \citep{leroy12, catalan2015, cluver17, belfiore23, calz24, calz25}, though with caveats related to metallicity, radiation field hardness, and contributions from the diffuse ISM that alter the underlying dust emission. Furthermore, the fact that dust is well mixed with gas in the ISM also makes mid-IR emission a useful tracer of gas surface density in galaxies. This emerges as a strong correlation between mid-IR intensities and molecular gas traced by CO luminosities at global \citep{regan06, gao19, chown21} and sub-kpc scales \citep{gao22, leroy23a, leroy23b, zhang23, chown25} in galaxies, provided that variations in dust-to-gas mass ratio, ISRF, and PAH abundance are modest.  

The expected behavior of mid-IR emission at near-solar metallicity for stochastically heated dust grains \citep{draine07, compiegne2010, leroy23b} can be expressed as:
\begin{align}\label{eq:1}
    I_{\rm MIR}&\propto D/G \times U \times q_{\rm PAH} \times \Sigma_{\rm gas}, 
\end{align}
where $I_{\rm MIR}$ is the mid-IR intensity, $D/G$ is the dust-to-gas mass ratio, $U$ is the strength of the local ISRF illuminating the dust relative
to that in the Solar neighborhood, $q_{\rm PAH}$ is the PAH mass fraction, and $\Sigma_{\rm gas}$ is the gas mass surface density. Other potentially important variables include the radiation field spectrum and  properties of the PAH population like their size and charge distribution \citep{draine21}. While $\Sigma_{\rm gas}$ represents both atomic and molecular hydrogen, in the molecular-dominated regime of the ISM, $\Sigma_{\rm gas}\approx\Sigma_{\rm mol}$, the molecular gas surface density. Since CO line emission traces molecular gas, assuming a CO-to-\ce{H2} conversion factor \citep{bolatto13}, Equation \ref{eq:1} provides a basis for the expected correlation between CO and mid-IR emission in systems at near-solar metallicity. 

The James Webb Space Telescope \citep[JWST,][]{reike15, gordon15} has transformed our view of the mid-IR emission in the ISM with its high resolution, separating star-forming regions from the diffuse ISM on $\lesssim$100 pc scales and leading to increasing observational evidence of PAH emission tracking CO emission (and the gas column in general) in galaxies. \cite{chown25}'s investigation of 70 nearby galaxies at $\sim$100~pc resolution showed an approximately linear scaling between \cotwo\ and PAH emission traced by F770W and F1130W, as well as continuum-subtracted F335M filter that captures the 3.3\micron\ PAH feature. Even in extreme conditions, like M82's outflow, CO clouds and PAH structures are strongly correlated \citep{villa25, sebastian25}. 
Galaxies between $z\sim$1--3 also show constant CO-to-PAH luminosity ratios that remain invariant under a range of luminosities, galaxy types, and redshifts \citep{shivaei24}. Furthermore, PAHs may also be used to probe the HI-dominated ISM and CO-dark \ce{H2} \citep{sandstrom23, chown25dwarf}. 

\cite{whitcomb23} demonstrated that the degree to which mid-IR emission traces CO and SFR varies with wavelength. PAH-dominated bands, such as IRAC-8~\micron\ and 12~\micron\ band from Wide-field Infrared Survey Explorer (WISE), tend to correlate more closely with CO. In contrast, bands like WISE-22~\micron\ and MIPS-24~\micron\, which are dominated by continuum emission from small dust grains, show a tighter correlation with SFR. Even when PAH and continuum components were separated in the IRAC-8~\micron\ and WISE-12~\micron\ bands, the PAH component was tightly linked to CO, whereas the continuum was better correlated to star formation. 
These results provided compelling observational evidence that mid-IR emission is shaped by both column density and local radiation fields in the ISM. Using JWST/MIRI F770W, F1000W, F1130W, and F2100W observations of 4 nearby galaxies, \cite{leroy23b} showed that $\sim50$\% of mid-IR emission arises from regions traced by CO and $\sim50$\% from star forming regions traced by \ha, with some variations between galaxies. Interestingly, in both \cite{whitcomb23, leroy23b}, the correlations of mid-IR with CO and \ha\ were stronger than the correlation between CO and \ha\ itself.

The dual sensitivity of mid-IR emission to heating and ISM column density makes it a powerful tracer of ISM conditions in galaxies. Current advances discussed above open the opportunity to address key underlying questions about how the mid-IR emission traces the ISM. Knowing that mid-IR emission is driven by a combination of ISM column density and local heating, we ask: {\emph {1) How do different ISM phases -- molecular, atomic, and ionized gas -- contribute to the observed mid-IR emission in a galaxy? 2) How do the relative roles of gas column density and local radiation field vary across the mid-IR wavelengths and galactic environments within a galaxy?}} 

We investigate these questions in the Whirlpool Galaxy, or M51, as part of the JWST Whirlpool Galaxy Treasury (Dale, Sandstrom et al. in prep). We exploit the full mid-IR coverage with JWST/MIRI imaging from 5.6-21 \micron\ spanning PAH-dominated and dust continuum-dominated bands, in combination with extensive ancillary data tracing molecular, atomic, and ionized gas at pc-scale spatial resolution. Firstly, we adopt a a distance of 7.59 Mpc \citep{csornyei2023}, where 1\arcsec\ corresponds to 46~pc spatial resolution. At these scales molecular gas clouds and ionized regions begin to spatially de-correlate \citep{kruijssen19}, allowing us to distinguish high $U$ environments around young star clusters from giant molecular clouds and probe variations in the ISRF that are typically averaged above kpc scales. 
Secondly, access to the set of eight MIRI filters is critical as the mid-IR bands differentially trace PAH features and the underlying dust continuum. While PAH emission dominates at shorter mid-IR wavelengths, warm dust emission increases toward longer wavelengths. The continuous coverage across this wavelength range therefore provides an empirical lever arm to disentangle variations in mid-IR emission caused by changes in dust mass and column density from those caused by changes in $U$. 

The paper is organized as follows: \S\ref{sec:data} describes the data and our fitting and correlation methodology. In \S\ref{sec:results} we describe  the correlations of mid-IR with \coone, \paa, \hi, and total gas (\hi+\ce{H2}) followed by \S\ref{sec:discuss} where we investigate the relative contribution of gas and star formation to mid-IR emission and  discuss the implications for using mid-IR emission as a tracer of gas and star formation. 

\begin{figure*}
    \includegraphics[trim =0.1cm 0cm 0.3cm 0cm, clip,scale=0.42]{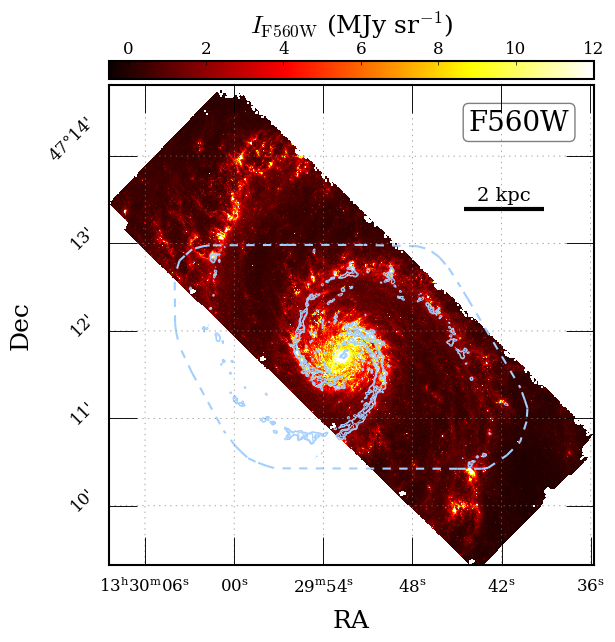}\hfill
    \includegraphics[trim = 2.5cm 0cm 0.3cm 0cm, clip,scale=0.42]{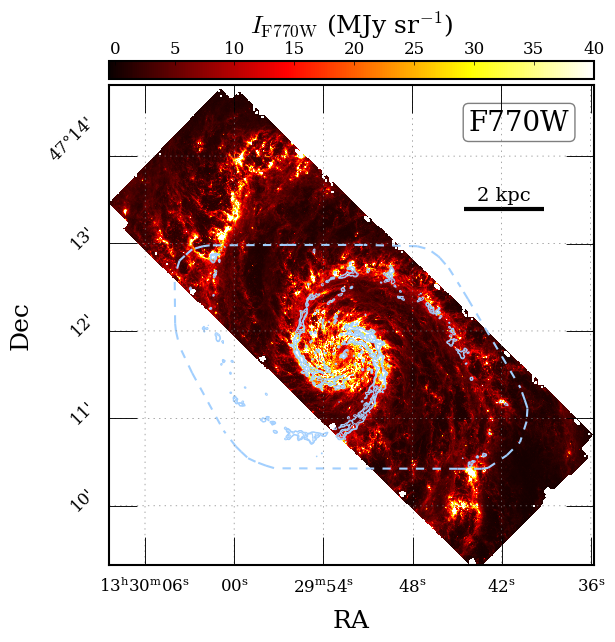}\hfill
    \includegraphics[trim=2.5cm 0cm 0.3cm 0cm, clip, width=0.3\textwidth, height=6.85cm]{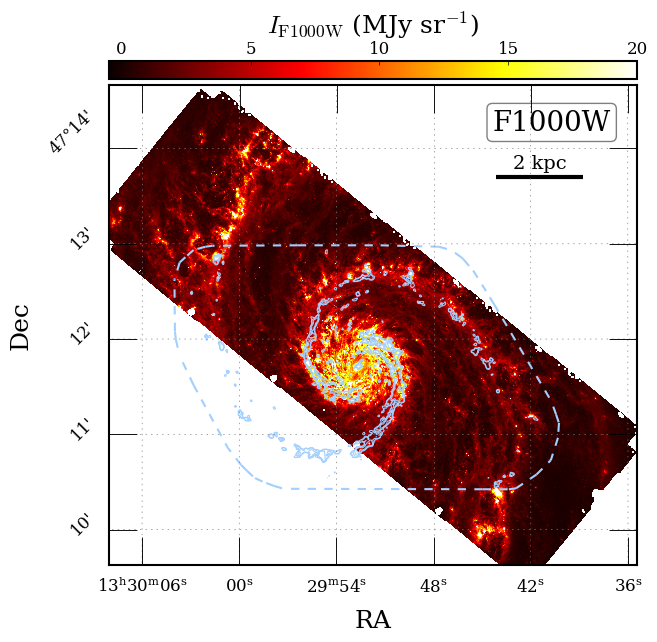}
    \includegraphics[trim=0.1cm 0cm 0.3cm 0cm, clip, width=0.355\textwidth, height=6.85cm]{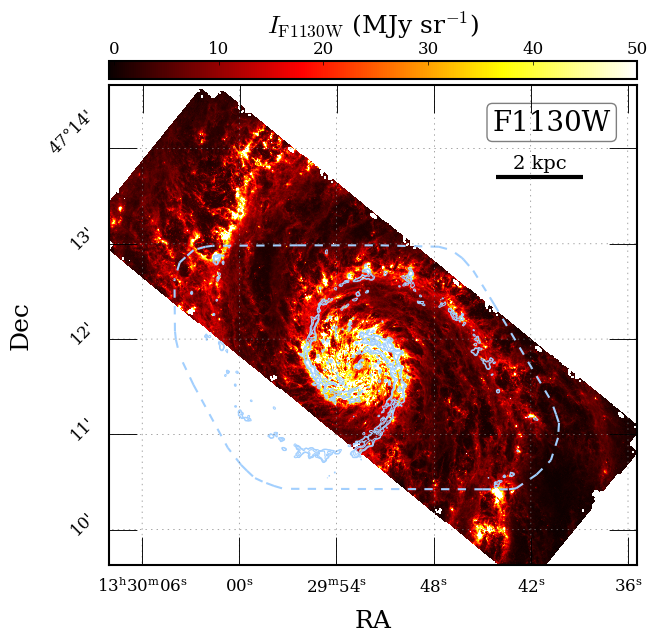}\hfill
    \includegraphics[trim=2.5cm 0cm 0.3cm 0cm, clip, width=0.3\textwidth, height=6.85cm]{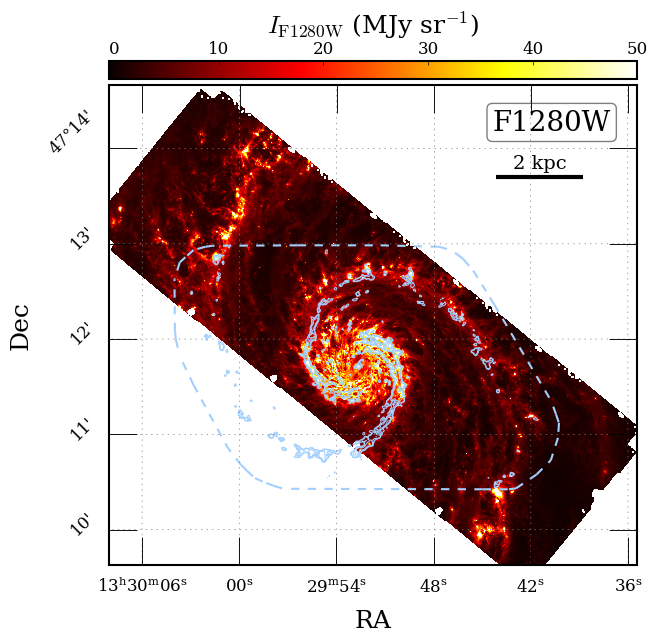}\hfill
    \includegraphics[trim=2.5cm 0cm 0.3cm 0cm, clip, width=0.3\textwidth, height=6.85cm]{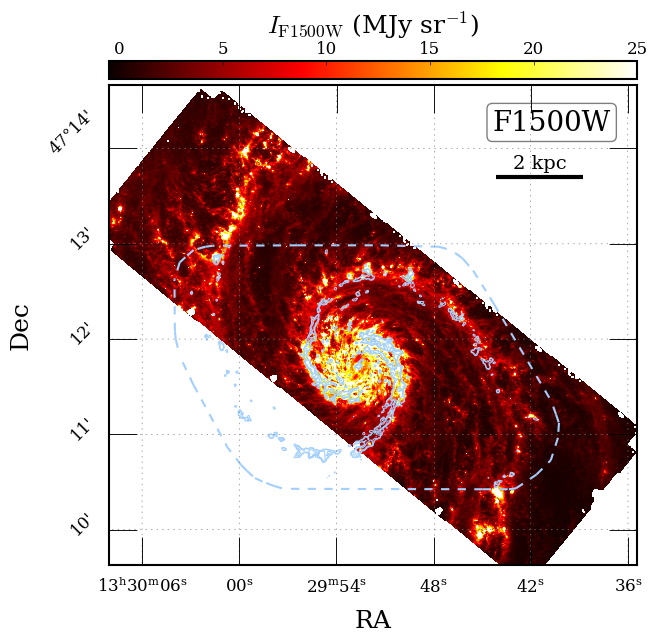}
   \includegraphics[trim=0.1cm 0cm 0.3cm 0cm, clip, width=0.355\textwidth, height=6.85cm]{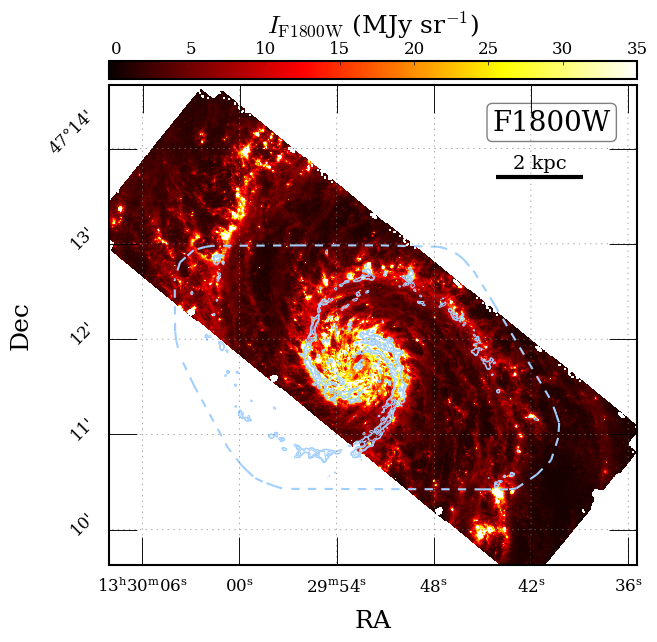}\hspace{0.3cm}
    \includegraphics[trim=2.5cm 0cm 0.3cm 0cm, clip, width=0.3\textwidth, height=6.85cm]{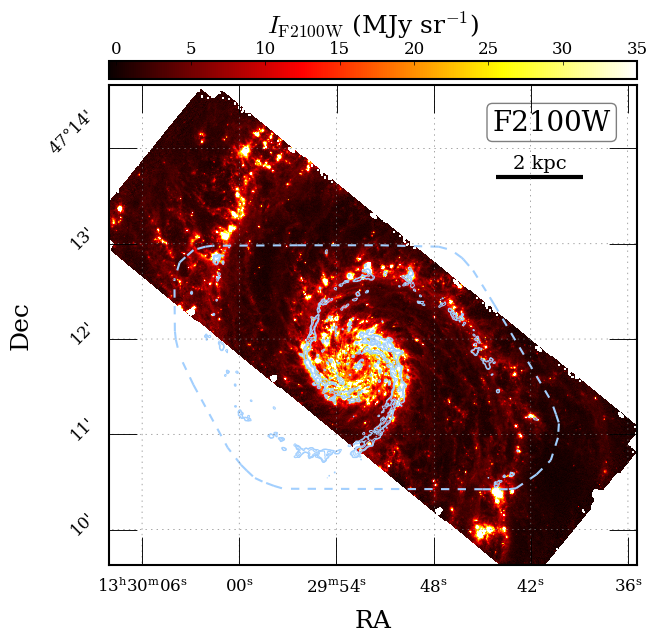}
    \caption{The reduced JWST/MIRI images of the M51 disk in (top panels) F560W, F770W, F1000W, (middle panels) F1130W, F1280W, 1500W, (bottom panels) F1800W, and F2100W. These map the central 4.4~kpc$\times$13.2~kpc. The footprint of the \coone\ map is marked by the dashed border with contours at 20$\sigma$ overlaid in blue. } \label{fig:miri} 
\end{figure*}

\section{Data \& Methods} \label{sec:data}
\subsection{Mid-IR, \coone, Pa$\alpha$, and \ion{H}{1} data}
We use data from eight JWST/MIRI bands -- F560W, F770W, F1000W, F1130W, F1280W, F1500W, F1800W, and F2100W in this work. Observations for bands from F1000W to F2100W were obtained as part of the Cycle 2 JWST program \#3435 (The JWST Whirlpool Galaxy Treasury, PI: K. Sandstrom \& D. Dale), while
F560W and F770W images were obtained with the Cycle 1 JWST program \#1783 (Feedback in Emerging extrAgalactic Star clusTers, JWST–FEAST, P.I.: A. Adamo). We also make use of the narrowband \paa\ map (described further in this section) and JWST/NIRCam F300M observed as part of the FEAST program to perform starlight subtraction of MIRI F560W and F770W (see \S\ref{sec:analysis}).
The observation strategy for these programs used a 1$\times$5 pointing mosaic and a four-point dither pattern covering a $\sim$6\arcmin$\times$2\arcmin\ footprint of the M51 disk. These data were reduced using {\tt pjpipe} \citep{will24} version 1.2.0, a wrapper around the official JWST Calibration Pipeline, and CRDS context 1322.pmap. 

Background subtraction for the MIRI filters was carried out by obtaining ``off'' observations immediately following the MIRI science observations, using a single nearby pointing devoid of bright foreground stars located $\sim$7.5$\arcmin$ west (program \#1783) and $\sim$8.0$\arcmin$ southeast (program \#3435) of the M51 galactic center. We used the strict per-pixel sigma clipping limit of 1.5$\sigma$ following \citet{will24} while combining our dithers to produce the master dark image. It should be noted that background subtraction was especially important for our longest wavelength MIRI filters (namely F2100W and F1800W) in removing a single large-scale banding artefact attributed to thermal noise. Additionally, the Lyot coronograph was masked for each MIRI observation and was not included in our final science images. Background subtraction was not carried out for our NIRCam observations.

Since our mosaic includes only a limited extent of the full galactic disk, we follow the procedure laid out in \citet{leroy23b}, Appendix B, to perform zero-point flux anchoring. Specifically, we compare our F1130W image to the 12$\mu$m (Channel 3) map of M51 from WISE, which includes a substantial off-galaxy component. By convolving to a common resolution of 7\farcs5, gridding to a common astrometrically-aligned pixel scale, and performing a linear fit between pixel intensities of the two bands, we interpret the intercept as the flux offset between our image and the true sky background level. We then perform a similar procedure between F1130W and the other MIRI bands to anchor each remaining filter. For our NIRCam images we anchor the F300M channel to WISE1 (3.4$\mu$m) and then adjust the remaining NIRCam filters to F300M using the same process.
The resulting maps have 1$\sigma$ noise levels of 0.04--0.4~\mjysr\ with $\sim$60--95\% pixels detected above 3$\sigma$ at native resolution. The statistical noise in each filter is reported in Table \ref{tab:noise} and the maps are shown in Figure \ref{fig:miri}. 

\begin{deluxetable}{lccc}
\setlength{\tabcolsep}{15pt}
\tablewidth{\columnwidth}
\tablecaption{RMS noise in MIRI maps at varying resolution \label{tab:noise}}
\tablehead{
\colhead{Filter} & \multicolumn{3}{c}{RMS noise [\mjysr]} \\
\cmidrule(lr){2-4}
\colhead{} & \colhead{Native} & \colhead{1\farcs25} & \colhead{12\farcs00}
}
\startdata
F560W & 0.0308 & 0.0022 & 0.0002\\
F770W & 0.0371 & 0.0027 & 0.0003\\
F1000W & 0.0652 & 0.0048 & 0.0005\\
F1130W & 0.1920 & 0.0140 & 0.0014\\
F1280W & 0.0962 & 0.0073 & 0.0007\\
F1500W & 0.1406 & 0.0109 & 0.0010\\
F1800W & 0.2376 & 0.0197 & 0.0017\\
F2100W & 0.3387 & 0.0298 & 0.0024
\enddata
\end{deluxetable}

\begin{figure*}[!ht]
    \includegraphics[trim =0.1cm 0cm 0.3cm 0cm, clip,scale=0.35]{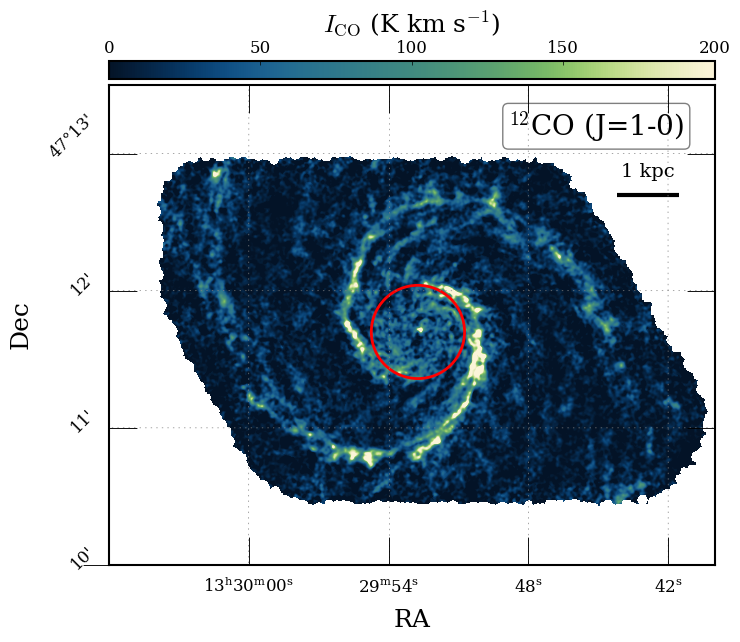}
    \includegraphics[trim = 2.5cm 0cm 0.0cm 0cm, clip,scale=0.35]{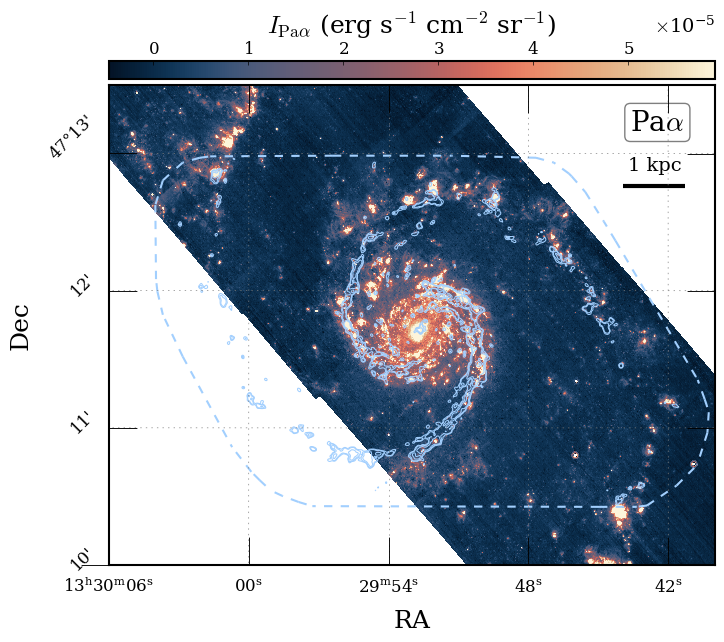}
    \includegraphics[trim = 2.8cm 0cm 0.0cm 0cm, clip,scale=0.35]{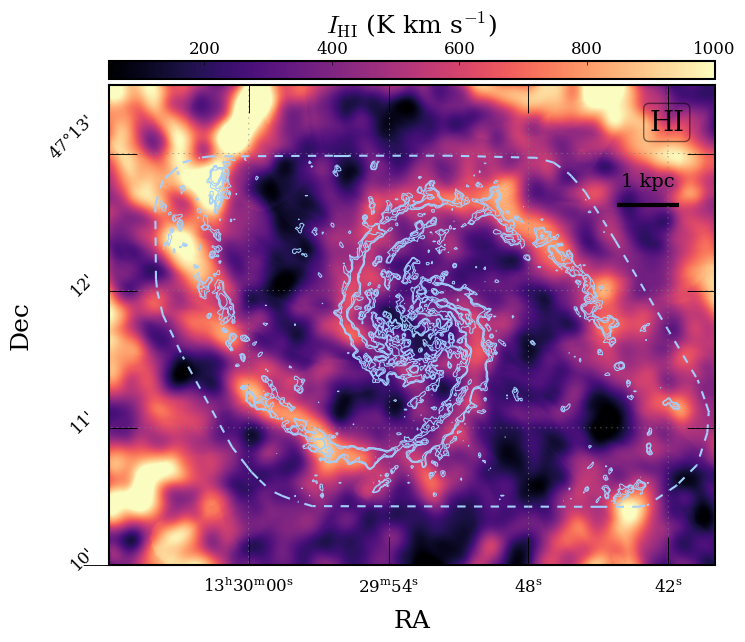}
    \caption{\coone, \paa, and \hi\ maps of M51. The footprint of the \coone\ map is marked on the \paa\ and \hi\ maps with contours at 20$\sigma$ overlaid in blue. The red circle in the \coone\ map corresponds to the central region that is masked in our study. } \label{fig:tracers} 
\end{figure*}

We use \coone\ line emission observed with the Plateau de Bure Interferometer (PdBI) as part of the PdBI Arcsecond Whirlpool Survey (PAWS) \citep{schin13} to trace the molecular gas distribution in M51. The field of view of this map covers the central 11~kpc$\times$7~kpc of the M51 disk. Details of the data reduction are described in \citet{pety13}, resulting in a cube with angular resolution of 1\farcs$\times$0\farcs97 and a mean rms of 0.4 K per 5~km~s$^{-1}$ wide channel. We make use of the broad weighted intensity map smoothed to a synthesized beam size of 1\farcs25$\times$1\farcs25 and 12\farcs00$\times$12\farcs00 with 0\farcs3 pixel$^{-1}$ and 6\farcs0 pixel$^{-1}$, respectively. The \coone\ map is shown in the left panel of Figure \ref{fig:tracers}.

The continuum-subtracted \paa\ map used in this work was created as part of JWST FEAST program using the F187N band that covers the \paa\ emission line. The emission line map is obtained by iteratively subtracting pivot stellar continuum estimated using F150W and F200W maps, detailed in \cite{calz25} following the procedure from \cite{gregg2024}. We also subtract a residual background gradient from the continuum-subtracted \paa\ following the method described in \cite{pedrini24}. We manually select empty sky regions and measure the mode and rms value in each. A background model is obtained by fitting a 2D plane to the mode values and subtracted from the \paa\ map to remove the gradient. The resulting 1$\sigma$ rms noise of the \paa\ map is 1.65$\times$10$^{-6}$~erg~s~cm$^{-2}$~sr$^{-1}$. The continuum-subtracted \paa\ map is shown in the middle panel of Figure \ref{fig:tracers}.

We use \hi-21cm data tracing atomic gas in M51, combining archival observations from VLA project IDs AW605 \citep[THINGS;][]{walter08} and 11A-142 (PI: A. Hughes). Together, these data span the A, B, C, and D configurations of the NSF's Karl G. Jansky Very Large Array \footnote{The National Radio Astronomy Observatory is a facility of the National Science Foundation operated under cooperative agreement by Associated Universities, Inc.} and thus are sensitive the full range of angular scales accessible to the VLA.
Details of the reduction and imaging are provided in Koch et al. 2026 (submitted). In brief, we calibrate the AW605 and 11A-142 observations following standard methods with CASA and adopt the PHANGS imaging pipeline \citep{phangs_alma_pipeline} to image and create final data products.
We use the image version from Koch et al. 2026 (submitted)  with Briggs robust weighting (0.5) and a uv-taper of $8\arcsec$ that, after circularizing, yields a final circular beam of $11\farcs5$.
This data cube has an rms noise of 1.8 K per 5.15~km/s channel, equivalent to an \hi\ column density sensitivity of $4.6\times10^{18}$~cm$^{-2}$ per 10 km/s.
We note that this data cube is moderately more sensitive than the naturally-weighted THINGS BCD data cube \citep{walter08} as the baseline distribution in A configuration overlaps with those in B configuration, and we take advantage of multi-scale deconvolution to better recover the large-scale emission.
In this work, we use the broad integrated intensity (moment 0) map (Koch et al. 2026 submitted) that fully covers the spatial areas of JWST and PdBI. The \hi\ map is shown in the right panel of Figure \ref{fig:tracers}.

\subsubsection{A note on the MIRI filters}
The MIRI filters capture contributions from stellar continuum, PAH emission, dust continuum as well as emission lines and the contribution of each component also varies with filter. Starlight contributions are most significant at shorter wavelengths, mainly affecting F560W and F770W, and decrease rapidly toward longer wavelengths. Details for removing starlight from these bands are presented in \S\ref{sec:analysis}. While we do not perform any decomposition of individual filters to separate out PAHs and dust continuum, we note that each MIRI filter includes both PAHs and dust continuum emission at some level. Using spectral decomposition with PAHFIT \citep{smith07} and synthetic photometry on Spitzer-IRS spectra from the SINGS survey \citep{kenn03}, \cite{whitcomb23b} investigated the relative contribution of each component in F560W, F770W, F1000W, F1130W, and F1280W. They find that the PAH contribution to F560W, F770W, F1130W, and F1280W filters which capture the 6.2, 7.7, 11.3, 12.8\micron\ PAH features can vary between $\sim$40-80\%, using PAHFIT's definition of continuum and PAH features. The dust continuum contribution to these bands increases with wavelength, contributing $\sim$10\% at the shorter wavelength filters and increasing to $\sim$40\% in F1280W. F1280W also includes $\sim$10\% contribution from emission lines, predominantly \ion{Ne}{2} emission. F1000W is primarily dust continuum-dominated ($\sim$60\%) with PAHs contributing $\sim$30\%. At longer wavelengths, the emission captured in F1500W, F1800W, and F2100W is dominated by dust continuum. Following a similar procedure to \citet{whitcomb23b}, Hands et al. (in prep.) find $\sim$20\% PAH contribution to F1800W which captures the 17\micron\ PAH feature \citep{smith07}. They also find that F1500W contains on $\sim$70\% dust continuum, increasing to $\sim$90\% in F2100W. 
Throughout the study, we therefore refer to F560W, F770W, F1130W, and F1280W as PAH-dominated filters, while the remaining filters are considered dust-continuum dominated. It is worth noting that the distinction between PAH emission and continuum varies between different spectral decomposition techniques and the nature of the mid-IR continuum itself is not fully settled in dust models \citep{smith07}.

\subsection{Analysis}\label{sec:analysis}

To investigate how the mid-IR emission correlates with \coone, \paa\, and \hi\ in M51, we generate resolution-matched datasets by convolving the JWST MIRI maps to match the angular resolutions of the \coone\ maps with FWHM 1\farcs25 and FWHM 12\farcs0. At our adopted distance of 7.59 Mpc, these correspond to 46 pc and 442 pc. We use JWST point spread functions (PSFs) generated with {\tt webbpsf}\footnote{\href{https://webbpsf.readthedocs.io/en/latest/}{https://webbpsf.readthedocs.io/en/latest/}} and construct convolution kernels to match the MIRI PSFs to target Gaussian PSFs of 1\farcs25 and 12\farcs0 using {\tt jwst\_kernels}\footnote{\href{https://github.com/francbelf/jwst_kernels}{https://github.com/francbelf/jwst\_kernels}}. The MIRI maps are convolved following the procedure outlined in \cite{aniano11}, and then reprojected onto the astrometric grids of the \coone\ maps, with pixel sizes of 0\farcs3 and 6\farcs0, respectively. For comparison at matched resolution, the \hi\ map is also convolved to 12\farcs0 and reprojected onto the \coone\ grid to assess the mid-IR and total gas correlations on a coarser scale. 

Starlight can contribute significantly to the F560W and F770W bands and must be removed to isolate the dust and PAH emission. To obtain stellar continuum-subtracted \iffive\ and \ifseven\ maps, we use the NIRCam F300M band to estimate and remove the stellar contribution. For \ifseven, we adopt a scaling factor of I$_{F770W_{\rm starlight}}$/I$_{F300M}$ = $0.22 \pm 0.08$ from \cite{sutter24}. To determine the appropriate scaling for \iffive, we use the Code for Investigating GALaxy Emission \citep[CIGALE;][]{boqu19}, following the model setup described in \cite{sutter24}. We construct a suite of stellar population models with ages spanning 0.5–13 Gyr and a delayed star formation history, assuming a \cite{chabrier03} initial mass function and applying \cite{calzetti00} attenuation. From these models, we compute synthetic photometry in the F560W and F300M bands to derive the average stellar flux ratio, yielding I$_{F560W_{\rm starlight}}$/I$_{F300M}$ = $0.36 \pm 0.10$. Throughout the analysis, we refer to the starlight-subtracted bands as F560W$_{\rm ss}$ and F770W$_{\rm ss}$, indicating the stellar continuum-corrected I$_{F560W}$ and I$_{F770W}$ intensities, respectively. 

We also make use of the environment mask from \cite{colombo14} to investigate our correlations. The environment mask divided the \coone\ field-of-view into 3 zones: center, spiral arms, and interarm. The center is further separated into nuclear bar and a molecular ring zone. We mask the nuclear bar ($r\leq0.63$~kpc) in our analysis across all maps to avoid contamination from the bright nuclear emission, which may include AGN-related effects and high ISRFs not representative of the star-forming disk. The masked central region is indicated on the \coone\ map in Figure \ref{fig:tracers}.
We also use the SINGS narrowband continuum-subtracted \ha\ map \citep{kenn03} to construct an \hii\ region mask. \hii\ regions are detected using \texttt{HIIPhot} \citep{hiiphot} package. \texttt{HIIPhot} identifies \hii\ regions by locating emission peaks in the \ha\ map and fitting Gaussian-based models to regions around each peak to determine the region geometry. These regions are then grown outward iteratively with growth limited by the local background or overlapping neighboring regions. In the \coone\ field of view of the \ha\ map, we detect $\sim$330 \hii\ regions not including the regions detected in the center. The lower resolution of the \ha\ map (FWHM $\sim$2\farcs0) makes it suitable for region segmentation and \hii\ region identification, which would be challenging with the \paa\ map which resolves complex small-scale spatial structure.

\subsubsection{Fitting Methods}
To assess the relationship of the mid-IR bands with \coone, \paa, and \hi, we measure the Spearman rank correlation coefficient (\spr) and fit a power-law relation for each pair. For all correlations, we restrict our analysis to the region of overlap between the \coone\ and MIRI maps in addition to excluding the central region. The \coone\ field of view is overlaid on each MIRI map in Figure \ref{fig:miri}. The same spatial restriction is applied to the \paa\ and \hi\ maps, such that only pixels included in the \coone–mid-IR correlations are used when computing the \paa–mid-IR and \hi–mid-IR correlations. Additionally, we retain only pixels with S/N$\geq$5 in each MIRI filter for our analysis. 

For each combination of a mid-IR filter and resolution, pixel grid-matched tracer (\coone, \paa, and \hi), we compute \spr\ and its uncertainties using a combination of non-parametric bootstrapping and Monte Carlo error propagation. In each of 500 iterations, we resample the dataset with replacement and perturb each point by adding Gaussian noise based on its uncertainties. We then estimate \spr\ for each realization and adopt the median and standard deviation of the resulting distribution as the final correlation coefficient and its associated uncertainty. We note that for estimating \spr\ at $\sim$40~pc scale, the mid-IR, \coone, and \paa\ maps are resampled to a 1\farcs25 pixel grid, corresponding approximately to one pixel per beam, in order to reduce the impact of spatial covariance introduced by the original 0\farcs3 pixel sampling.

To determine the best-fit power-law, with the mid-IR intensities as the independent variable, we bin the data and perform linear regression in log–log space using the {\tt linmix} Markov Chain Monte Carlo (MCMC) algorithm \citep{kelly07}. The data are binned uniformly in $\log_{10}(I_{\rm MIR})$ with a fixed bin width of 0.1 dex. For each bin, we compute the 16th, 50th, and 84th percentiles of the tracer intensity distribution. The median (50th percentile) values of each bin are input to {\tt linmix}. Additionally, individual pixel uncertainties within each bin are propagated as $\sqrt{\sum_i\sigma_{i}^2}/N$ where $\sigma_i$ are 1$\sigma$ uncertainties in the mid-IR or our tracers and N is the number of pixels per bin and passed to {\tt linmix}. 
The linear regression is modeled as $\log_{10} y=m(\log_{10}x - x_0) +  b$, where pivot $x_0$ is the median $x$ to minimize covariance between slope $m$, and intercept $b$. This approach accounts for errors in both variables.
We use 100 MCMC chains in {\tt linmix}, and the best-fit slope and its uncertainty are taken as the median and the 16–84 percentile range, respectively, of the posterior distribution. 
We note that we fit the correlation statistics and regression over the range $-0.5\leq\log$~I$_{MIR}\leq2$ (or I$_{MIR}$ = 0.31--100 MJy/sr) in all bands except F560W$_{\rm ss}$. For F560W$_{\rm ss}$, the intensities are approximately 0.4 dex lower than other bands, and the linear regression is fit between 0.1-30 MJy/sr. These cuts exclude both low surface brightness pixels where uncertainties in anchoring may dominate and extremely high star-forming environments that represent distinct ISM conditions from the disk. 

We also independently estimate the intrinsic scatter, $\sigma_{\rm int}$ from the full, unbinned data and best-fit $m$ and $b$ estimated using {\tt linmix}. The total variance for each pixel $i$ in the convolved maps is defined as, $\sigma_{{\rm tot},i}^2 = \sigma_{y,i}^2 + m^2 \sigma_{x,i}^2 + \sigma_{\rm int}^2$, where $\sigma_{x,i}$ and $\sigma_{y,i}$ are the 1$\sigma$ uncertainties in the mid-IR data and our tracers, respectively. The intrinsic scatter is estimated by minimizing the reduced $\chi^2$ of the residuals, $\chi^2_{\rm red} = \sum_i \Delta_i^2 / \sigma_{{\rm tot},i}^2 /(N-2)$, where $\Delta_i = y_i - (m x_i + b)$, and $N$ is the total number of valid pixels in the unbinned data.

\begin{figure*}[!htb]
\centering
\includegraphics[trim =0cm 0cm 0cm 0cm, clip,scale=0.24]{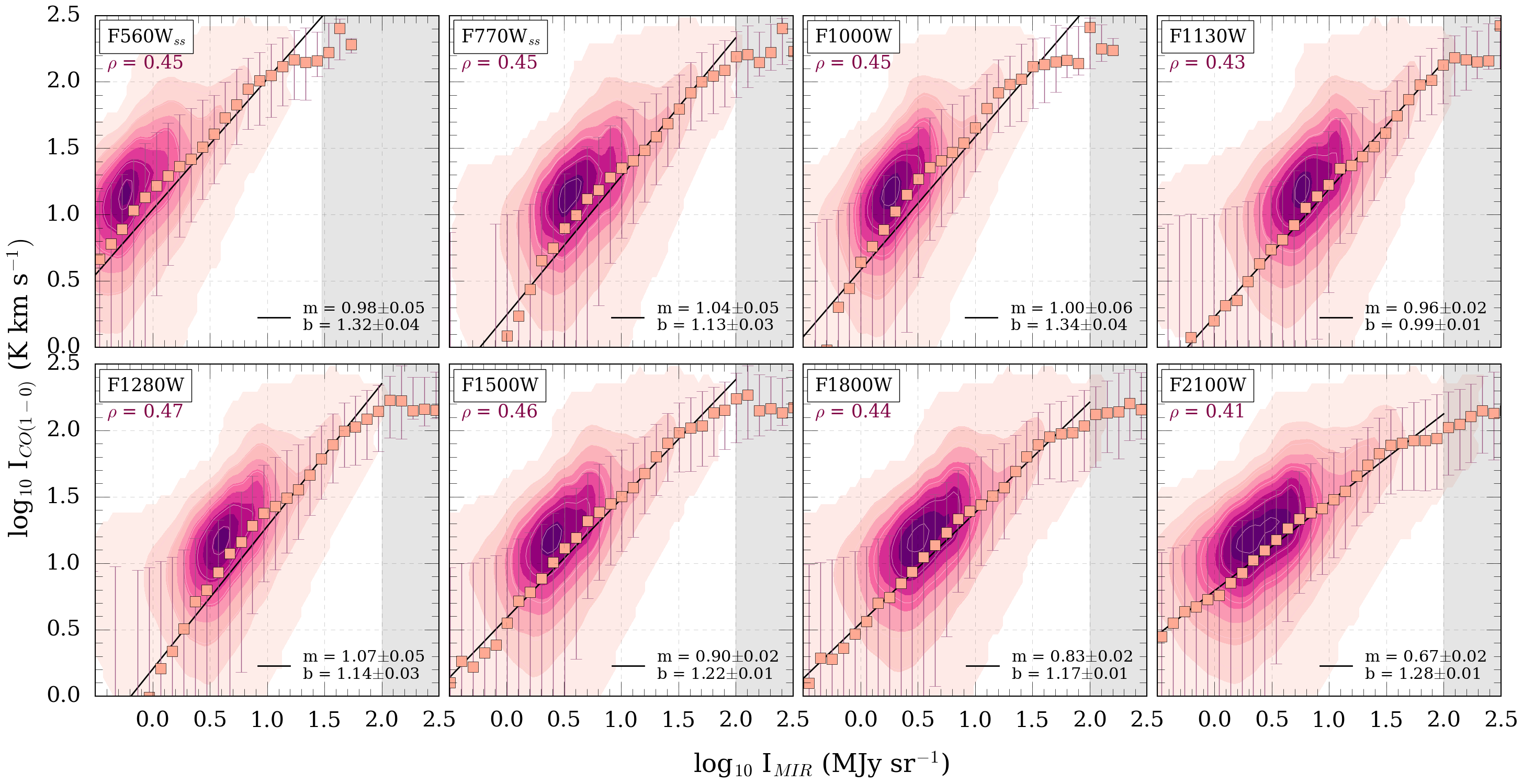}
\caption{\coone\ intensity, I$_{\rm CO}$ as a function of I$_{\rm MIR}$ at 1.25\arcsec\ resolution for a given MIRI band. The shaded contours represent data density contours enclosing the densest 15, 25, 50, 75, and 95\% of the data points. The MIRI filter and Spearman rank correlation coefficient, \spr\ are noted on the top left corner of each panel. Pink squares and error bars represent the median and the 16-84\% range of \coone\ intensity after binning the mid-IR intensity. The black line shows the line of best-fit for the binned data with the slope $m$, and intercept $b$ noted in the bottom right corner. The shaded region in each panel is excluded from any statistical analysis. } \label{fig:co10-miri}
\end{figure*}

\section{Results}\label{sec:results}
\subsection{Molecular Gas: CO(1-0) correlates near-linearly with PAH emission and sub-linearly with dust continuum emission} \label{sec:co-mir}

Figure~\ref{fig:co10-miri} shows \coone\ intensity, I$_{\rm CO(1-0)}$ as a function of mid-IR intensities, I$_{\rm MIR}$ at $\sim$40~pc resolution. We observe moderate monotonic correlations with Spearman rank coefficients, \spr\ of $\sim$0.42--0.47 across all filters. The CO and mid-IR distribution are well described by a power law relation, I$_{CO(1-0)}\propto$~I$_{\rm MIR}^{m_{CO}}$, with the best fit slope $m_{CO}$ ranging from 0.67 to 1.04. This indicates an approximately linear to mildly sublinear relationship over $\sim$two orders of magnitude in both CO and mid-IR intensity. The observed rms scatter about the best-fit relation spans 0.46--0.50 dex with intrinsic scatter of $\sigma_{\rm int}\approx$~0.31--0.40 dex. A summary of the correlation coefficients, best-fit parameters, and CO-to-mid-IR ratios is provided in Table~\ref{tab:corr-miri}. 

\begin{figure*}[!t]
\centering
\includegraphics[trim =0cm 0cm 0cm 0cm, clip,scale=0.31]{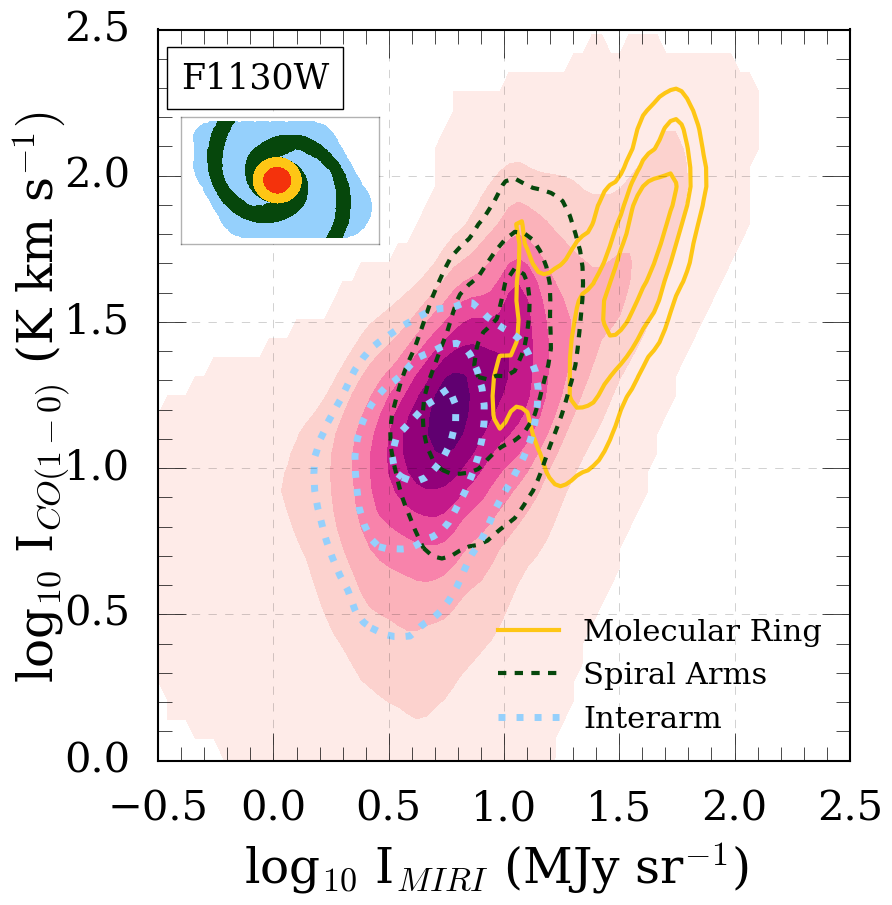}
\includegraphics[trim =0cm 0cm 0cm 0cm, clip,scale=0.335]{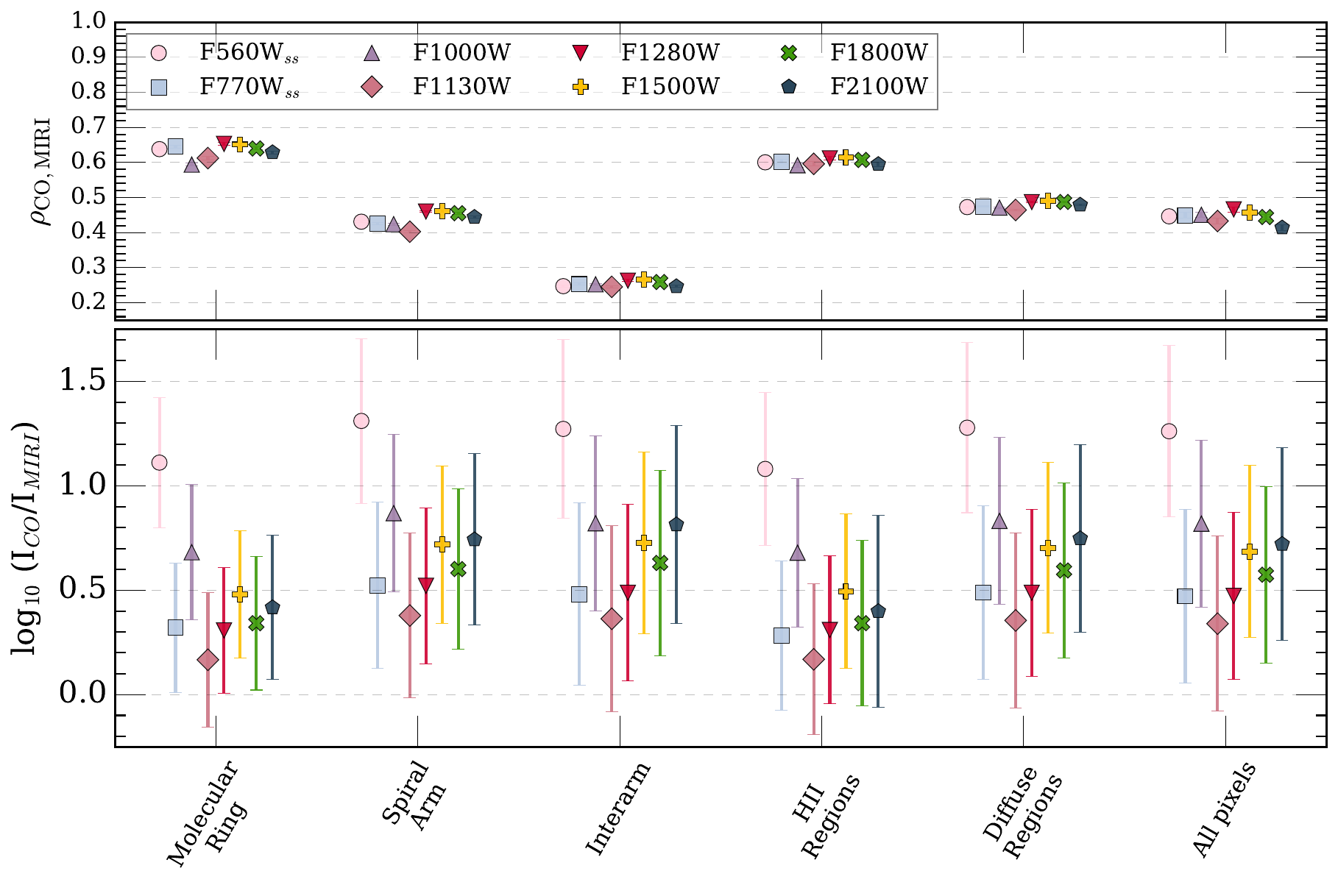}
\caption{{\it Left:} $I_{\rm CO}$ as a function of \ifeleven, with contours highlighting the distribution of pixels associated with distinct environments in M51: the molecular ring, spiral arms, and interarm regions. These environments are defined by the mask from \citet{colombo14}, shown in the inset. {\it Right:} Variation in \spr\ (top) and CO-to-mid-IR median ratios (bottom) for each CO–mid-IR filter pair across environments. Each mid-IR filter is represented by a unique symbol within each environment. Error bars denote the uncertainties associated with each quantity. The filters are arbitrarily spaced along the horizontal axis in order of increasing wavelength. }\label{fig:env}
\end{figure*}

We observe systematic filter-to-filter variations in the power law slopes of the I$_{\rm CO(1-0)}$-I$_{\rm MIR}$ relations. The PAH dominated bands, F560W$_{\rm ss}$, F770W$_{\rm ss}$, F1130W, F1280W exhibit a near-linear scaling ($m_{CO}\sim0.98-1.04$). We note that we tested the correlations of \coone\ and F560W$_{\rm ss}$, F770W$_{\rm ss}$ where the starlight subtraction was performed using NIRcam F200W following similar procedure noted in \S\ref{sec:analysis} instead of F300M and we find the derived statistics to be comparable.   
Interestingly, the F1000W filter also shows a slope of $\sim 1$ similar to the PAH-dominated filters. The F1500W filter also shows $m_{CO}\sim0.9$, with the slopes becoming progressively shallower for the longer wavelength F1800W and F2100W filters  where the emission is increasingly dominated by dust continuum.  
As the F1800W filter captures the 17\micron\ PAH feature, we attempted to separate PAHs and dust continuum in the band using recipes calibrated based on Spitzer spectroscopy decomposed with PAHFIT (Hands et al. 2026 in prep). We, however, do not find the power law indices for F1800W$_{\rm PAH}$ ($m_{CO}$=0.85) to differ significantly from that derived using the full F1800W filter.
The shallowest relation is observed for F2100W with $m_{CO}=0.67$. While the power-law scaling between I$_{\rm CO(1-0)}$-I$_{\rm MIR}$ shows a clear wavelength dependence, flattening towards longer wavelengths, all mid-IR filters show similar correlation strengths and $\sigma_{\rm int}$.  

We also investigate $I_{\rm CO}/I_{\rm MIR}$ ratios across distinct environments within M51 and further explore the wavelength-dependent variations in the power-law scaling. We do not fit power-law relations separately within individual environments as each subset spans a narrow intensity range (typically $\lesssim$1 dex), resulting in poorly constrained slopes with substantially larger intrinsic scatter. Instead, we compare CO-to-mid-IR ratios, which provide a more robust measure of environmental variations. Figure \ref{fig:env} shows the distribution, correlation strengths and CO-mid-IR ratios in different environments within the disk of M51. The left panel shows the I$_{\rm CO(1-0)}$ and I$_{\rm F1130W}$ distribution with smoothed colored contours highlighting various environments in Figure \ref{fig:env}.

The Spearman rank correlation coefficients, \spr, vary modestly between environments but remain nearly identical ($\Delta\rho=\pm0.1$) within a given environment across all filters. The correlation between $I_{\rm CO}$ and I$_{\rm MIR}$ is weakest in the interarm regions ($\rho\sim0.25$), where gas surface densities are low and mid-IR emission is dominated by heating from the diffuse radiation field. In the spiral arms, the correlation is moderate ($\rho\approx0.45$–0.5), reflecting increased gas surface densities and ongoing star formation that boosts both $I_{\rm CO}$ and $I_{\rm MIR}$. Interestingly, the diffuse regions also show a $\rho\approx0.5$, but these encompass all areas outside \hii\ regions and are therefore distinct from the interarm regions.
The strongest correlation is observed in the molecular ring and within \hii\ regions ($\rho\approx0.62$–0.63), where both molecular gas and star formation activity are concentrated. 

Overall, the $\log_{10} (I_{\rm CO}/I_{\rm MIR})$ ratios for a given mid-IR filter vary by $\lesssim 0.5$ dex across the different environments in M51. The higher $\log_{10} (I_{\rm CO}/I_{\rm F560W_{ss}})$ ratios primarily reflect the lower absolute surface brightness of the starlight-subtracted F560W filter relative to other MIRI filters. For the PAH-dominated bands, $I_{\rm CO}/I_{\rm MIR}$ typically ranges from $\sim$2.1-4.8 in the molecular ring, and $\sim$3.3-7.4 in the spiral arms, with comparable values in the diffuse and interarm regions. The low variation in the CO-to-mid-IR ratios for the PAH-dominated bands is consistent with the near-linear scaling seen in Figure \ref{fig:co10-miri}, indicating that PAHs remain well-mixed with gas and largely trace the bulk molecular gas distribution. 

\begin{figure*}[!ht]
\centering
\includegraphics[trim =0cm 0cm 0cm 0cm, clip,scale=0.24]{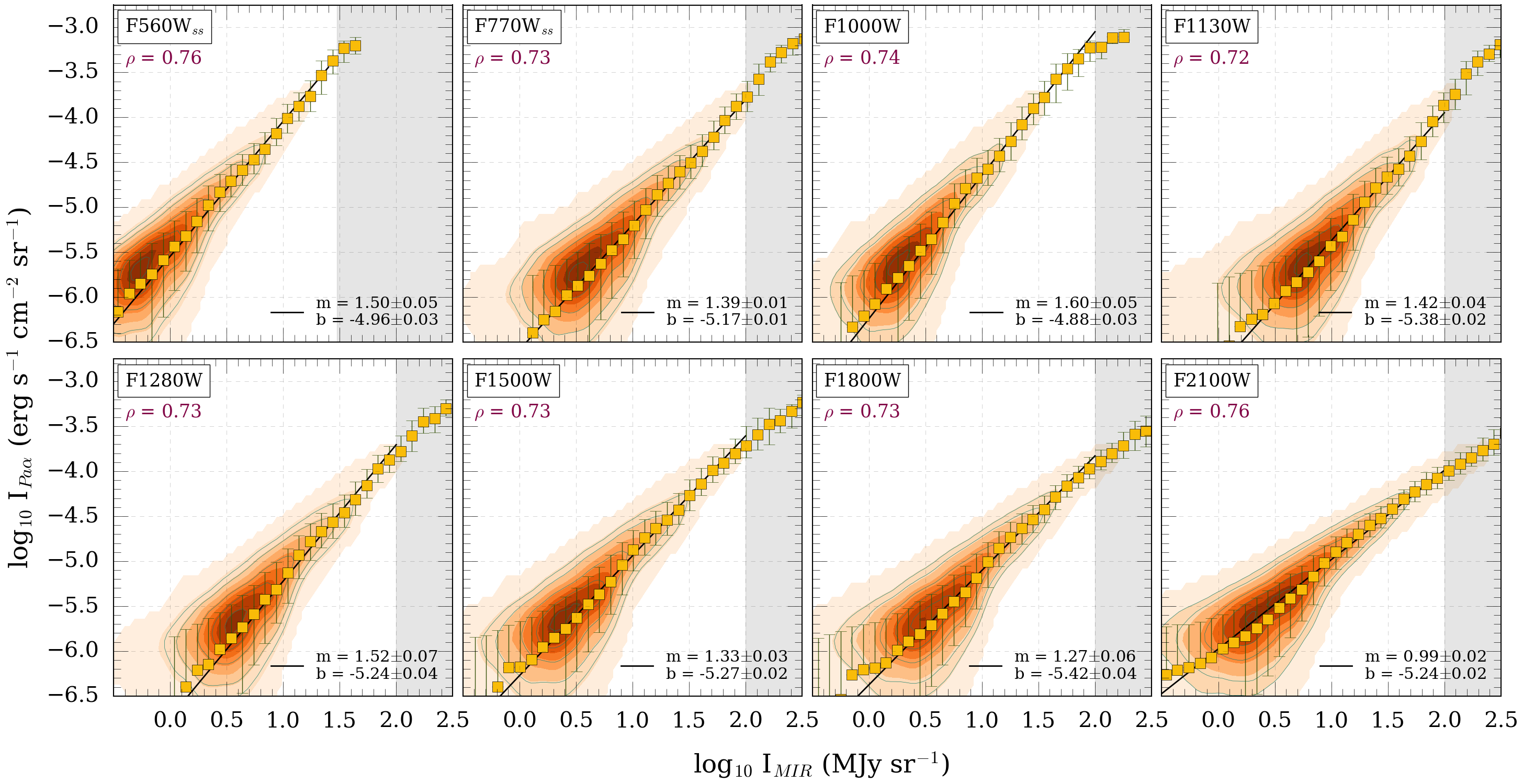}
\caption{\paa\ intensity as a function of mid-IR intensity at 1.25\arcsec\ resolution for a given MIRI band. The shaded contours represent data density contours enclosing the densest 15, 25, 50, 75, and 95\% of the data points. The MIRI filter and Spearman rank correlation coefficient, \spr\ are noted on the top left corner of each panel. Orange squares and error bars represent the median and the 16-84\% range of \paa\ intensity after binning the mid-IR intensity. The black line shows the line of best-fit for the binned data with the slope, $m$ and intercept, $b$ noted in the bottom right corner. The shaded region in each panel is excluded from any statistical analysis.} \label{fig:pa-miri}
\end{figure*}

The dust continuum filters, particularly F2100W, show a systematic environmental dependence relative to the PAH-dominated filters. In high-$U$, star-forming environments in M51, like the molecular ring, and \hii\ regions, the $I_{\rm CO}$/$I_{\rm F2100W}$ ratios are lower by $\sim0.4$ dex than for other environments, reflecting the increased contribution of mid-IR dust continuum emission heated by intense radiation fields. In the spiral arms, interarm, and diffuse regions, the $I_{\rm CO}$/$I_{\rm F2100W}$ ratios are higher ($\sim$2.5–7.0), exceeding the corresponding CO-to-PAH-dominated filter ratios. This behavior can explain the wavelength-dependent flattening of the CO-to-mid-IR power law slopes seen in Figure \ref{fig:co10-miri}. While the dust continuum-dominated filters remain correlated with \coone, they are increasingly weighted toward high $U$, bright star-forming regions. As a result, at a given gas column, $I_{MIR}$ increases with increasing $U$, thereby producing shallower slopes. The PAH-dominated filters, on the other hand, maintain near-linear scaling as they remain well-mixed with gas across environments. Further, PAH emission may also be suppressed in high $U$ environments due to destruction \citep{madden06, egorov23}, reducing the contribution of high $U$ regions in the CO-mid-IR scaling relation for PAH-dominated filters. We discuss these in \S\ref{sec:discuss}. 

\subsection{Ionized Gas: Pa$\alpha$ correlates super-linearly with PAH emission and near-linearly with dust continuum emission} \label{sec:pa-mir}

While we find that all mid-IR filters correlate with the molecular gas distribution, increasing contributions from high $U$ regions to dust heating at longer wavelengths are evident in the CO-mid-IR scaling relations. UV photons from massive stars power much of the local radiation field and heat the dust grains \citep{draine07, aniano20}. To directly investigate the connection between mid-IR emission and recent star formation, we compare I$_{\rm MIR}$ to \paa\ intensity (I$_{\rm Pa\alpha}$) in this section. We note that we do not correct I$_{\rm Pa\alpha}$ for extinction, as \paa\ is intrinsically much less affected by dust than \ha. Young stellar clusters in M51 show typical color excesses $E(B\!-\!V)\lesssim1$, with a tail toward more embedded regions reaching $E(B\!-\!V)\gtrsim4$ \citep{pedrini25}. Additionally, at our resolution of $\sim$40~pc, each beam would encompass multiple stellar and diffuse regions with a range of extinction, such that the contribution from highly embedded sources is spatially diluted within a mix of lower-extinction environments, reducing their impact on the integrated \paa\ signal. This implies that attenuation at \paa\ remains modest for the bulk of the young massive star population and I$_{\rm Pa\alpha}$, thus, provides a low extinction measure of the ionizing photon rate and the young massive star population.

Figure \ref{fig:pa-miri} shows I$_{\rm Pa\alpha}$ as a function of  I$_{\rm MIR}$ for all eight bands at $\sim$40 pc scale. We find a strong monotonic correlation  with \spr$\sim$0.73--0.76. The \paa\ and mid-IR intensities are also described by a single power law relation, I$_{\rm Pa\alpha}\propto$~I$_{\rm MIR}^{m_{Pa\alpha}}$ with best-fit power law slopes ranging from $\sim$0.99--1.60. The median rms scatter about the best-fit relation is $\sigma\sim0.45$–0.49 dex with $\sigma_{\rm int}\sim0.38$-0.47. These are detailed in Table \ref{tab:corr-miri}. 
 
Similar to the CO-to-mid-IR relations, we observe wavelength-dependent variations in the power-law slopes between I$_{\rm Pa\alpha}$ and I$_{\rm MIR}$ while \spr\ stays relatively the same for all filters. While all mid-IR filters show comparable correlation with star formation, the power law slopes between I$_{\rm Pa\alpha}$ and I$_{\rm MIR}$ are super-linear ($m\gtrsim$1.2) for all filters except F2100W. The slopes also flatten with wavelength for the dust continuum filters with F2100W showing the shallowest and a near-linear relation with \paa. Conversely, in the shorter-wavelength and PAH-dominated filters, we find that \paa\ emission increases more compared to $I_{MIR}$ in regions of strong ionizing radiation, suggesting that these bands may receive a larger fractional contribution from lower-intensity radiation fields or diffuse ISM heating.
\setlength{\tabcolsep}{8pt} 
\startlongtable
\begin{deluxetable*}{lllllllll}
\tablewidth{\textwidth}
\tablecaption{Statistics for \coone\ and \paa\ correlations with mid-IR intensities at $\sim40$~pc scale. }\label{tab:corr-miri}
\tablehead{
\colhead{Filter} & \colhead{N$_{\rm pix}$} & \colhead{\spr} & \colhead{m} & \colhead{b} & \colhead{x$_0$} & \colhead{$\log_{10}\!\left(\frac{I_Y}{I_{\mathrm{\rm MIR}}}\right)$} &\colhead{$\sigma_{\rm int}$} & \colhead{$\sigma_{\rm obs}$}}
\colnumbers
\startdata
\multicolumn{8}{c}{\bf{\coone}} \\
\tableline
F560W$_{\rm ss}$ & 245369 & $0.45 \pm 0.01$ & $0.98 \pm 0.052$ & $1.32 \pm 0.035$ & 0.28 & $1.26 \pm 0.411$ & 0.39 & 0.50 \\
F770W$_{\rm ss}$ & 246861 & $0.45 \pm 0.01$ & $1.05 \pm 0.047$ & $1.13 \pm 0.027$ & 0.85 & $0.47 \pm 0.417$ & 0.39 & 0.50 \\
F1000W & 253037 & $0.45 \pm 0.01$ & $1.00 \pm 0.059$ & $1.34 \pm 0.040$ & 0.75 & $0.82 \pm 0.400$ & 0.40 & 0.50 \\
F1130W & 244716 & $0.43 \pm 0.01$ & $0.96 \pm 0.015$ & $0.99 \pm 0.010$ & 0.79 & $0.34 \pm 0.420$ & 0.36 & 0.49 \\
F1280W & 253021 & $0.47 \pm 0.01$ & $1.08 \pm 0.047$ & $1.14 \pm 0.029$ & 0.87 & $0.47 \pm 0.401$ & 0.40 & 0.50 \\
F1500W & 243815 & $0.46 \pm 0.01$ & $0.90 \pm 0.020$ & $1.22 \pm 0.015$ & 0.71 & $0.69 \pm 0.412$ & 0.35 & 0.48 \\
F1800W & 237789 & $0.44 \pm 0.01$ & $0.83 \pm 0.017$ & $1.17 \pm 0.012$ & 0.75 & $0.57 \pm 0.423$ & 0.34 & 0.47 \\
F2100W & 224690 & $0.41 \pm 0.01$ & $0.67 \pm 0.015$ & $1.28 \pm 0.011$ & 0.74 & $0.72 \pm 0.461$ & 0.31 & 0.46 \\
\tableline
\multicolumn{8}{c}{\bf{\paa}}\\
\tableline
F560W$_{\rm ss}$ & 302050 & $0.76 \pm 0.005$ & $1.49 \pm 0.039$ & $-4.82 \pm 0.027$ & 0.49 & $-5.40 \pm 0.325$ & 0.42 & 0.45 \\
F770W$_{\rm ss}$ & 313444 & $0.73 \pm 0.005$ & $1.39 \pm 0.014$ & $-5.17 \pm 0.008$ & 1.02 & $-6.19 \pm 0.344$ & 0.42 & 0.45 \\
F1000W & 310882 & $0.74\pm 0.005$ & $1.60 \pm 0.046$ & $-4.88 \pm 0.031$ & 0.85 & $-5.85 \pm 0.348$ & 0.43 & 0.46 \\
F1130W & 296464 & $0.72 \pm 0.005$ & $1.42 \pm 0.040$ & $-5.38 \pm 0.025$ & 0.99 & $-6.32 \pm 0.361$ & 0.47 & 0.49 \\
F1280W & 311169 & $0.73 \pm 0.005$ & $1.52 \pm 0.067$ & $-5.24 \pm 0.039$ & 0.99 & $-6.19 \pm 0.347$ & 0.45 & 0.48 \\
F1500W & 294225 & $0.73 \pm 0.005$ & $1.33 \pm 0.031$ & $-5.27 \pm 0.022$ & 0.76 & $-5.97 \pm 0.346$ & 0.45 & 0.47 \\
F1800W & 284943 & $0.73 \pm 0.005$ & $1.27 \pm 0.055$ & $-5.42 \pm 0.040$ & 0.75 & $-6.08 \pm 0.332$ & 0.44 & 0.47 \\
F2100W & 265057 & $0.76 \pm 0.005$ & $0.99 \pm 0.021$ & $-5.24 \pm 0.015$ & 0.74 & $-5.95 \pm 0.303$ & 0.38 & 0.41 \\
\enddata
\tablecomments{Column 1: JWST MIRI band. Subscript ``ss" denotes starlight subtraction (see \S\ref{sec:analysis}). Column 2: Number of pixels entering the analysis. Column 3: Spearman rank correlation coefficient, \spr\ between \coone\ (or \paa) and the respective mid-IR band intensities. Column 4-6: Slope, intercept, and pivot value describing the best fit power law relation between \coone\ (or \paa) and the respective mid-IR band. Column 7: Logarithmic CO-to-mid-IR and \paa-to-mid-IR ratios. Column 8-9: Intrinsic and observed RMS scatter in the data about the best fit line, noted in dex. }
\end{deluxetable*}

We note that a fraction of the \paa\ emission can arise from leaked ionizing photons that contribute to the diffuse ionized gas (DIG). In M51, \cite{calz25} find $\sim$50\% of the ionizing photons are in DIG outside the \hii\ regions. Since our footprint is restricted to the \coone\ map and we subtracted a gradient to account for background variation, thus reducing sensitivity to extended low surface brightness emission, we do not separate \hii\ regions from DIG in the \paa\ map. Given the presence of DIG, the \paa-mid-IR correlations may also reflect spatially extended UV heating rather than just contributions from star-forming regions. 

 
We show the comparison between \paa–mid-IR and CO–mid-IR power-law slopes estimated at $\sim$40~pc scale in Figure \ref{fig:cosf_slopes}. These clearly highlights how different mid-IR bands relate differently to gas column versus local radiation fields. The PAH-dominated F560W$_{\rm ss}$, F770W$_{\rm ss}$, F1130W, and F1280W bands show m$_{\rm CO}$ slopes close to unity, indicating that PAH emission rises roughly in proportion to molecular gas column in M51. In contrast, the same PAH bands show super-linear m$_{\rm Pa\alpha}\gtrsim1.5$, demonstrating that, as the ionizing photon rate increases, PAH emission does not keep up, possibly due to PAH destruction in \hii\ regions. This diverging behavior suggests that emission in the PAH-dominated filters arises mainly from regions exposed to relatively uniform $U$ where PAHs are well-mixed with gas. 

At longer wavelengths, the dust continuum-dominated bands (F1500W, F1800W, F2100W) tilt the other way. Their m$_{\rm CO}$ become progressively sublinear with shallower slopes, while their m$_{\rm Pa\alpha}$ move toward unity. This reflects that emission in these filters is increasingly arising from dust heating in intense radiation fields, associated with star formation while also tracing the underlying gas column (and dust mass). F2100W, in particular, exhibits the shallowest CO-mid-IR and a near-linear \paa-mid-IR scaling, showing that its emission is dominated by bright star-forming regions, scaling almost directly with the ionizing photon rate and therefore acting as a reliable tracer of star formation activity (as $\Sigma_{\rm SFR}\propto$~I$_{\rm Pa\alpha}$) at these scales. 

We assessed the robustness of the measured CO–mid-IR and \paa–mid-IR power-law slopes by repeating the analysis using different binning schemes, percentile-based intensity cuts, and a common spatial mask for all MIRI filters. The tests using different binning schemes and a common spatial mask result in changes in the fitted slopes of $\lesssim 0.02$ for all filters, while the percentile-based intensity cuts result in slope variations of $\lesssim 0.2$. In all tests, the wavelength dependence of the slopes remains unchanged.
We also conducted the regression analysis on maps resampled to 1\farcs25 grid without binning. These fits yield systematically shallower values compared to the fiducial values, with $m_{CO}$ varying by $\lesssim0.2$ and $m_{Pa\alpha}$ by $\lesssim0.4$, likely resulting from individual pixels being weighted equally. 

Despite the different physical drivers of mid-IR emission seen in the 
CO-mid-IR and Pa$\alpha$–mid-IR scaling relations, both \coone\ and \paa\ show strong correlations across all mid-IR bands. While the Pa$\alpha$–mid-IR relations exhibit higher $\rho\sim0.7$ than the CO–mid-IR relations with $\rho\sim0.5$, the intrinsic scatter in both relations are similar ($\sigma_{\rm int}\sim0.3$–0.4 dex). The lower $\rho$ values in the CO–mid-IR relations could be attributed more to noise in the \coone\ map than a fundamentally weaker correlation between molecular gas and mid-IR emission. At $\sim$40~pc scales, mid-IR traces both molecular gas and recent star formation with the scaling relations encoding how each mid-IR filter partitions its sensitivity between the gas column and heating by local star-forming regions. We discuss this in \S\ref{sec:discuss}.

\begin{figure}[!t]
\centering
\includegraphics[trim = 0cm 0cm 0cm 0cm, clip,scale=0.35]{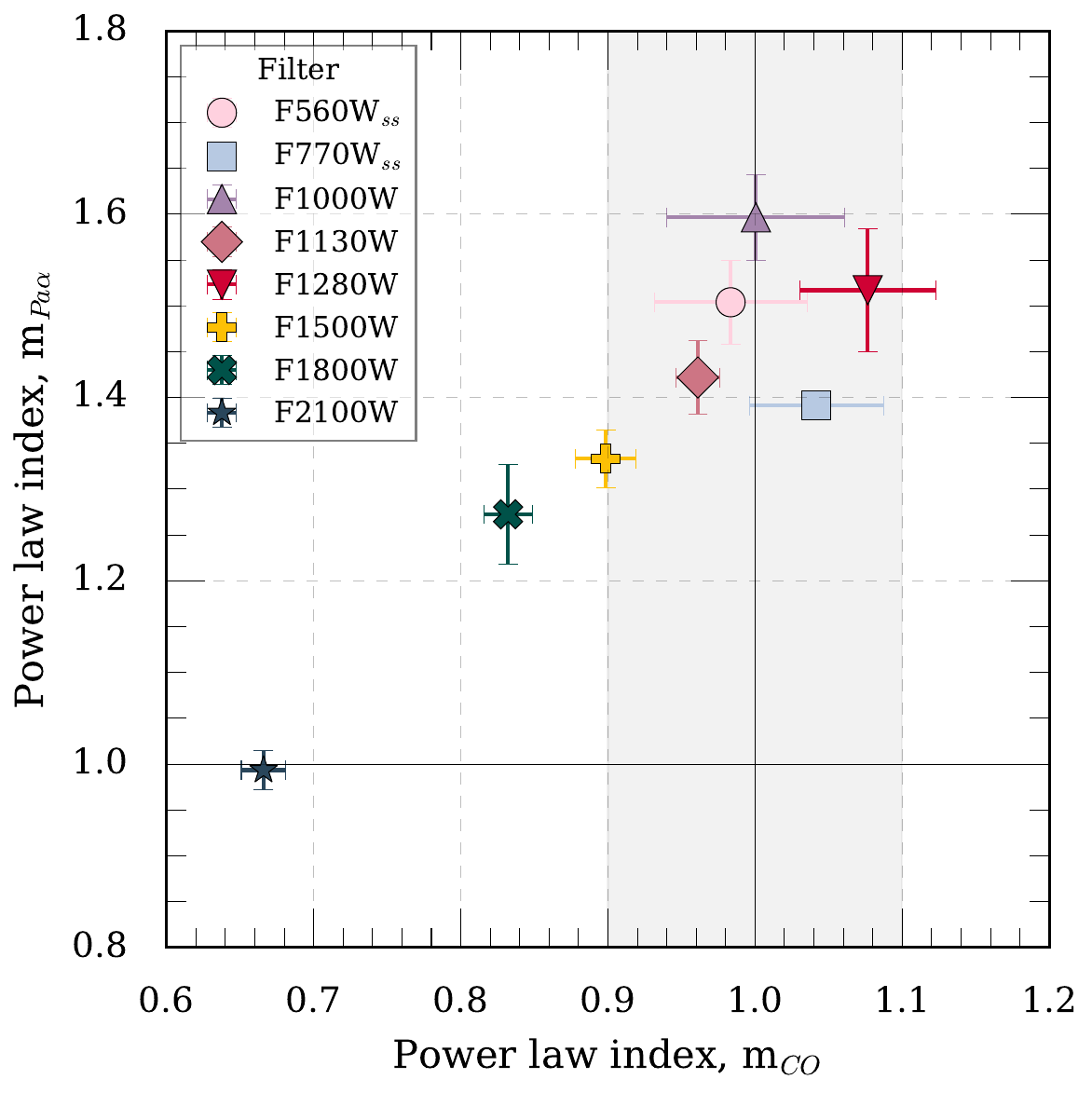}
\caption{Comparison between I$_{\rm CO}$-I$_{\rm MIR}$ slopes and I$_{\rm Pa\alpha}$-I$_{\rm MIR}$ power law slopes. Each symbol represents a mid-IR band. The solid lines indicate a slope of 1. The shaded area highlights the near-linear regime for the I$_{\rm CO}$-I$_{\rm MIR}$ relation where the PAH-dominated bands reside.} \label{fig:cosf_slopes}
\end{figure}

\begin{figure*}[!htb]
\centering
\includegraphics[trim =0cm 0cm 0cm 0cm, clip,scale=0.22]{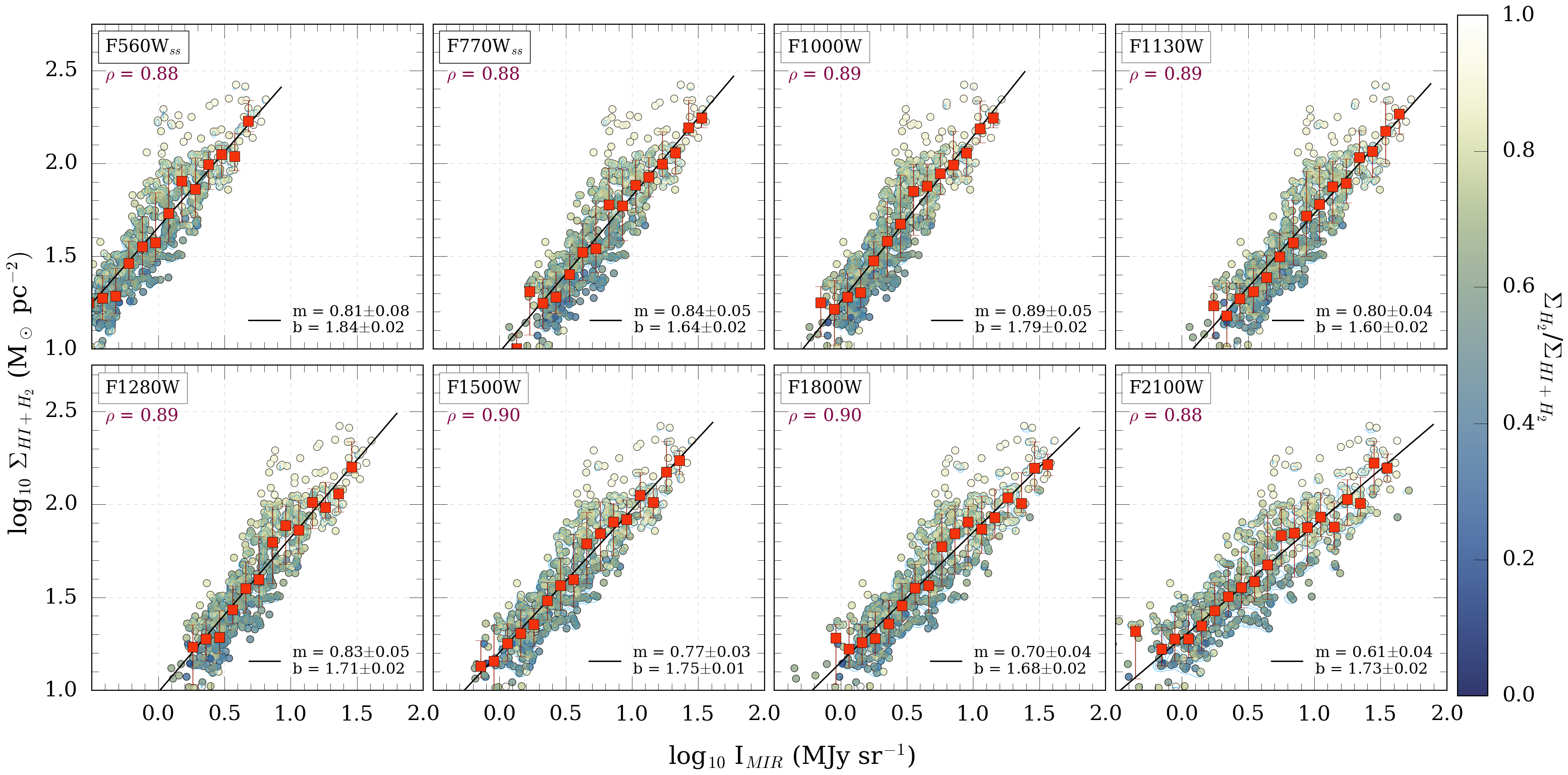}
\caption{\ion{H}{1}+\ce{H2} surface density as a function of  mid-IR intensity  at 12\arcsec\ resolution. The JWST mid-IR filter is noted on the top left corner in each panel. Red squares and error bars represent the median and the 16-84\% range of $\Sigma_{\rm HI+H_2}$ after binning the mid-IR intensity. The black line shows the best-fit power law for the binned data. Scatter point colors show the molecular gas fraction.} \label{fig:gas-miri}
\end{figure*}

\subsection{$\Sigma_{\rm gas}$: Testing linear scaling of mid-IR emission with \hi+\ce{H2} surface density}

The CO-mid-IR correlations presented in \S\ref{sec:co-mir} motivate a more direct test of the fundamental assumption underlying Equation \ref{eq:1}: mid-IR emission scales linearly with the total dust-bearing gas column, i.e., I$_{\rm MIR}\propto  \Sigma_{\rm gas}=\Sigma_{\rm HI + H_2}$, under the assumption of roughly constant D/G ratio and $U$. This expectation is well posed in M51, where mid-IR bands show tight correlation with \coone\ at $\sim40$~pc in the molecular-gas dominated part of the disk. We now construct a total gas map ($\Sigma_{\rm HI+H_2}$) using $\Sigma_{\rm HI}$ and $\Sigma_{\rm H_2}$, where $\Sigma_{\rm H_2}$ is inferred from \coone\ with $\alpha_{\rm CO} = 2.4$~M$_\odot$~pc$^{-2}$/K~km s$^{-1}$ \citep[from][]{denbrok25}. Due to the coarser angular resolution of the \hi\ map compared to \coone, the I$_{\rm MIR}$ and $\Sigma_{\rm HI+H_2}$ relations are examined at $\sim$440~pc.

In Figure \ref{fig:gas-miri}, we show the correlations between I$_{\rm MIR}$ and $\Sigma_{\rm HI+H_2}$. All mid-IR bands exhibit strong monotonic relations with $\Sigma_{\rm HI+H_2}$ with \spr$\sim$0.87--0.90. Power-law fits yield slopes in the range of $\sim$0.61--0.84 and an rms scatter of 0.11-0.13 dex. For comparison, we estimate CO-mid-IR and \paa-mid-IR slopes at $\sim$440~pc. Notably, the $\Sigma_{\rm HI+H_2}$-mid-IR slopes are comparable to CO-mid-IR scaling at similar scale and \spr\ values are comparable to both both CO-mid-IR and \paa-mid-IR correlations. The correlations and best fit scaling relations are tabulated in Table \ref{tab:allgas}. Once again, we note the PAH-dominated filters show steeper, near-linear relations with the total-gas distribution, while the long-wavelength continuum filters appear slightly shallower. Taken together, the CO–mid-IR and $\Sigma_{\rm gas}$–mid-IR relations at $\sim$440 pc resolution present a coherent picture in which mid-IR emission traces the total dust-bearing gas column, with similar wavelength-dependent variation in its sensitivity to local heating as seen in \S\ref{sec:co-mir}. The modest sub-linearity observed in all bands at this resolution likely reflects averaging bright star forming regions with diffuse emission within each $\sim$440 pc beam, and blending multiple environments spanning a range of radiation fields and gas phases. We note a $\sim$10\%  suppression in our slopes at $\sim$440 pc relative to measurements on cloud scales at $\sim$40 pc of CO-mid-IR. Nevertheless, the uniformly high Spearman coefficients demonstrate that mid-IR emission remains an effective tracer of the underlying gas distribution. 

\subsection{Atomic gas: H~{\small{I}} also correlates near-linearly with mid-IR emission}\label{sec:hi}

Our analysis above shows that mid-IR emission correlates strongly with the total dust-bearing gas column, $\Sigma_{\rm gas} = \Sigma_{\rm HI} + \Sigma_{\rm H_2}$, across M51’s disk. However, dust grains responsible for mid-IR emission may not be confined to molecular-dominated regions in the ISM but are well mixed with both atomic and molecular gas \citep{bohlin78, sandstrom13, galliano18}. The $\Sigma_{\rm gas}$ term in Equation \ref{eq:1} also contains contributions from atomic gas. It is natural to ask whether the mid-IR emission tracks the atomic component, and if so, whether that relation is similar to the one seen for \coone. In the diffuse ISM, where atomic gas dominates, the same framework that explains the CO-to-mid-IR relation also predicts that mid-IR emission should respond to atomic gas \citep{sandstrom23}.   

\begin{figure*}[!ht]
\centering
\includegraphics[trim =0cm 0cm 0cm 0cm, clip,scale=0.22]{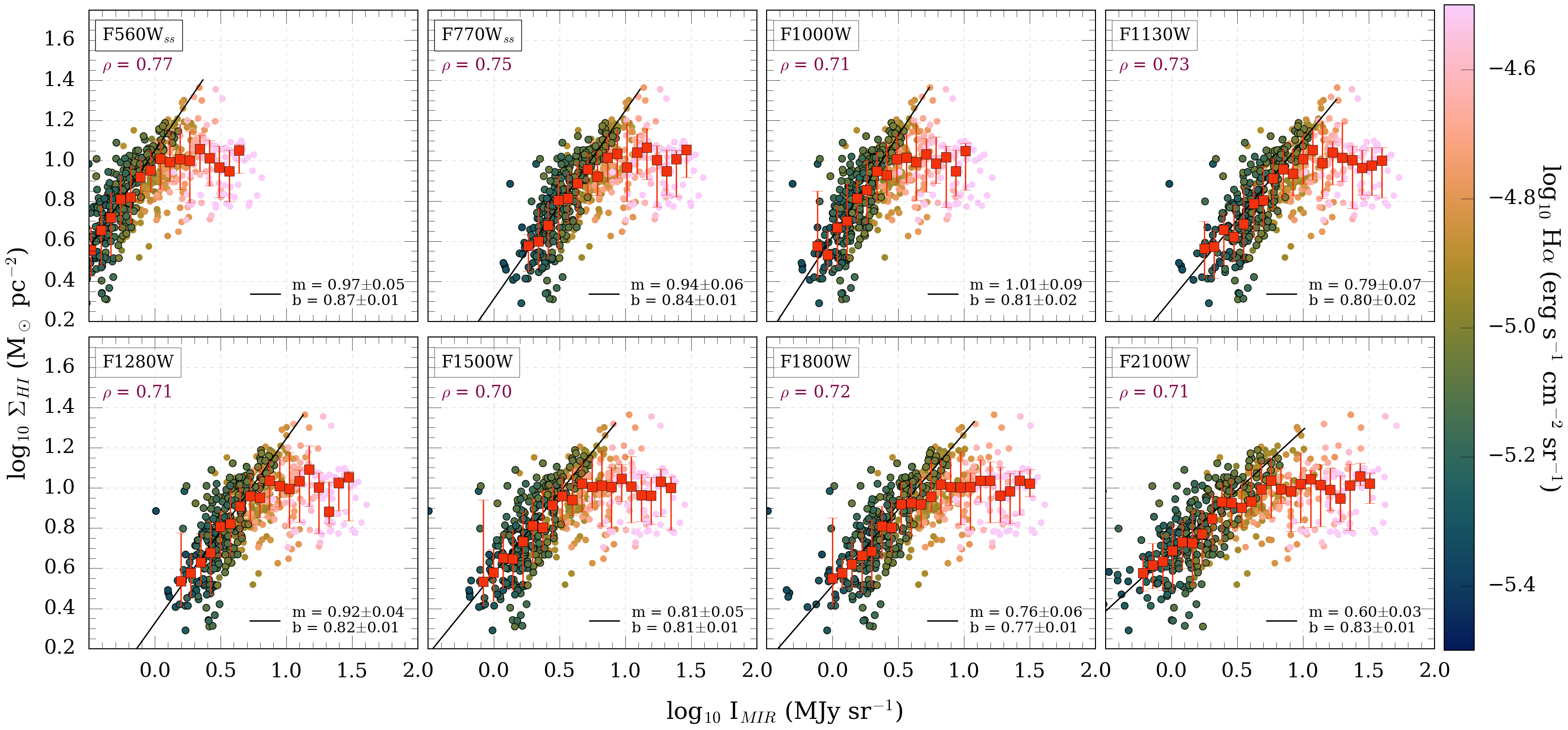}
\caption{\hi\ surface density, $\Sigma_{\rm HI}$ as a function of I$_{\rm MIR}$ at 12\arcsec\ resolution. The scatter points are colored based on the \ha\ intensity. The MIRI band is noted on the top left corner in each panel. Red squares and error bars represent the median and the 16-84\% range of \sighi\ after binning the mid-IR intensity. The black line represents the best-fit line to the points where the \ha\ intensity is $\leq10^{-5}$~erg~s$^{-1}$~cm$^{-2}$~sr$^{-1}$ that are highlighted with black edges. The correlation coefficient \spr\ is estimated for these points and noted on the top left and the best-fit parameters are noted at the bottom of each panel.} \label{fig:hi-miri}
\end{figure*}

To investigate how well mid-IR emission traces the atomic gas phase of the ISM, we examine the correlations between I$_{\rm MIR}$ and atomic gas surface density, \sighi\ in M51 at $\sim$440~pc resolution (12\arcsec), as shown in Figure \ref{fig:hi-miri}. Although the \hi\ map of M51 extends further out, we limit our analysis to the footprint of the \coone\ map to allow for a direct comparison between mid-IR's response to the molecular and atomic gas phases of the ISM. One can see from these plots that \sighi\ initially increases with I$_{\rm MIR}$, but above I$_{\rm MIR}\approx 2$--10 \mjysr\ (depending on the mid-IR band), it begins to plateau or decline. The declining part of the distribution shows high \ha\ intensities. To quantify how mid-IR emission relates to the neutral \hi\ column, we limit our analysis to regions with $I_{\rm H\alpha}\leq10^{-5}$~erg~s$^{-1}$~cm$^{-2}$~sr$^{-1}$ to exclude bright star forming and mainly probe the diffuse ISM. In this regions, we find the \hi-to-mid-IR distribution shows a \spr\ of 0.70--0.75. We find the best-fit power law slopes to range from 0.60--1.01. The wavelength-dependent variation in the power law relations with PAH-dominated filters exhibiting near-linear slopes and flattening noted for the continuum-dominated filter is once again similar to those observed the previous sections. 

We note that the power-law scaling of the mid-IR filters with \hi\ is comparable to that with \coone\ and $\Sigma_{\rm gas}$ for all filters, highlighting the fact that dust is well mixed with atomic gas as well. Our observations clearly show that at low and intermediate I$_{\rm MIR}$, \sighi\ rises because PAHs and continuum-emitting dust grains are well mixed with diffuse atomic gas. However, once I$_{\rm MIR}$ traces regions of higher column density and stronger radiation fields, the ISM crosses into the regime where the atomic-to-molecular transition occurs, typically near $\Sigma_{\rm gas}\sim10$~M$_\odot$~pc$^{-2}$ at solar metallicity \citep{krum09}, and \sighi\ saturates. This naturally leads to a turnover at high mid-IR intensity where \hi\ becomes an ever smaller fraction of the total gas column in the ISM. As this turnover corresponds to regions where the ISM is \ce{H2}-dominated, \sighi\ no longer increases even as dust heating and mid-IR emission rise. Regardless, extracting mainly the diffuse, atomic-gas dominated regions, we find that the \hi-mid-IR correlations are strong (\spr$\sim$0.7) for all filters indicating that mid-IR emission can trace \hi\ where it dominates the dust-bearing column.   

\setlength{\tabcolsep}{14pt} 
\begin{deluxetable*}{llllllll}
\tablewidth{\textwidth}
\tablecaption{Statistics for \coone, total gas, $\Sigma_{\rm HI}$, and \paa\ correlations with mid-IR intensities at $\sim440$~pc scale \label{tab:allgas}}
\tablehead{
\colhead{Filter} & \colhead{N$_{\rm pix}$} & \colhead{\spr} & \colhead{m} & \colhead{b} & \colhead{x$_0$} & \colhead{$\sigma_{\rm int}$} & \colhead{$\sigma$}}
\colnumbers
\startdata
\multicolumn{7}{c}{\bf{\coone}}\\
\tableline
F560W$_{\rm ss}$ & 562 & $0.86 \pm 0.010$ & $0.92 \pm 0.081$ & $1.12 \pm 0.023$ & 0.23 & 0.19 & 0.28 \\
F770W$_{\rm ss}$ & 562 & $0.85 \pm 0.010$ & $0.93 \pm 0.056$ & $0.90 \pm 0.022$ & 0.78 & 0.20 & 0.28 \\
F1000W & 562 & $0.87 \pm 0.009$ & $1.00 \pm 0.050$ & $1.07 \pm 0.017$ & 0.60 & 0.19 & 0.28 \\
F1130W & 562 & $0.87 \pm 0.010$ & $0.87 \pm 0.063$ & $0.86 \pm 0.023$ & 0.84 & 0.19 & 0.28 \\
F1280W & 562 & $0.88 \pm 0.010$ & $0.95 \pm 0.049$ & $0.97 \pm 0.018$ & 0.86 & 0.18 & 0.28 \\
F1500W & 562 & $0.88 \pm 0.009$ & $0.86 \pm 0.034$ & $1.02 \pm 0.014$ & 0.71 & 0.17 & 0.28 \\
F1800W & 562 & $0.88 \pm 0.009$ & $0.79 \pm 0.037$ & $0.95 \pm 0.016$ & 0.76 & 0.18 & 0.28 \\
F2100W & 562 & $0.86 \pm 0.010$ & $0.68 \pm 0.038$ & $1.00 \pm 0.016$ & 0.75 & 0.20 & 0.28 \\
\tableline
\multicolumn{7}{c}{\bf{$\Sigma_{\rm HI+H_2}$}} \\
\tableline
F560W$_{\rm ss}$ & 562 & $0.88 \pm 0.008$ & $0.81 \pm 0.077$ & $1.84 \pm 0.022$ & 0.23 & 0.16& 0.28 \\
F770W$_{\rm ss}$ & 562 & $0.88 \pm 0.009$ & $0.84 \pm 0.047$ & $1.64 \pm 0.019$ & 0.78 & 0.16& 0.28 \\
F1000W & 562 & $0.89 \pm 0.008$ & $0.89 \pm 0.049$ & $1.79 \pm 0.017$ & 0.60 & 0.15& 0.28 \\
F1130W & 562 & $0.89 \pm 0.008$ & $0.80 \pm 0.040$ & $1.60 \pm 0.015$ & 0.84 & 0.15& 0.28 \\
F1280W & 562 & $0.89 \pm 0.008$ & $0.83 \pm 0.051$ & $1.71 \pm 0.019$ & 0.86 & 0.15& 0.28 \\
F1500W & 562 & $0.90 \pm 0.007$ & $0.77 \pm 0.034$ & $1.75 \pm 0.014$ & 0.71 & 0.14& 0.28 \\
F1800W & 562 & $0.90 \pm 0.008$ & $0.70 \pm 0.039$ & $1.68 \pm 0.017$ & 0.76 & 0.15& 0.28 \\
F2100W & 562 & $0.88 \pm 0.009$ & $0.61 \pm 0.037$ & $1.73 \pm 0.016$ & 0.75 & 0.16& 0.28 \\
\tableline
\multicolumn{7}{c}{\bf{$\Sigma_{\rm HI}$}} \\
\tableline
F560W$_{\rm ss}$ & 258 & $0.77 \pm 0.033$ & $0.97 \pm 0.052$ & $0.87 \pm 0.010$ & -0.19 & 0.15 & 0.29 \\
F770W$_{\rm ss}$ & 258 & $0.75 \pm 0.033$ & $0.94 \pm 0.060$ & $0.84 \pm 0.012$ & 0.56 & 0.15 & 0.28 \\
F1000W & 258 & $0.70 \pm 0.039$ & $1.01 \pm 0.088$ & $0.81 \pm 0.017$ & 0.19 & 0.16 & 0.30 \\
F1130W & 258 & $0.73 \pm 0.034$ & $0.79 \pm 0.067$ & $0.80 \pm 0.016$ & 0.62 & 0.16 & 0.26 \\
F1280W & 258 & $0.71 \pm 0.036$ & $0.92 \pm 0.045$ & $0.82 \pm 0.010$ & 0.53 & 0.16 & 0.29 \\
F1500W & 258 & $0.69 \pm 0.039$ & $0.81 \pm 0.046$ & $0.81 \pm 0.011$ & 0.30 & 0.17 & 0.28 \\
F1800W & 258 & $0.72 \pm 0.035$ & $0.76 \pm 0.057$ & $0.77 \pm 0.013$ & 0.34 & 0.17 & 0.29 \\
F2100W & 258 & $0.71 \pm 0.037$ & $0.60 \pm 0.032$ & $0.83 \pm 0.009$ & 0.23 & 0.18 & 0.28 \\
\tableline
\multicolumn{7}{c}{\bf{\paa}} \\
\tableline
F560W$_{\rm ss}$ & 562 & $0.91 \pm 0.011$ & $1.30 \pm 0.122$ & $-4.85 \pm 0.024$ & 0.38 & 0.21 & 0.28 \\
F770W$_{\rm ss}$ & 562 & $0.90 \pm 0.013$ & $1.33 \pm 0.047$ & $-5.22 \pm 0.016$ & 0.88 & 0.21 & 0.28 \\
F1000W & 562 & $0.90 \pm 0.012$ & $1.39 \pm 0.069$ & $-5.13 \pm 0.023$ & 0.60 & 0.21 & 0.28 \\
F1130W & 562 & $0.90 \pm 0.013$ & $1.31 \pm 0.061$ & $-5.31 \pm 0.019$ & 0.94 & 0.23 & 0.28 \\
F1280W & 562 & $0.90 \pm 0.013$ & $1.28 \pm 0.061$ & $-5.24 \pm 0.023$ & 0.86 & 0.22 & 0.28 \\
F1500W & 562 & $0.90 \pm 0.013$ & $1.17 \pm 0.046$ & $-5.20 \pm 0.019$ & 0.71 & 0.22 & 0.28 \\
F1800W & 562 & $0.91 \pm 0.012$ & $1.07 \pm 0.053$ & $-5.28 \pm 0.023$ & 0.76 & 0.21 & 0.28 \\
F2100W & 562 & $0.91 \pm 0.012$ & $0.98 \pm 0.027$ & $-5.23 \pm 0.012$ & 0.75 & 0.21 & 0.28 \\
\enddata
\tablecomments{Column 1: JWST MIRI band. Subscript ``ss" denotes starlight subtraction (see \S\ref{sec:analysis}). Column 2: Number of pixels entering the analysis. Column 3: Spearman rank correlation coefficient, \spr\ between the tracer and the respective mid-IR band intensities. Column 4-6: Slope, intercept, and pivot value describing the best fit power law relation between the tracer and the respective mid-IR band. Column 7-8: Intrinsic and observed RMS scatter in the data about the best fit line, noted in dex.}
\end{deluxetable*}

\section{Discussion}\label{sec:discuss}

It is clear that the correlations examined above highlight the effect of both mixing of dust with gas and dust heating by the radiation field on mid-IR emission. 
The near-linearity observed for PAH-dominated filters relative to \coone\ indicates that PAH emission primarily traces the amount of PAH-bearing dust mixed with the gas reservoir in conditions where $U$ is approximately constant \citep{leroy23b, chown25}. The stochastic heating of small dust grains leads to emission at $\lambda<20$\micron\ that is linearly dependent on $U$ in low $U$ environments \citep[$\lesssim10^3$;][]{draine07}. At longer wavelengths where the mid-IR filters become increasingly dominated by larger grains approaching thermal equilibrium, the emission becomes more strongly weighted toward regions of intense star formation and high $U$, while still mainly responding linearly to $U$. This produces elevated $I_{\mathrm{MIR}}$ at fixed $I_{\mathrm{CO}}$ and yields shallower CO–mid-IR slopes. In these high $U$ environments, PAH emission can be suppressed due to destruction by intense UV radiation \citep{helou04, lebou11, chastenet19, chastenet23, egorov23, egorov25}, limiting the contribution of these regions to PAH-dominated filters. Consistent with this picture, the correlations with \paa\ show that F2100W seems to directly trace the ionizing photon rate and thus recent star formation. Further, the super-linear scaling of PAH-dominated filters with \paa\ indicates a suppression of PAH emission at high $U$. Contributions from PAHs heated in diffuse, low $U$ regions could also elevate $I_{\mathrm{MIR}}$ at low \paa\ and steepen the observed relation. 

Our study clearly indicates that mid-IR emission is not a pure tracer of either the gas column or star formation in the ISM. In fact, in the molecular gas-dominated disk of M51, we find that mid-IR emission across 5.6-21 \micron\ correlates with both \coone\ and \paa\ on $\sim$40~pc scales, with their relations well described by power laws. The correlations also strengthen when averaged to $\sim$440~pc as we now probe averaged ISM properties over multiple physical regions. We show a summary of the power law slopes and correlation coefficients for all tracer-mid-IR pairs in Figure \ref{fig:ism_slopes}, all estimated here at $\sim$440~pc. Two key trends are evident here: 1) the gas and star formation tracers exhibit a comparable \spr\ with mid-IR at all wavelengths, 2) the slopes display a systematic, wavelength-dependent trend that is remarkably consistent across all tracers. 

\begin{figure}[!t]
\centering
\includegraphics[trim = 0cm 3.75cm 0cm 0cm, clip,scale=0.42]{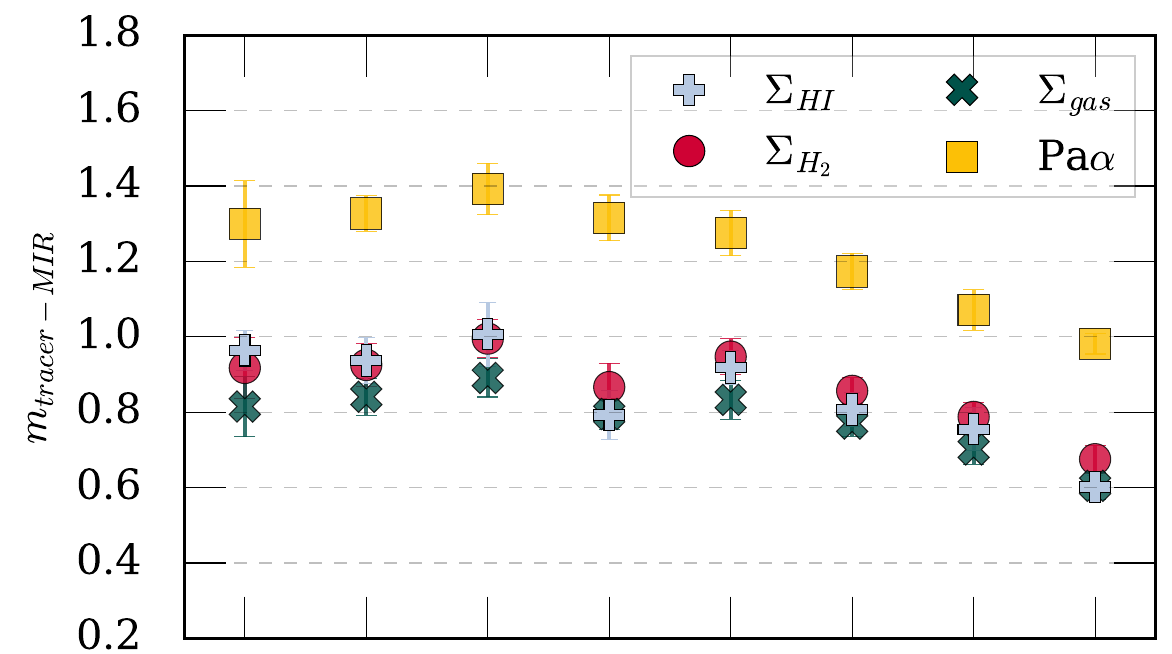}
\includegraphics[trim = 0cm 0cm 0cm 0cm, clip,scale=0.42]{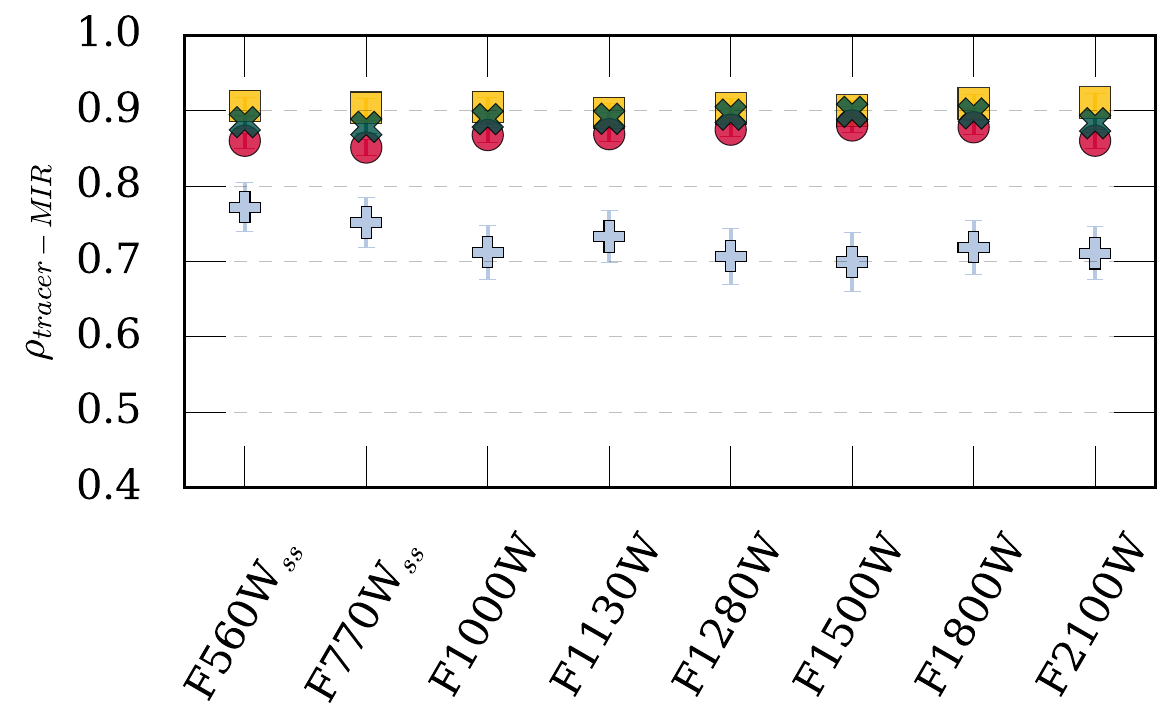}
\caption{Best-fit power law slope $m$ ({\it top}) and correlation coefficient ({\it bottom}) between different tracers and mid-IR bands estimated at 12\arcsec. } \label{fig:ism_slopes}
\end{figure}

The wavelength-dependent behavior of the emission captured by mid-IR filters to gas column and star formation has been explored in previous studies \citep{bendo08, bendo10, crocker13, boquien16, whitcomb23, leroy23b}. Using WISE and Spitzer data of SINGS \hii\ regions at $\sim$100s of pc scales, \cite{whitcomb23} showed that the correlation strengths of PAH- and dust continuum-dominated bands with CO emission and star formation vary with wavelength, with PAHs showing stronger correlations with CO emission than with star formation. At $\sim100$~pc scales, \cite{leroy23b} demonstrated that mid-IR emission can be decomposed into a component associated with dust mixed with gas traced by \cotwo\ and heated by a relatively uniform ISRF, and a component associated with dust heated by intense radiation fields traced by \ha. Our results extend these findings across the full mid-IR range from 5.6-21\micron, and show that the systematic change in slopes reflects a gradual shift from gas-linked PAH emission at shorter wavelengths to heating-weighted dust continuum emission at longer wavelengths, while the near constancy in the correlation strengths indicates that mid-IR emission remains tightly correlated to both gas and star formation. This behavior reflects a shift in the dominant physical drivers of mid-IR emission with wavelength, rather than a decoupling from either the gas reservoir or recent star formation.  

\subsection{Relative Contributions of Gas Column Density and Local Radiation Field to mid-IR emission} \label{sec:twocomp}

\begin{figure*}[!t]
\centering
\includegraphics[trim = 0cm 0cm 0cm 0cm, clip,scale=0.42]{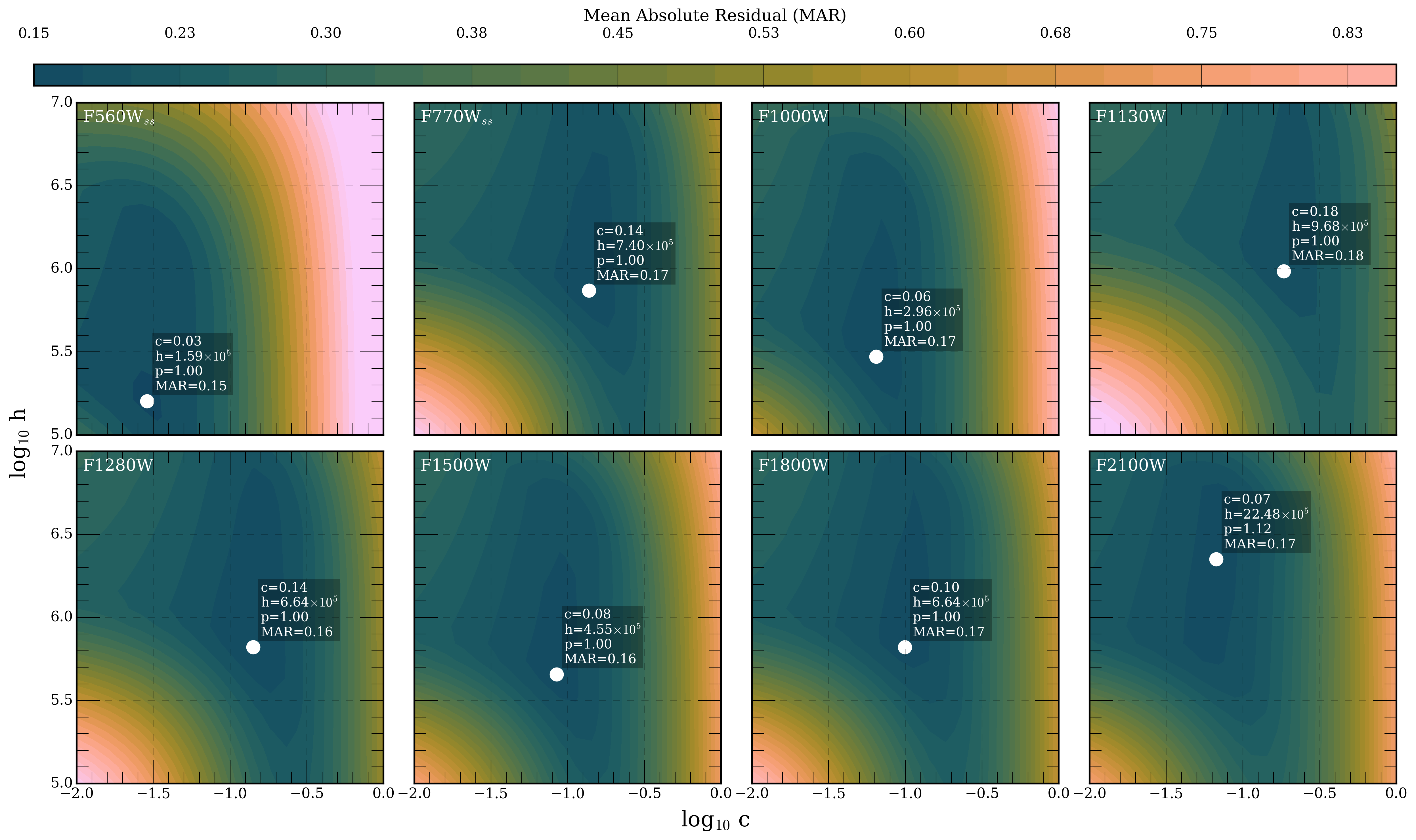}
\caption{Template-matching model fits as a function of \coone\ and \paa\ scaling coefficients, $c$ and $h$. The color shows the mean absolute logarithmic residuals for different combinations of $\log_{10} c$ and $\log_{10} h$ for the best-fit $p$. The best-fit $c$, $h$, and $p$ for the model minimize the mean absolute residual (MAR) and are indicated by a circle. The best-fit parameters and MAR are noted in each panel.} \label{fig:contoursres}
\end{figure*}

The relative contribution of the gas column density and the local radiation field to mid-IR emission has been quantitatively investigated in \citet{leroy23b}. They explored a linear model using scaled \cotwo\ and extinction-corrected \ha\ maps of four nearby galaxies, expressing the mid-IR emission as a combination of gas- and star formation–related components. In this framework, the mid-IR emission is described as a superposition of 1) a gas-linked component arising from dust mixed with gas and heated by the diffuse ISRF, and 2) a star formation-linked component associated with dust exposed to intense radiation fields. This two-component model reproduces the bright mid-IR emission in their sample galaxies to within $\pm$0.2 dex in F770W, F1000W, F1130W, and F2100W over a wide range of intensities. On average, they find that roughly half of the mid-IR emission arises from the CO-tracing component and half from the \ha-tracing component.

Physically, this resolved, template-based model is analogous to dust population synthesis models of \cite{draine07} that represent integrated dust emission as a combination of dust heated by a uniform ISRF and dust exposed to intense radiation fields in photodissociation regions. These theoretical models also predict that the mid-IR response to the local radiation field is not strictly linear, especially at U~$\gtrsim10^4$. Motivated by this, we extend the linear formulation from \cite{leroy23b} using \coone\ and \paa\ maps, allowing the mid-IR emission to scale non-linearly with the ionized tracer, while retaining a linear dependence on \coone.  Specifically, we model the emission in each MIRI filter as
\begin{align}\label{eq:2}
    I_{\rm MIR}^{\rm model} = c\times I_{\rm CO} + h\times I_{\rm Pa\alpha}^p
 \end{align} 
where $c$, $h$, $p$ are free parameters to be fit, with $c$ and $h$ being associated with molecular gas and \paa, respectively, while 
$p$ quantifies any non-linear response of the mid-IR emission to $U$. When $p=1$, the model reduces to the linear formulation from \cite{leroy23b}.

\setlength{\tabcolsep}{10pt} 
\begin{deluxetable*}{llllcll}
\tablewidth{\textwidth}
\tablecaption{Two component template fitting model results  \label{tab:2comp}}
\tablehead{
\colhead{Filter} & \colhead{$c$} & \colhead{$h\times10^5$} & \colhead{$p$} & \colhead{Mean $|\log_{10}\frac{I_{\rm model}}{I_{\rm obs}}|$} & \colhead{$f_c$} & \colhead{$f_h$}}
\colnumbers
\startdata
\tableline
F560W$_{ss}$ & 0.029 $\pm$ 0.000 & 1.593 $\pm$ 0.000 & 1.000 & 0.155 & 0.379 $\pm$ 0.001 & 0.621 $\pm$ 0.001 \\
F770W$_{ss}$ & 0.137 $\pm$ 0.003 & 7.395 $\pm$ 0.297 & 1.000 & 0.166 & 0.385 $\pm$ 0.014 & 0.615 $\pm$ 0.014 \\
F1000W & 0.065 $\pm$ 0.001 & 2.960 $\pm$ 0.138 & 1.000 & 0.165 & 0.423 $\pm$ 0.017 & 0.577 $\pm$ 0.017 \\
F1130W & 0.185 $\pm$ 0.003 & 9.680 $\pm$ 0.155 & 1.000 & 0.181 & 0.391 $\pm$ 0.007 & 0.609 $\pm$ 0.007 \\
F1280W & 0.141 $\pm$ 0.001 & 6.641 $\pm$ 0.064 & 1.000 & 0.159 & 0.411 $\pm$ 0.004 & 0.589 $\pm$ 0.004 \\
F1500W & 0.085 $\pm$ 0.001 & 4.554 $\pm$ 0.060 & 1.000 & 0.165 & 0.384 $\pm$ 0.006 & 0.616 $\pm$ 0.006 \\
F1800W & 0.099 $\pm$ 0.002 & 6.641 $\pm$ 0.087 & 1.000 & 0.166 & 0.340 $\pm$ 0.007 & 0.660 $\pm$ 0.007 \\
F2100W & 0.067 $\pm$ 0.001 & 22.485 $\pm$ 0.539 & 1.118 & 0.169 & 0.257 $\pm$ 0.008 & 0.743 $\pm$ 0.008 \\
\enddata
\tablecomments{Column 1: JWST MIRI filter. Subscript ``ss" denotes starlight subtraction (see \S\ref{sec:analysis}). Column 2, 3, 4: Best fit model parameters, $c$, $h$, and $p$ to fit the observed mid-IR intensities using Equation \ref{eq:2}. Column 5: Mean absolute logarithmic residual between the observed and modeled mid-IR intensities. Column 6, 7: Fraction of total flux associated with the \coone\ and \paa\ components, respectively.}
\end{deluxetable*}

The parameter space is explored using a grid-based search. First, we evaluate the model over a coarse grid spanning a broad range of plausible values for $c$, $h$, and $p$, and compute the mean absolute logarithmic residual between the observed and modeled mid-IR intensities as $\mathrm{MAR} = \left\langle \left| \log_{10} \left( \frac{I_{\mathrm{obs}}}{I_{\mathrm{model}}} \right) \right| \right\rangle$. For each mid-IR filter, we determine the best-fitting parameters that minimize the $\mathrm{MAR}$.
We then iteratively refine the search by constructing a finer grid around the initial values, and repeat till the MAR converges to within 5\% between iterations. For each filter, we retain the best-fitting parameters and the corresponding residual surface in the $(c, h)$ plane at the optimal value of $p$. We also estimate the fraction of total flux in the model associated with \coone\ and \paa, 
\begin{align*}
    f_c = \frac{\Sigma c\times I_{\rm CO}}{\Sigma I_{\rm MIR}^{\rm model}},\\
    f_h = \frac{\Sigma h\times I_{\rm Pa\alpha}^p}{\Sigma I_{\rm MIR}^{\rm model}},
\end{align*} 
where $f_c$ and $f_h$ are the flux fractions associated with \coone\ and \paa, respectively. We estimate uncertainties in the fitted parameters and flux fractions using 100 bootstrap realizations. In each realization, valid pixels are resampled with replacement and observed I$_{MIR}$ are perturbed using their measured uncertainties assuming Gaussian errors. The model is refit over a parameter grid spanning $\pm1$ dex around the best-fit $(c, h)$ with $p$ fixed at its best fitting value. We adopt the standard deviation of the resulting distributions of the parameters as their uncertainties. The best-fitting $c$, $h$, and $p$ along with the fractional contribution of each component are reported in Table \ref{tab:2comp}. 

\begin{figure*}[!t]
\centering
\includegraphics[trim = 0cm 0cm 0cm 0cm, clip,scale=0.225]{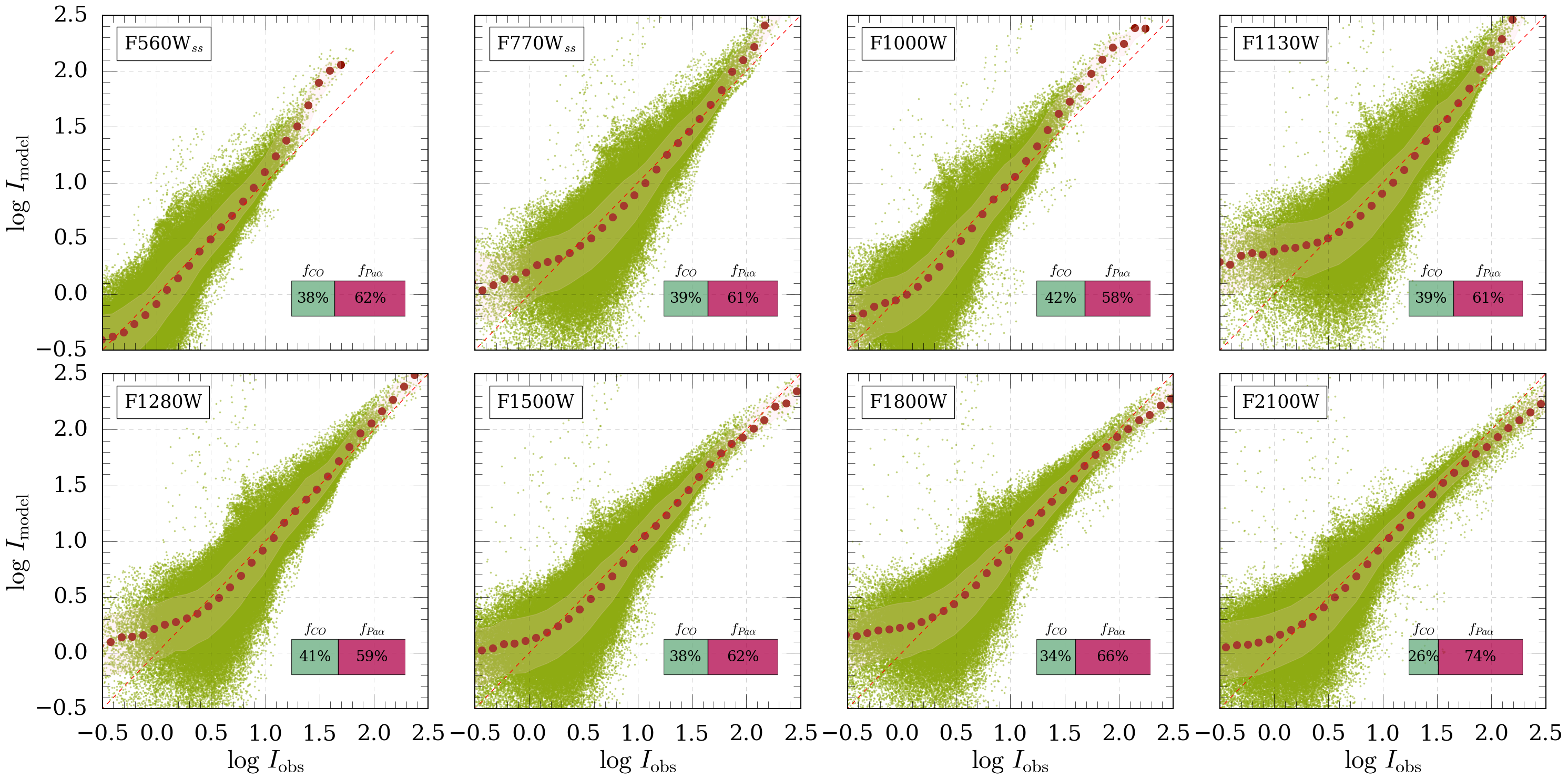}
\caption{Modeled mid-IR intensity (I$_{\rm model}$) prediction from a combination of \coone\ and \paa\ as a function of observed mid-IR intensities (I$_{\rm obs}$). The dashed line represent a one-to-one relation between I$_{\rm model}$ and I$_{\rm obs}$. The red points show the median I$_{\rm model}$ in a given I$_{\rm obs}$ bin. The bar on the bottom right shows the fractional contribution of \coone\ and \paa\ intensities to each mid-IR band. Each bar represents the total modeled IR flux and is subdivided into two sections: \coone\ flux fraction (left) and \paa\ flux fraction (right). Percentages noted on each bar state the fraction represented by each of the two components. } \label{fig:res}
\end{figure*}

Figure \ref{fig:contoursres} shows the mean absolute residual in $\log_{10} h$ vs. $\log_{10} c$ space for the best fit $p$ for all filters and the the modeled mid-IR intensity, $I_\mathrm{model}$ as a function of $I_{\mathrm{obs}}$ are depicted in Figure \ref{fig:res}. Our best-fit parameters minimize the residuals at $\pm0.15-0.18$ dex and the resulting model follows the observed data fairly well over a wide range of intensities. Interestingly, we find the best-fit $p$ is consistent with unity for all MIRI filters with no significant wavelength variation except F2100W with $p=1.12$. This indicates that, over the range of radiation field strengths probed in M51, the mid-IR emission associated with \paa\ scales approximately linearly with the ionizing photon rate. Although theoretical models predict non-linear mid-IR responses at very high $U$, such regimes are not strongly sampled in the environments investigated here. As a result, the observed mid-IR intensities do not require an explicit non-linear response to $U$ to reproduce $I_{\mathrm{obs}}$ with the two-component model capturing the dominant structure in the data. 

Comparing the $I_\mathrm{model}$ and $I_{\mathrm{obs}}$ in Figure \ref{fig:res} reveals systematic over-predictions at high $I_{\mathrm{obs}}$ for shorter wavelength filters (F560W$_{\rm ss}$, F770W$_{\rm ss}$, F1000W, and F1130W). These discrepancies are most pronounced in regions where the \paa-associated component dominates the model, indicating that a simple linear scaling with \paa\ overestimates PAH emission in intense star-forming environments. This behavior is consistent with the known suppression of PAH emission in \hii\ regions due to grain destruction in harsh radiation fields, and mirrors the trends reported by \cite{leroy23b} for F770W and F1130W. In contrast, F1280W exhibits the most linear behavior, even at high intensities. Contribution of nebular 12.8\micron\ [\ion{Ne}{2}] emission line could play some role as it is strongly correlated with star formation \citep{whitcomb23}. For F1500W, F1800W, and F2100W, the model increasingly underpredicts $I_\mathrm{obs}$ at high $I_\mathrm{obs}$. This behavior shows the growing contribution of warm dust continuum from bright star-forming regions \citep{pathak24}, as well as environmental variations in the PAH fraction, rather than a breakdown of the overall linear scaling with \coone\ or \paa.

We find our empirical two-component model attributes $\sim$60\% to 75\% of the modeled mid-IR emission to the \paa-associated component, with the corresponding \coone-associated component decreasing from $\sim$40\% to 25\%. Within the model, the \paa\ contribution remains near $\sim$60\% for the PAH-dominated bands (F560W$_{\rm ss}$–F1280W), but rises to $\sim$75\% in the dust continuum–dominated F2100W filter. 
Taken together, these empirical results indicate that roughly half of the mid-IR emission at 5–13 \micron\ in M51 arises from dust mixed with the molecular gas and heated by the diffuse interstellar radiation field, while the remainder originates from dust in the immediate environments of star forming regions exposed to higher $U$ traced by \paa. At longer wavelengths, the increasing dominance of the \paa\ component reflects the fact that dust continuum is produced in regions where $U$ is elevated. The wavelength-dependent shift in the dominant heating sources provides a physical explanation for the changing CO–MIR and \paa\ mid-IR slopes: PAH-dominated bands more faithfully trace the gas distribution, while dust continuum bands increasingly reflect high-$U$ star-forming regions.

\subsection{Behavior of F1000W}
The F1000W filter, centered at 10\micron, is nominally considered to be tracing warm dust continuum,  but the mid-IR emission in this regime is complex. This band also includes contributions from the wings of adjacent PAH features along with the broad silicate absorption feature at 9.7\micron\ \citep{smith07, tielens08, brandl06}. The filter consists of $\sim$30\% emission from PAHs with the continuum definition from PAHFIT \citep{whitcomb23b}. In our analysis, F1000W exhibits a correlation with \coone\ and relative contributions from CO- and \paa-tracing components in the template-fitting model that are similar to the PAH-dominated filters. This was also observed by \cite{leroy23b} in other nearby galaxies. The observed behavior suggests that the mid-IR emission in F1000W has significant contributions from the adjacent PAH feature wings or that the dust grain population carrying the 10\micron\ emission behaves very similarly to PAHs. 
 
\begin{figure*}[!ht]
\centering
\includegraphics[trim =0cm 0cm 0cm 0cm, clip,scale=0.45]{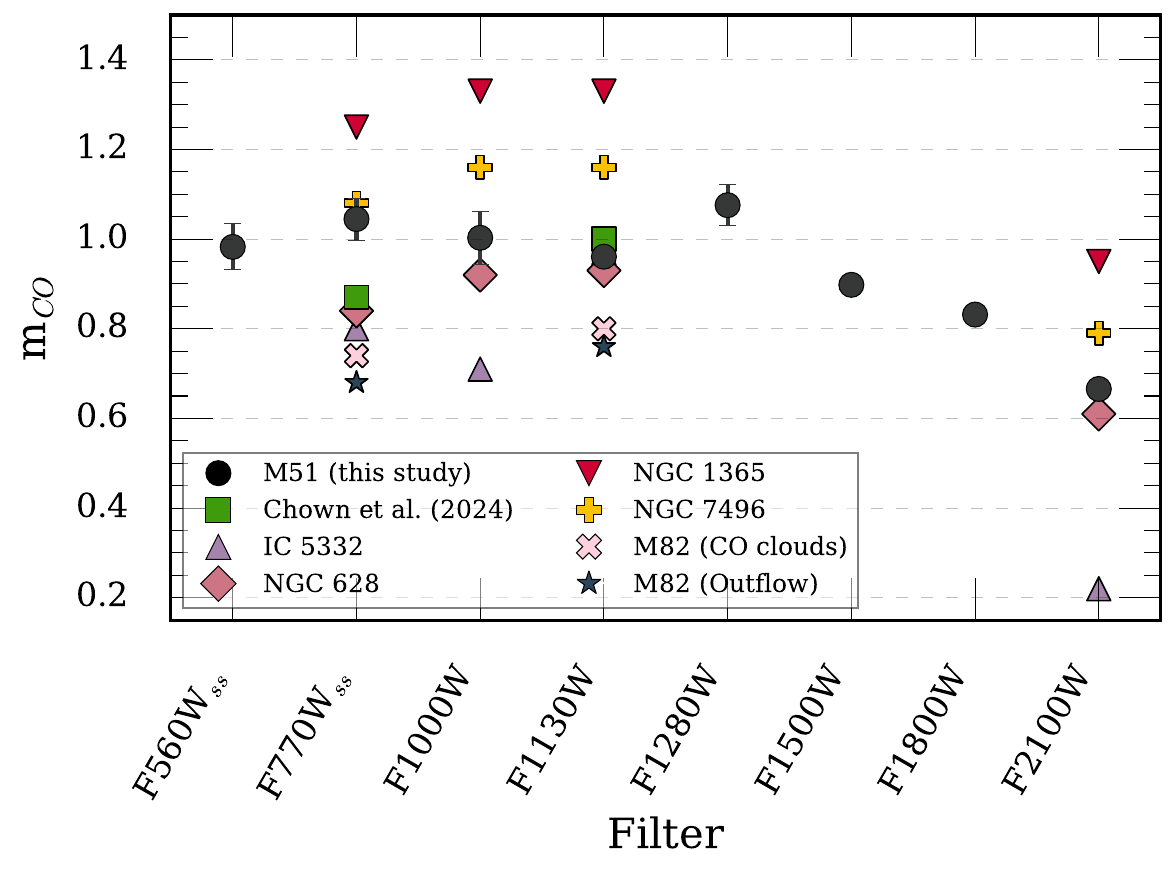}
\includegraphics[trim =0cm 0cm 0cm 0cm, clip,scale=0.45]{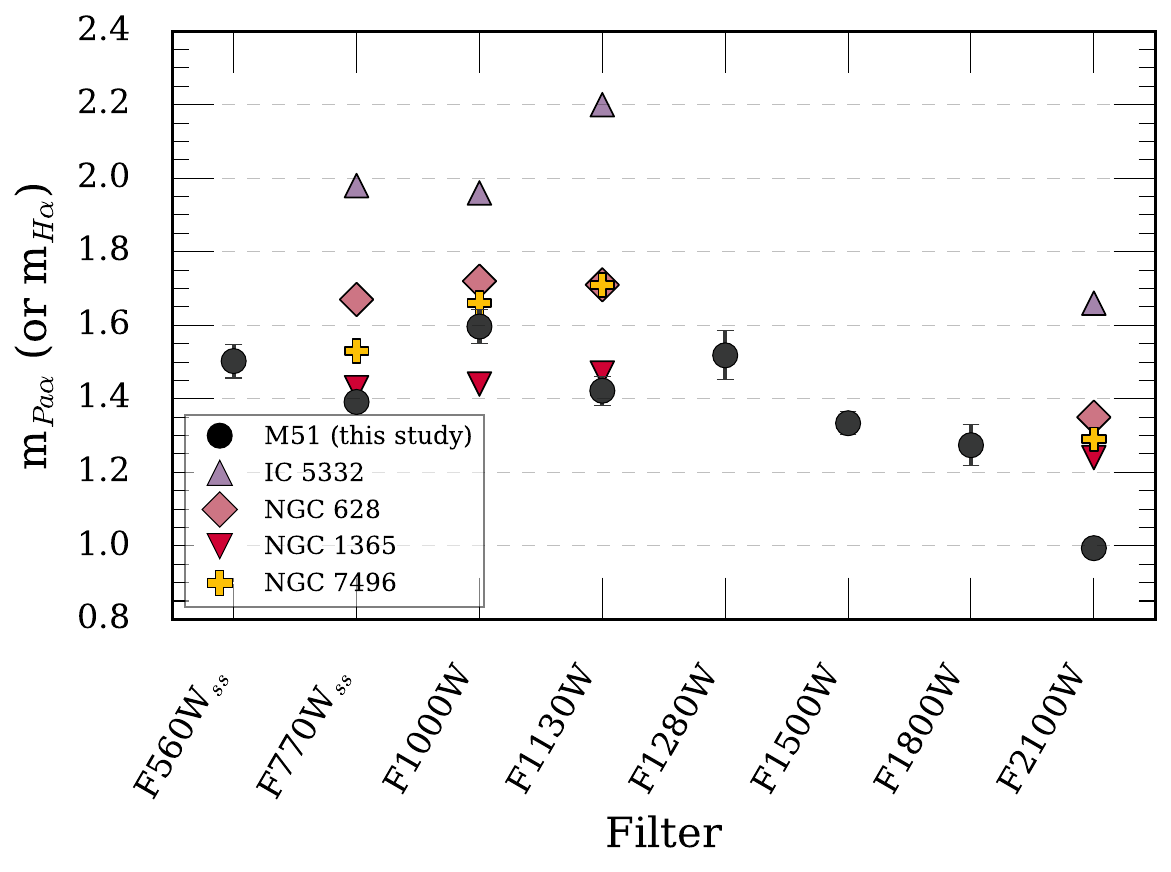}
\caption{{\it Left:} Power law indices between CO and mid-IR intensity for different MIRI filters. {\it Right:} Similar for \paa\ (or \ha\ for literature) and mid-IR intensity. The black circles correspond to slopes estimated in this study. For comparison, we plot the slopes from recent studies: NGC 628, IC 5332,  NGC 1365 and NGC 7496 from \cite{leroy23b}, M82 from \cite{villa25} \cite{chown25} for 70 PHANGS galaxies 
} \label{fig:slopes}
\end{figure*}

\subsection{Mid-infrared (PAH) emission as a gas tracer}
PAH emission in the mid-IR is emerging as a robust tracer of molecular gas in galaxies. Several studies have shown strong spatial correlations between PAH and CO emission on sub-kpc scales and galaxy-integrated measurements \citep{regan06, gao19, cortzen19, chown21, leroy23a, whitcomb23, zhang23, shivaei24} across galaxy samples. JWST observations have extended these correlations to $\sim$50-100 pc scales using PAH-dominated filters like F335M, F770W, and F1130W, revealing a near-linear relationship between these filters and CO intensity \citep{leroy23b, chown25}. 
Even in M51, we note slopes close to unity for the PAH-dominated MIRI filters at both $\sim$40 and 440 pc scales. 

In the left panel of Figure \ref{fig:slopes}, we show a qualitative comparison of CO-to-mid-IR slopes derived using JWST MIRI filters from several recent studies \citep{leroy23b, chown25, villa25, sebastian25} with our observations in M51. Evidently, these show significant variations in slopes between galaxies. This may arise because the literature measurements are not fully homogeneous, differing in spatial resolution, fitting methodology, and masking choices. Additionally, the galaxy properties, such as stellar mass, star formation rate, metallicity, and morphology, can influence the ISRF, dust-to-gas ratio, and $q_{\rm PAH}$, which are assumed to be relatively uniform in Equation \ref{eq:1}. Despite this diversity, the slopes from the \cite{chown25} investigation of $\sim$70 nearby galaxies are 0.91$\pm$0.07 and 1.02$\pm$0.09 between \cotwo\ and F770W and F1130W, respectively. They also find the stellar-continuum subtracted F335M, which captures the 3.3\micron\ PAH feature to show a linear relation (slope $\sim$1.01) with \cotwo. These findings indicate that PAH luminosities scale approximately linearly with CO luminosities across a wide range of environments and galaxy types, supporting the use of PAH emission as a proxy for the molecular gas reservoir under typical conditions. We also find the CO-to-PAH ratios (Figure \ref{fig:env}) show little to no variation across environments, indicating limited evolution of PAH population across environments in M51. Several recent studies have found modest spatial variations in PAH band ratios and ionization diagnostics on sub-kpc scale in nearby galaxies \citep[e.g.,][H. Koziol et al. submitted, D. Pathak et al. in prep]{sutter24, baron25}.

The near-linear behavior of PAH emission is also recovered clearly when correlated to $\Sigma_{\rm HI+H_2}$, indicating that PAHs are primarily tracing the overall ISM gas column rather than a single gas phase. As dust is present throughout the ISM, the mid-IR emission does not solely arise from dust associated with CO-bright molecular gas. In M51, we also observe a correlation between PAH-dominated filters and \hi, although the PAH-\hi\ relation is slightly weaker compared (\spr$\sim$0.7) to PAH-CO and PAH-$\Sigma_{\rm HI+H_2}$ (\spr$\sim$0.9), likely because the regions analyzed are strongly molecular-dominated. Nevertheless, the measurable contribution from both atomic and molecular gas suggests that PAHs do not preferentially reside in a single neutral phase of the ISM and may perhaps be sensitive to molecular gas not traced by \coone \citep{sandstrom23}. This behavior is consistent with PAHs being well mixed with the neutral gas phases and primarily responding to the local radiation field conditions \citep{galliano18, bendo10}. Further, dust evolution models predict that small grain populations, including PAHs, are maintained through a balance between grain growth in dense regions and shattering in diffuse, turbulent media \citep{asano13, hirashita15, galliano18} operating across the neutral ISM. The similarity of the scaling relations across \hi\ and \ce{H2} therefore suggests that the PAH population is relatively stable within the neutral ISM.  

Departures from linearity between CO (or gas) and PAH emission can arise as PAH emission is sensitive to metallicity, grain processing, and radiation hardness. Galaxies with low $q_{\rm PAH}$ or intense $U$ can exhibit suppressed PAH emission relative to CO \citep[e.g.,][]{engel05, smith07, hunt10}, contributing to the scatter and galaxy-to-galaxy variations in PAH-gas relations.
Metallicity is a key factor affecting $q_{\rm PAH}$, dust-to-gas ratio, and hence, the use of mid-IR as a gas tracer in galaxies. PAH emission has been observed to weaken in low-metallicity environments, such as dwarf galaxies, M33, and the Magellanic Clouds \citep{engel05, hunt10, sandstrom10, chastenet19, chown25dwarf}. Metallicity also affects the CO-to-\ce{H2} conversion factor \citep{bolatto13}, further complicating PAH-gas relations. Particularly, in low-extinction or low-metallicity regions where CO is photodissociated while \ce{H2} survives through self-shielding, CO no longer remains a good tracer of molecular gas \citep{bohlin78, grenier05, wolfire10, pineda13}. Observations of extremely metal-poor dwarf galaxy, Sextans A \citep{tarantino26} with $Z\sim0.07Z_\odot$ show that PAH emission is confined to compact, dust-shielded clumps, indicating that at low metallicity PAH survival, like CO, is restricted to the most shielded regions of the ISM.

Just like PAHs, the dust continuum emission also carries information about the gas column in the ISM. Far-IR and submillimeter continuum have been used to trace the total gas mass under the assumption of roughly constant dust-to-gas ratio \citep{israel97, leroy11, eales12, groves15}. However, the interpretation of mid-IR as a tracer of gas would require caution, given the increase in contributions from bright star-forming regions. This appears as shallow slopes in our CO-mid-IR relations with contributions from massive star-forming regions reaching 75\% in the F2100W filter. 

\subsection{Implications for using mid-infrared emission as a star formation tracer}

Over the past decades, mid-IR emission has become a widely used tracer of recent star formation \citep{calz07, kenn12}. Studies using Spitzer observation demonstrated that the 8\micron\ band, dominated by PAH emission, and the 24\micron\ band, dominated by dust continuum, both show empirical correlations with \ha, \paa, and UV emission \citep{calzetti05, calz07, relano09, kenn09}. The 24\micron\ continuum, in particular, has been shown to trace obscured star formation, leading to its widespread use as a star formation rate indicator on both sub-kpc and galaxy-integrated scales \citep{calz07, rieke09, hao11, binder18}. To account for dust attenuation affecting UV and optical tracers, hybrid SFR calibrations combining hydrogen recombination lines or UV emission with mid-IR luminosities were also developed \citep[e.g., \ha+24\micron or FUV+24\micron; ][]{calz07, leroy12, cluver17}, enabling recovery of both obscured and unobscured star formation across a wide range of environments. Furthermore, with JWST, recent studies have provided extensive evidence that the F2100W filter serves as an effective tracer of star formation \citep{leroy23b, belfiore23, calz24, calz25}.  

In M51, we also find F2100W to show a near-linear relation with \paa. At the same time, the CO-to-F2100W relation shows a shallow slope, indicating how F2100W is weighted towards bright star-forming regions. In Figure \ref{fig:slopes}, the right panel shows the slopes (m$_{Pa\alpha}$ or m$_{H\alpha}$) of \paa\ or \ha\ intensities as a function of mid-IR emission from several recent JWST studies. We note shallower slopes for M51 compared to other galaxies for most bands. Interestingly, within the \hii\ regions in M51, \cite{calz25} find an even shallower slope of 0.89 at 21\micron\ compared to our slope of 1.01. This difference likely reflects the fact that measurements focused on compact \hii\ regions isolate the most intensely heated dust associated with young stars, whereas our spatially resolved analysis also includes dust heated by the diffuse ISRF ($\sim$25\% CO-tracing component contribution seen at 21\micron). 

Despite the association between mid-IR continuum emission and recent star formation, contributions from the dust heated by diffuse ISRF are also present. This is the CO-tracing part in our two-component template model, which at F2100W contributes to about $\sim$25\% in M51. This extended ``cirrus” component has been identified in several studies, generally attributed to dust heated by older stellar populations \citep{walterbos87, calzetti05, groves12, leroy12, crocker13, boquein16, calapa14, belfiore23}. However, radiative transfer modeling of M51 by \citep{delooze14} finds that $\sim$94\% of the 24\micron\ emission is heated by young stars. \citet{kenn09} showed that SFR calibrations based on H$\alpha$+IR differ between \hii\ regions and whole galaxies, with the discrepancy driven by an additional diffuse dust component unrelated to ongoing star formation. This diffuse emission highlights the fundamental duality when considering mid-IR as a tracer of star formation: even continuum bands effective at tracing embedded star formation include a non-negligible contribution from dust heated by diffuse ISRF.

PAHs have also been extensively used to trace star formation \citep{peeters04, calz07, kenn09, shipley16}. In M51, we find PAH-to-\paa\ relation to be superlinear with slopes $\sim$1.4--1.6 at $\sim$40 pc scales. This reduction in PAH emission for regions with more intense \paa\ emission reflects their destruction in strongly irradiated regions. Observations of \hii\ regions have shown PAHs to be confined to the surrounding photodissociation region, while 21 \micron\ peaks in the center \citep{helou04, bendo08, relano09, egorov25}. Our slopes are comparable to results from the FEAST investigation of M51 reported in \cite{gregg25}, showing the relation of 3.3 and 7.7 \micron\ PAH emission with \paa\ in young star clusters. The key distinction between our studies is the physical scales probed: at star cluster scales, the emission is dominated by individual \hii\ regions and their immediate photodissociation regions, whereas at $\sim$40~pc scales considered here, the measurement includes both compact \hii\ regions and the diffuse component. 

Diffuse heating from the ISRF also contributes to PAH emission, with our and \cite{leroy23b} two-component decomposition showing roughly comparable contributions from CO-tracing and star-formation-tracing components in the PAH dominated bands. On larger, galaxy-wide scales, these contributions are seen to decrease; for example, studies of 8~\micron\ emission in NGC 628 suggest only $\sim$20–30\% of PAH emission originates from regions unrelated to ongoing star formation \citep{crocker13}. In addition, several factors complicate the use of PAHs as quantitative SFR tracers: metallicity strongly affects PAH abundance and emission, with low-metallicity galaxies showing significant deficits due to inhibited formation or enhanced destruction \citep{engel05, calz07, sandstrom10, smith07, gregg25}; radiation hardness and local UV intensity can destroy small PAH molecules in extreme star-forming environments or near AGN \citep{madden06, egorov23, egorov25}.

\section{Conclusions}

We examine how mid-infrared emission at 5.6–21\micron\ relates to molecular gas, total gas, and atomic gas surface density, and recent star formation in the central disk of M51. We use JWST MIRI observations of M51 in 8 filters along with \coone, \hi-21cm, and \paa. Our analysis spans spatial scales from $\sim$40 pc to $\sim$440 pc and probes both PAH-dominated and dust continuum–dominated MIRI bands. Our main conclusions are as follows:
\begin{enumerate}
    \item {{\it{Mid-IR emission correlates with both gas column density and star formation.}}
Across all MIRI bands, mid-IR intensity shows a moderate correlation with \coone\ and \paa\ on $\sim$40~pc scales with \spr$\sim$0.5. When averaged to $\sim$440~pc, these correlations become stronger (\spr$\sim$0.87-0.89) for both tracers and are similarly strong with the total gas surface density ($\Sigma_{\rm HI+H_2}$). These results suggest that mid-IR emission is not a pure tracer of just the ISM column or star formation but effectively traces both, albeit depending on wavelength}. 

\item{{\it{Atomic gas also shows a strong, near linear correlation to mid-IR emission.}} At $\sim$440~pc scales, \hi\ also shows a tight correlation (\spr$\sim$0.7) with mid-IR with power-law indices close to unity for PAH-dominated filters for regions where (I$_{\rm H\alpha}\le10^{-5}$ erg s$^{-1}$ cm$^{-2}$ sr$^{-1}$). At higher mid-IR intensities, corresponding to molecular gas-dominated regions, \sighi\ saturates as the atomic-to-molecular transition occurs and produces a turnover. These results indicate that mid-IR emission can effectively trace atomic gas where it contributes significantly to the dust-bearing ISM, but \hi\ alone cannot drive the mid-IR emission in molecular gas-dominated regions.}

\item{{\it{PAH-dominated bands scale linearly with gas but super-linearly with star formation.}} For the PAH-dominated filters (F560W$_{ss}$, F770W$_{ss}$, F1130W, and F1280W), the CO-to-mid-IR relations show slopes $\sim$0.98-1.06, indicating that PAHs are well mixed with the cold neutral gas and closely follow the dust-bearing gas column. The power-law indices and correlations are similar at $\sim$440 pc scales with \hi\ and the total gas surface densities suggesting that PAHs do not preferentially reside in a specific phase of the ISM. In contrast, \paa\ scales super-linearly with the PAH-dominated mid-IR filters ($\sim$1.4-1.6) indicating a possible suppression of PAH emission in star-forming regions or high-$U$ environments.  }

\item {{\it{Dust continuum bands scale sub-linearly with gas and near-linearly with star formation.}}
At longer wavelengths (F1500W, F1800W, F2100W), CO–to-mid-IR relations become progressively sublinear ($m\sim$0.6–0.8), while \paa–mid-IR slopes approach close to unity. This indicates that mid-IR continuum emission is increasingly weighted toward regions of strong radiation fields, such that variations in $U$ dominate over changes in gas column density and trace local heating.}

\item{{\it{The wavelength-dependent behavior of \coone-to-mid-IR and \paa-to-mid-IR slopes reflects a shift from gas-tracing to heating-weighted mid-IR emission.}} We note a wavelength-dependent transition in what drives the mid-IR emission: shorter wavelength PAH-dominated filters near-linearly tracing the gas column and dust heated by the diffuse ISRF while the longer wavelength dust continuum filters increasingly reflecting dust heating young stars in star-forming regions. This trend is corroborated with an empirical two-component model comprising \coone- and \paa-associated components, decomposing mid-IR into gas-tracing and star-formation-tracing components across all mid-IR filters. Within this empirical decomposition, we find the PAH-dominated bands have relatively similar contributions from both components, while the \paa-tracing component contributes $\sim$75\% of the total mid-IR emission at F2100W.}

\end{enumerate}

\begin{acknowledgments}
We thank the anonymous referee for their suggestions. This work is based on observations made with the 
NASA/ESA/CSA James Webb Space Telescope. The data were obtained from the Mikulski Archive for Space Telescopes at the Space Telescope Science Institute, which is operated by the Association of Universities for Research in Astronomy, Inc., under NASA contract NAS 5-03127 for JWST. These observations are associated with program \#3435 and \#1783. Support for program \#3435 was provided by NASA through a grant from the Space Telescope Science Institute, which is operated by the Association of Universities for Research in Astronomy, Inc., under NASA contract NAS 5-03127.
M.P., K.S., L.H., and H.K. acknowledge funding support from grant JWST-GO-03435.001.
AA acknowledges support from the Swedish National Space Agency (SNSA) through the grants 2023-00260 and 2021-00108.
S.K.S is supported by an International Research Fellowship of the Japan Society for the Promotion of Science (JSPS).
SCOG acknowledges support from the European Research Council via the ERC Synergy Grant ``ECOGAL'' (project ID 855130) and from the German Excellence Strategy via the Heidelberg Cluster of Excellence (EXC 2181 - 390900948) ``STRUCTURES''.
DN is grateful for funding from NASA via ATP grant 80NSSC22K0716.
BALG is supported by the German Research Foundation (DFG) in the form of an Emmy Noether Research Group - DFG project \#542802847 (GA 3170/3-1).
IDL acknowledges funding from the Belgian Science Policy Office (BELSPO) through the PRODEX project ``JWST/MIRI Science exploitation'' (C4000142239), funding from the European Research Council (ERC) under the European Union’s Horizon 2020 research and innovation program DustOrigin (ERC-2019-StG-851622) and funding from the Flemish Fund for Scientific Research (FWO-Vlaanderen) through the research project 3G023821.
D.C. gratefully acknowledges the Collaborative Research Center 1601 (SFB 1601 sub-project B3) funded by the Deutsche Forschungsgemeinschaft (DFG, German Research Foundation) – 500700252. 
\end{acknowledgments}




%
\facilities{JWST, PdBI, VLA}

\software{astropy \citep{2013A&A...558A..33A,2018AJ....156..123A}}


\appendix


\bibliography{gas-miri}{}

\begin{thebibliography}{}
\expandafter\ifx\csname natexlab\endcsname\relax\def\natexlab#1{#1}\fi
\providecommand{\url}[1]{\href{#1}{#1}}
\providecommand{\dodoi}[1]{doi:~\href{http://doi.org/#1}{\nolinkurl{#1}}}
\providecommand{\doeprint}[1]{\href{http://ascl.net/#1}{\nolinkurl{http://ascl.net/#1}}}
\providecommand{\doarXiv}[1]{\href{https://arxiv.org/abs/#1}{\nolinkurl{https://arxiv.org/abs/#1}}}

\bibitem[{L.~J. {Allamandola} {et~al.}(1989){Allamandola}, {Tielens}, \& {Barker}}]{alla89}
{Allamandola}, L.~J., {Tielens}, A.~G.~G.~M., \& {Barker}, J.~R. 1989, \bibinfo{title}{{Interstellar Polycyclic Aromatic Hydrocarbons: The Infrared Emission Bands, the Excitation/Emission Mechanism, and the Astrophysical Implications},} \apjs, 71, 733, \dodoi{10.1086/191396}

\bibitem[{G. {Aniano} {et~al.}(2011){Aniano}, {Draine}, {Gordon}, \& {Sandstrom}}]{aniano11}
{Aniano}, G., {Draine}, B.~T., {Gordon}, K.~D., \& {Sandstrom}, K. 2011, \bibinfo{title}{{Common-Resolution Convolution Kernels for Space- and Ground-Based Telescopes},} \pasp, 123, 1218, \dodoi{10.1086/662219}

\bibitem[{G. {Aniano} {et~al.}(2020){Aniano}, {Draine}, {Hunt}, {Sandstrom}, {Calzetti}, {Kennicutt}, {Dale}, {Galametz}, {Gordon}, {Leroy}, {Smith}, {Roussel}, {Sauvage}, {Walter}, {Armus}, {Bolatto}, {Boquien}, {Crocker}, {De Looze}, {Donovan Meyer}, {Helou}, {Hinz}, {Johnson}, {Koda}, {Miller}, {Montiel}, {Murphy}, {Rela{\~n}o}, {Rix}, {Schinnerer}, {Skibba}, {Wolfire}, \& {Engelbracht}}]{aniano20}
{Aniano}, G., {Draine}, B.~T., {Hunt}, L.~K., {et~al.} 2020, \bibinfo{title}{{Modeling Dust and Starlight in Galaxies Observed by Spitzer and Herschel: The KINGFISH Sample},} \apj, 889, 150, \dodoi{10.3847/1538-4357/ab5fdb}

\bibitem[{R.~S. {Asano} {et~al.}(2013){Asano}, {Takeuchi}, {Hirashita}, \& {Nozawa}}]{asano13}
{Asano}, R.~S., {Takeuchi}, T.~T., {Hirashita}, H., \& {Nozawa}, T. 2013, \bibinfo{title}{{What determines the grain size distribution in galaxies?},} \mnras, 432, 637, \dodoi{10.1093/mnras/stt506}

\bibitem[{ {Astropy Collaboration} {et~al.}(2013){Astropy Collaboration}, {Robitaille}, {Tollerud}, {Greenfield}, {Droettboom}, {Bray}, {Aldcroft}, {Davis}, {Ginsburg}, {Price-Whelan}, {Kerzendorf}, {Conley}, {Crighton}, {Barbary}, {Muna}, {Ferguson}, {Grollier}, {Parikh}, {Nair}, {Unther}, {Deil}, {Woillez}, {Conseil}, {Kramer}, {Turner}, {Singer}, {Fox}, {Weaver}, {Zabalza}, {Edwards}, {Azalee Bostroem}, {Burke}, {Casey}, {Crawford}, {Dencheva}, {Ely}, {Jenness}, {Labrie}, {Lim}, {Pierfederici}, {Pontzen}, {Ptak}, {Refsdal}, {Servillat}, \& {Streicher}}]{2013A&A...558A..33A}
{Astropy Collaboration}, {Robitaille}, T.~P., {Tollerud}, E.~J., {et~al.} 2013, \bibinfo{title}{{Astropy: A community Python package for astronomy},} \aap, 558, A33, \dodoi{10.1051/0004-6361/201322068}

\bibitem[{ {Astropy Collaboration} {et~al.}(2018){Astropy Collaboration}, {Price-Whelan}, {Sip{\H{o}}cz}, {G{\"u}nther}, {Lim}, {Crawford}, {Conseil}, {Shupe}, {Craig}, {Dencheva}, {Ginsburg}, {VanderPlas}, {Bradley}, {P{\'e}rez-Su{\'a}rez}, {de Val-Borro}, {Aldcroft}, {Cruz}, {Robitaille}, {Tollerud}, {Ardelean}, {Babej}, {Bach}, {Bachetti}, {Bakanov}, {Bamford}, {Barentsen}, {Barmby}, {Baumbach}, {Berry}, {Biscani}, {Boquien}, {Bostroem}, {Bouma}, {Brammer}, {Bray}, {Breytenbach}, {Buddelmeijer}, {Burke}, {Calderone}, {Cano Rodr{\'\i}guez}, {Cara}, {Cardoso}, {Cheedella}, {Copin}, {Corrales}, {Crichton}, {D'Avella}, {Deil}, {Depagne}, {Dietrich}, {Donath}, {Droettboom}, {Earl}, {Erben}, {Fabbro}, {Ferreira}, {Finethy}, {Fox}, {Garrison}, {Gibbons}, {Goldstein}, {Gommers}, {Greco}, {Greenfield}, {Groener}, {Grollier}, {Hagen}, {Hirst}, {Homeier}, {Horton}, {Hosseinzadeh}, {Hu}, {Hunkeler}, {Ivezi{\'c}}, {Jain}, {Jenness}, {Kanarek}, {Kendrew}, {Kern}, {Kerzendorf}, {Khvalko}, {King}, {Kirkby}, {Kulkarni},
  {Kumar}, {Lee}, {Lenz}, {Littlefair}, {Ma}, {Macleod}, {Mastropietro}, {McCully}, {Montagnac}, {Morris}, {Mueller}, {Mumford}, {Muna}, {Murphy}, {Nelson}, {Nguyen}, {Ninan}, {N{\"o}the}, {Ogaz}, {Oh}, {Parejko}, {Parley}, {Pascual}, {Patil}, {Patil}, {Plunkett}, {Prochaska}, {Rastogi}, {Reddy Janga}, {Sabater}, {Sakurikar}, {Seifert}, {Sherbert}, {Sherwood-Taylor}, {Shih}, {Sick}, {Silbiger}, {Singanamalla}, {Singer}, {Sladen}, {Sooley}, {Sornarajah}, {Streicher}, {Teuben}, {Thomas}, {Tremblay}, {Turner}, {Terr{\'o}n}, {van Kerkwijk}, {de la Vega}, {Watkins}, {Weaver}, {Whitmore}, {Woillez}, {Zabalza}, \& {Astropy Contributors}}]{2018AJ....156..123A}
{Astropy Collaboration}, {Price-Whelan}, A.~M., {Sip{\H{o}}cz}, B.~M., {et~al.} 2018, \bibinfo{title}{{The Astropy Project: Building an Open-science Project and Status of the v2.0 Core Package},} \aj, 156, 123, \dodoi{10.3847/1538-3881/aabc4f}

\bibitem[{D. {Baron} {et~al.}(2025){Baron}, {Sandstrom}, {Sutter}, {Hassani}, {Groves}, {Leroy}, {Schinnerer}, {Boquien}, {Brazzini}, {Chastenet}, {Dale}, {Egorov}, {Glover}, {Klessen}, {Pathak}, {Rosolowsky}, {Bigiel}, {Chevance}, {Grasha}, {Hughes}, {M{\'e}ndez-Delgado}, {Pety}, {Williams}, {Hannon}, \& {Sarbadhicary}}]{baron25}
{Baron}, D., {Sandstrom}, K.~M., {Sutter}, J., {et~al.} 2025, \bibinfo{title}{{PHANGS-ML: The Universal Relation between PAH Band and Optical Line Ratios across Nearby Star-forming Galaxies},} \apj, 978, 135, \dodoi{10.3847/1538-4357/ad972a}

\bibitem[{F. {Belfiore} {et~al.}(2023){Belfiore}, {Leroy}, {Williams}, {Barnes}, {Bigiel}, {Boquien}, {Cao}, {Chastenet}, {Congiu}, {Dale}, {Egorov}, {Eibensteiner}, {Emsellem}, {Glover}, {Groves}, {Hassani}, {Klessen}, {Kreckel}, {Neumann}, {Neumann}, {Querejeta}, {Rosolowsky}, {Sanchez-Blazquez}, {Sandstrom}, {Schinnerer}, {Sun}, {Sutter}, \& {Watkins}}]{belfiore23}
{Belfiore}, F., {Leroy}, A.~K., {Williams}, T.~G., {et~al.} 2023, \bibinfo{title}{{Calibrating mid-infrared emission as a tracer of obscured star formation on H II-region scales in the era of JWST},} \aap, 678, A129, \dodoi{10.1051/0004-6361/202347175}

\bibitem[{G.~J. {Bendo} {et~al.}(2008){Bendo}, {Draine}, {Engelbracht}, {Helou}, {Thornley}, {Bot}, {Buckalew}, {Calzetti}, {Dale}, {Hollenbach}, {Li}, \& {Moustakas}}]{bendo08}
{Bendo}, G.~J., {Draine}, B.~T., {Engelbracht}, C.~W., {et~al.} 2008, \bibinfo{title}{{The relations among 8, 24 and 160 {\ensuremath{\mu}}m dust emission within nearby spiral galaxies},} \mnras, 389, 629, \dodoi{10.1111/j.1365-2966.2008.13567.x}

\bibitem[{G.~J. {Bendo} {et~al.}(2010){Bendo}, {Wilson}, {Warren}, {Brinks}, {Butner}, {Chanial}, {Clements}, {Courteau}, {Irwin}, {Israel}, {Knapen}, {Leech}, {Matthews}, {M{\"u}hle}, {Petitpas}, {Serjeant}, {Tan}, {Tilanus}, {Usero}, {Vaccari}, {van der Werf}, {Vlahakis}, {Wiegert}, \& {Zhu}}]{bendo10}
{Bendo}, G.~J., {Wilson}, C.~D., {Warren}, B.~E., {et~al.} 2010, \bibinfo{title}{{The JCMT Nearby Galaxies Legacy Survey - III. Comparisons of cold dust, polycyclic aromatic hydrocarbons, molecular gas and atomic gas in NGC 2403},} \mnras, 402, 1409, \dodoi{10.1111/j.1365-2966.2009.16043.x}

\bibitem[{B.~A. {Binder} \& M.~S. {Povich}(2018){Binder} \& {Povich}}]{binder18}
{Binder}, B.~A., \& {Povich}, M.~S. 2018, \bibinfo{title}{{A Multiwavelength Look at Galactic Massive Star-forming Regions},} \apj, 864, 136, \dodoi{10.3847/1538-4357/aad7b2}

\bibitem[{R.~C. {Bohlin} {et~al.}(1978){Bohlin}, {Savage}, \& {Drake}}]{bohlin78}
{Bohlin}, R.~C., {Savage}, B.~D., \& {Drake}, J.~F. 1978, \bibinfo{title}{{A survey of interstellar H I from Lalpha absorption measurements. II.},} \apj, 224, 132, \dodoi{10.1086/156357}

\bibitem[{A.~D. {Bolatto} {et~al.}(2013){Bolatto}, {Wolfire}, \& {Leroy}}]{bolatto13}
{Bolatto}, A.~D., {Wolfire}, M., \& {Leroy}, A.~K. 2013, \bibinfo{title}{{The CO-to-H$_{2}$ Conversion Factor},} \araa, 51, 207, \dodoi{10.1146/annurev-astro-082812-140944}

\bibitem[{M. {Boquien} {et~al.}(2019){Boquien}, {Burgarella}, {Roehlly}, {Buat}, {Ciesla}, {Corre}, {Inoue}, \& {Salas}}]{boqu19}
{Boquien}, M., {Burgarella}, D., {Roehlly}, Y., {et~al.} 2019, \bibinfo{title}{{CIGALE: a python Code Investigating GALaxy Emission},} \aap, 622, A103, \dodoi{10.1051/0004-6361/201834156}

\bibitem[{M. {Boquien} {et~al.}(2016{\natexlab{a}}){Boquien}, {Kennicutt}, {Calzetti}, {Dale}, {Galametz}, {Sauvage}, {Croxall}, {Draine}, {Kirkpatrick}, {Kumari}, {Hunt}, {De Looze}, {Pellegrini}, {Rela{\~n}o}, {Smith}, \& {Tabatabaei}}]{boquien16}
{Boquien}, M., {Kennicutt}, R., {Calzetti}, D., {et~al.} 2016{\natexlab{a}}, \bibinfo{title}{{Towards universal hybrid star formation rate estimators},} \aap, 591, A6, \dodoi{10.1051/0004-6361/201527759}

\bibitem[{M. {Boquien} {et~al.}(2016{\natexlab{b}}){Boquien}, {Kennicutt}, {Calzetti}, {Dale}, {Galametz}, {Sauvage}, {Croxall}, {Draine}, {Kirkpatrick}, {Kumari}, {Hunt}, {De Looze}, {Pellegrini}, {Rela{\~n}o}, {Smith}, \& {Tabatabaei}}]{boquein16}
{Boquien}, M., {Kennicutt}, R., {Calzetti}, D., {et~al.} 2016{\natexlab{b}}, \bibinfo{title}{{Towards universal hybrid star formation rate estimators},} \aap, 591, A6, \dodoi{10.1051/0004-6361/201527759}

\bibitem[{B.~R. {Brandl} {et~al.}(2006){Brandl}, {Bernard-Salas}, {Spoon}, {Devost}, {Sloan}, {Guilles}, {Wu}, {Houck}, {Weedman}, {Armus}, {Appleton}, {Soifer}, {Charmandaris}, {Hao}, {Higdon}, {Marshall}, \& {Herter}}]{brandl06}
{Brandl}, B.~R., {Bernard-Salas}, J., {Spoon}, H.~W.~W., {et~al.} 2006, \bibinfo{title}{{The Mid-Infrared Properties of Starburst Galaxies from Spitzer-IRS Spectroscopy},} \apj, 653, 1129, \dodoi{10.1086/508849}

\bibitem[{M.~D. {Calapa} {et~al.}(2014){Calapa}, {Calzetti}, {Draine}, {Boquien}, {Kramer}, {Xilouris}, {Verley}, {Braine}, {Rela{\~n}o}, {van der Werf}, {Israel}, {Hermelo}, \& {Albrecht}}]{calapa14}
{Calapa}, M.~D., {Calzetti}, D., {Draine}, B.~T., {et~al.} 2014, \bibinfo{title}{{The Heating of Mid-infrared Dust in the Nearby Galaxy M33: A Testbed for Tracing Galaxy Evolution},} \apj, 784, 130, \dodoi{10.1088/0004-637X/784/2/130}

\bibitem[{D. {Calzetti} {et~al.}(2000){Calzetti}, {Armus}, {Bohlin}, {Kinney}, {Koornneef}, \& {Storchi-Bergmann}}]{calzetti00}
{Calzetti}, D., {Armus}, L., {Bohlin}, R.~C., {et~al.} 2000, \bibinfo{title}{{The Dust Content and Opacity of Actively Star-forming Galaxies},} \apj, 533, 682, \dodoi{10.1086/308692}

\bibitem[{D. {Calzetti} {et~al.}(2005){Calzetti}, {Kennicutt}, {Bianchi}, {Thilker}, {Dale}, {Engelbracht}, {Leitherer}, {Meyer}, {Sosey}, {Mutchler}, {Regan}, {Thornley}, {Armus}, {Bendo}, {Boissier}, {Boselli}, {Draine}, {Gordon}, {Helou}, {Hollenbach}, {Kewley}, {Madore}, {Martin}, {Murphy}, {Rieke}, {Rieke}, {Roussel}, {Sheth}, {Smith}, {Walter}, {White}, {Yi}, {Scoville}, {Polletta}, \& {Lindler}}]{calzetti05}
{Calzetti}, D., {Kennicutt}, Jr., R.~C., {Bianchi}, L., {et~al.} 2005, \bibinfo{title}{{Star Formation in NGC 5194 (M51a): The Panchromatic View from GALEX to Spitzer},} \apj, 633, 871, \dodoi{10.1086/466518}

\bibitem[{D. {Calzetti} {et~al.}(2007){Calzetti}, {Kennicutt}, {Engelbracht}, {Leitherer}, {Draine}, {Kewley}, {Moustakas}, {Sosey}, {Dale}, {Gordon}, {Helou}, {Hollenbach}, {Armus}, {Bendo}, {Bot}, {Buckalew}, {Jarrett}, {Li}, {Meyer}, {Murphy}, {Prescott}, {Regan}, {Rieke}, {Roussel}, {Sheth}, {Smith}, {Thornley}, \& {Walter}}]{calz07}
{Calzetti}, D., {Kennicutt}, R.~C., {Engelbracht}, C.~W., {et~al.} 2007, \bibinfo{title}{{The Calibration of Mid-Infrared Star Formation Rate Indicators},} \apj, 666, 870, \dodoi{10.1086/520082}

\bibitem[{D. {Calzetti} {et~al.}(2024){Calzetti}, {Adamo}, {Linden}, {Gregg}, {Krumholz}, {Bajaj}, {Bik}, {Cignoni}, {Correnti}, {Elmegreen}, {Faustino Vieira}, {Gallagher}, {Grasha}, {Gutermuth}, {Johnson}, {Messa}, {Melinder}, {{\"O}stlin}, {Pedrini}, {Sabbi}, {Smith}, \& {Tosi}}]{calz24}
{Calzetti}, D., {Adamo}, A., {Linden}, S.~T., {et~al.} 2024, \bibinfo{title}{{JWST-FEAST: Feedback in Emerging extrAgalactic Star clusTers: Calibration of Star Formation Rates in the Mid-infrared with NGC 628},} \apj, 971, 118, \dodoi{10.3847/1538-4357/ad53c0}

\bibitem[{D. {Calzetti} {et~al.}(2025){Calzetti}, {Kennicutt}, {Adamo}, {Sandstrom}, {Dale}, {Elmegreen}, {Gallagher}, {Gregg}, {Bajaj}, {B{\"o}ker}, {Bortolini}, {Boyer}, {Correnti}, {De Looze}, {Draine}, {Duarte-Cabral}, {Faustino Vieira}, {Grasha}, {Hunt}, {Johnson}, {Klessen}, {Krumholz}, {Lai}, {Lapeer}, {Linden}, {Messa}, {{\"O}stlin}, {Pedrini}, {Rela{\~n}o}, {Sabbi}, {Schinnerer}, {Skillman}, {Smith}, {Tosi}, {Walter}, \& {Weinbeck}}]{calz25}
{Calzetti}, D., {Kennicutt}, R.~C., {Adamo}, A., {et~al.} 2025, \bibinfo{title}{{Quantification of the Age Dependence of Mid-infrared Star Formation Rate Indicators},} \apj, 991, 198, \dodoi{10.3847/1538-4357/adfbe0}

\bibitem[{C. {Catal{\'a}n-Torrecilla} {et~al.}(2015){Catal{\'a}n-Torrecilla}, {Gil de Paz}, {Castillo-Morales}, {Iglesias-P{\'a}ramo}, {S{\'a}nchez}, {Kennicutt}, {P{\'e}rez-Gonz{\'a}lez}, {Marino}, {Walcher}, {Husemann}, {Garc{\'\i}a-Benito}, {Mast}, {Gonz{\'a}lez Delgado}, {Mu{\~n}oz-Mateos}, {Bland-Hawthorn}, {Bomans}, {Del Olmo}, {Galbany}, {Gomes}, {Kehrig}, {L{\'o}pez-S{\'a}nchez}, {Mendoza}, {Monreal-Ibero}, {P{\'e}rez-Torres}, {S{\'a}nchez-Bl{\'a}zquez}, {Vilchez}, \& {CALIFA Collaboration}}]{catalan2015}
{Catal{\'a}n-Torrecilla}, C., {Gil de Paz}, A., {Castillo-Morales}, A., {et~al.} 2015, \bibinfo{title}{{Star formation in the local Universe from the CALIFA sample. I. Calibrating the SFR using integral field spectroscopy data},} \aap, 584, A87, \dodoi{10.1051/0004-6361/201526023}

\bibitem[{G. {Chabrier}(2003){Chabrier}}]{chabrier03}
{Chabrier}, G. 2003, \bibinfo{title}{{Galactic Stellar and Substellar Initial Mass Function},} \pasp, 115, 763, \dodoi{10.1086/376392}

\bibitem[{J. {Chastenet} {et~al.}(2019){Chastenet}, {Sandstrom}, {Chiang}, {Leroy}, {Utomo}, {Bot}, {Gordon}, {Draine}, {Fukui}, {Onishi}, \& {Tsuge}}]{chastenet19}
{Chastenet}, J., {Sandstrom}, K., {Chiang}, I.-D., {et~al.} 2019, \bibinfo{title}{{The Polycyclic Aromatic Hydrocarbon Mass Fraction on a 10 pc Scale in the Magellanic Clouds},} \apj, 876, 62, \dodoi{10.3847/1538-4357/ab16cf}

\bibitem[{J. {Chastenet} {et~al.}(2023){Chastenet}, {Sutter}, {Sandstrom}, {Belfiore}, {Egorov}, {Larson}, {Leroy}, {Liu}, {Rosolowsky}, {Thilker}, {Watkins}, {Williams}, {Barnes}, {Bigiel}, {Boquien}, {Chevance}, {Chiang}, {Dale}, {Kruijssen}, {Emsellem}, {Grasha}, {Groves}, {Hassani}, {Hughes}, {Kreckel}, {Meidt}, {Rickards Vaught}, {Sardone}, \& {Schinnerer}}]{chastenet23}
{Chastenet}, J., {Sutter}, J., {Sandstrom}, K., {et~al.} 2023, \bibinfo{title}{{PHANGS-JWST First Results: Variations in PAH Fraction as a Function of ISM Phase and Metallicity},} \apjl, 944, L11, \dodoi{10.3847/2041-8213/acadd7}

\bibitem[{R. {Chown} {et~al.}(2021){Chown}, {Li}, {Parker}, {Wilson}, {Li}, \& {Gao}}]{chown21}
{Chown}, R., {Li}, C., {Parker}, L., {et~al.} 2021, \bibinfo{title}{{A new estimator of resolved molecular gas in nearby galaxies},} \mnras, 500, 1261, \dodoi{10.1093/mnras/staa3288}

\bibitem[{R. {Chown} {et~al.}(2025{\natexlab{a}}){Chown}, {Leroy}, {Sandstrom}, {Chastenet}, {Sutter}, {Koch}, {Koziol}, {Neumann}, {Sun}, {Williams}, {Baron}, {Anand}, {Barnes}, {Bazzi}, {Belfiore}, {Bigiel}, {Bolatto}, {Boquien}, {Cao}, {Chevance}, {Colombo}, {Dale}, {den Brok}, {Egorov}, {Eibensteiner}, {Emsellem}, {Hassani}, {Henshaw}, {He}, {Kim}, {Klessen}, {Kreckel}, {Larson}, {Lee}, {Meidt}, {Murphy}, {Oakes}, {Ostriker}, {Pan}, {Pathak}, {Rosolowsky}, {Sarbadhicary}, {Schinnerer}, {Teng}, {Thilker}, {Weinbeck}, \& {Watkins}}]{chown25}
{Chown}, R., {Leroy}, A.~K., {Sandstrom}, K., {et~al.} 2025{\natexlab{a}}, \bibinfo{title}{{Polycyclic Aromatic Hydrocarbon and CO(2{\textendash}1) Emission at 50{\textendash}150 pc Scales in 70 Nearby Galaxies},} \apj, 983, 64, \dodoi{10.3847/1538-4357/adbd40}

\bibitem[{R. {Chown} {et~al.}(2025{\natexlab{b}}){Chown}, {Leroy}, {Bolatto}, {Chastenet}, {Glover}, {Indebetouw}, {Koch}, {Donovan Meyer}, {Pingel}, {Rosolowsky}, {Sandstrom}, {Sutter}, {Tarantino}, {Bigiel}, {Boquien}, {Chiang}, {Dale}, {Dalcanton}, {Egorov}, {Eibensteiner}, {Grasha}, {Hassani}, {He}, {Kim}, {Meidt}, {Pathak}, {Sarbadhicary}, {Stanimirovic}, {Villanueva}, \& {Williams}}]{chown25dwarf}
{Chown}, R., {Leroy}, A.~K., {Bolatto}, A.~D., {et~al.} 2025{\natexlab{b}}, \bibinfo{title}{{Relationships between PAHs, Small Dust Grains, H$_2$, and HI in Local Group Dwarf Galaxies NGC 6822 and WLM Using JWST, ALMA, and the VLA},} arXiv e-prints, arXiv:2504.08069, \dodoi{10.48550/arXiv.2504.08069}

\bibitem[{M.~E. {Cluver} {et~al.}(2017){Cluver}, {Jarrett}, {Dale}, {Smith}, {August}, \& {Brown}}]{cluver17}
{Cluver}, M.~E., {Jarrett}, T.~H., {Dale}, D.~A., {et~al.} 2017, \bibinfo{title}{{Calibrating Star Formation in WISE Using Total Infrared Luminosity},} \apj, 850, 68, \dodoi{10.3847/1538-4357/aa92c7}

\bibitem[{D. {Colombo} {et~al.}(2014){Colombo}, {Hughes}, {Schinnerer}, {Meidt}, {Leroy}, {Pety}, {Dobbs}, {Garc{\'\i}a-Burillo}, {Dumas}, {Thompson}, {Schuster}, \& {Kramer}}]{colombo14}
{Colombo}, D., {Hughes}, A., {Schinnerer}, E., {et~al.} 2014, \bibinfo{title}{{The PdBI Arcsecond Whirlpool Survey (PAWS): Environmental Dependence of Giant Molecular Cloud Properties in M51},} \apj, 784, 3, \dodoi{10.1088/0004-637X/784/1/3}

\bibitem[{M. {Compi{\`e}gne} {et~al.}(2010){Compi{\`e}gne}, {Flagey}, {Noriega-Crespo}, {Martin}, {Bernard}, {Paladini}, \& {Molinari}}]{compiegne2010}
{Compi{\`e}gne}, M., {Flagey}, N., {Noriega-Crespo}, A., {et~al.} 2010, \bibinfo{title}{{Dust in the Diffuse Emission of the Galactic Plane: The Herschel/Spitzer Spectral Energy Distribution Fitting},} \apjl, 724, L44, \dodoi{10.1088/2041-8205/724/1/L44}

\bibitem[{I. {Cortzen} {et~al.}(2019){Cortzen}, {Garrett}, {Magdis}, {Rigopoulou}, {Valentino}, {Pereira-Santaella}, {Combes}, {Alonso-Herrero}, {Toft}, {Daddi}, {Elbaz}, {G{\'o}mez-Guijarro}, {Stockmann}, {Huang}, \& {Kramer}}]{cortzen19}
{Cortzen}, I., {Garrett}, J., {Magdis}, G., {et~al.} 2019, \bibinfo{title}{{PAHs as tracers of the molecular gas in star-forming galaxies},} \mnras, 482, 1618, \dodoi{10.1093/mnras/sty2777}

\bibitem[{A.~F. {Crocker} {et~al.}(2013){Crocker}, {Calzetti}, {Thilker}, {Aniano}, {Draine}, {Hunt}, {Kennicutt}, {Sandstrom}, \& {Smith}}]{crocker13}
{Crocker}, A.~F., {Calzetti}, D., {Thilker}, D.~A., {et~al.} 2013, \bibinfo{title}{{Quantifying Non-star-formation-associated 8 {\ensuremath{\mu}}m Dust Emission in NGC 628},} \apj, 762, 79, \dodoi{10.1088/0004-637X/762/2/79}

\bibitem[{G. {Cs{\"o}rnyei} {et~al.}(2023){Cs{\"o}rnyei}, {Anderson}, {Vogl}, {Taubenberger}, {Blondin}, {Leibundgut}, \& {Hillebrandt}}]{csornyei2023}
{Cs{\"o}rnyei}, G., {Anderson}, R.~I., {Vogl}, C., {et~al.} 2023, \bibinfo{title}{{Reeling in the Whirlpool galaxy: Distance to M 51 clarified through Cepheids and the type IIP supernova 2005cs},} \aap, 678, A44, \dodoi{10.1051/0004-6361/202346971}

\bibitem[{I. {De Looze} {et~al.}(2014){De Looze}, {Fritz}, {Baes}, {Bendo}, {Cortese}, {Boquien}, {Boselli}, {Camps}, {Cooray}, {Cormier}, {Davies}, {De Geyter}, {Hughes}, {Jones}, {Karczewski}, {Lebouteiller}, {Lu}, {Madden}, {R{\'e}my-Ruyer}, {Spinoglio}, {Smith}, {Viaene}, \& {Wilson}}]{delooze14}
{De Looze}, I., {Fritz}, J., {Baes}, M., {et~al.} 2014, \bibinfo{title}{{High-resolution, 3D radiative transfer modeling. I. The grand-design spiral galaxy M 51},} \aap, 571, A69, \dodoi{10.1051/0004-6361/201424747}

\bibitem[{J. {den Brok} {et~al.}(2025){den Brok}, {Jim{\'e}nez-Donaire}, {Leroy}, {Schinnerer}, {Bigiel}, {Pety}, {Petitpas}, {Usero}, {Teng}, {Humire}, {Koch}, {Rosolowsky}, {Sandstrom}, {Liu}, {Zhang}, {Stuber}, {Chevance}, {Dale}, {Eibensteiner}, {Gali{\'c}}, {Glover}, {Pan}, {Querejeta}, {Smith}, {Williams}, {Wilner}, \& {Zhang}}]{denbrok25}
{den Brok}, J., {Jim{\'e}nez-Donaire}, M.~J., {Leroy}, A., {et~al.} 2025, \bibinfo{title}{{CO Isotopologue-derived Molecular Gas Conditions and CO-to-H$_{2}$ Conversion Factors in M51},} \aj, 169, 18, \dodoi{10.3847/1538-3881/ad888a}

\bibitem[{B.~T. {Draine}(2011){Draine}}]{draine11}
{Draine}, B.~T. 2011, {Physics of the Interstellar and Intergalactic Medium}

\bibitem[{B.~T. {Draine} \& A. {Li}(2001){Draine} \& {Li}}]{draine01}
{Draine}, B.~T., \& {Li}, A. 2001, \bibinfo{title}{{Infrared Emission from Interstellar Dust. I. Stochastic Heating of Small Grains},} \apj, 551, 807, \dodoi{10.1086/320227}

\bibitem[{B.~T. {Draine} \& A. {Li}(2007){Draine} \& {Li}}]{draine07}
{Draine}, B.~T., \& {Li}, A. 2007, \bibinfo{title}{{Infrared Emission from Interstellar Dust. IV. The Silicate-Graphite-PAH Model in the Post-Spitzer Era},} \apj, 657, 810, \dodoi{10.1086/511055}

\bibitem[{B.~T. {Draine} {et~al.}(2021){Draine}, {Li}, {Hensley}, {Hunt}, {Sandstrom}, \& {Smith}}]{draine21}
{Draine}, B.~T., {Li}, A., {Hensley}, B.~S., {et~al.} 2021, \bibinfo{title}{{Excitation of Polycyclic Aromatic Hydrocarbon Emission: Dependence on Size Distribution, Ionization, and Starlight Spectrum and Intensity},} \apj, 917, 3, \dodoi{10.3847/1538-4357/abff51}

\bibitem[{S. {Eales} {et~al.}(2012){Eales}, {Smith}, {Auld}, {Baes}, {Bendo}, {Bianchi}, {Boselli}, {Ciesla}, {Clements}, {Cooray}, {Cortese}, {Davies}, {De Looze}, {Galametz}, {Gear}, {Gentile}, {Gomez}, {Fritz}, {Hughes}, {Madden}, {Magrini}, {Pohlen}, {Spinoglio}, {Verstappen}, {Vlahakis}, \& {Wilson}}]{eales12}
{Eales}, S., {Smith}, M. W.~L., {Auld}, R., {et~al.} 2012, \bibinfo{title}{{Can Dust Emission be Used to Estimate the Mass of the Interstellar Medium in Galaxies{\textemdash}A Pilot Project with the Herschel Reference Survey},} \apj, 761, 168, \dodoi{10.1088/0004-637X/761/2/168}

\bibitem[{O.~V. {Egorov} {et~al.}(2023){Egorov}, {Kreckel}, {Sandstrom}, {Leroy}, {Glover}, {Groves}, {Kruijssen}, {Barnes}, {Belfiore}, {Bigiel}, {Blanc}, {Boquien}, {Cao}, {Chastenet}, {Chevance}, {Congiu}, {Dale}, {Emsellem}, {Grasha}, {Klessen}, {Larson}, {Liu}, {Murphy}, {Pan}, {Pessa}, {Pety}, {Rosolowsky}, {Scheuermann}, {Schinnerer}, {Sutter}, {Thilker}, {Watkins}, \& {Williams}}]{egorov23}
{Egorov}, O.~V., {Kreckel}, K., {Sandstrom}, K.~M., {et~al.} 2023, \bibinfo{title}{{PHANGS-JWST First Results: Destruction of the PAH Molecules in H II Regions Probed by JWST and MUSE},} \apjl, 944, L16, \dodoi{10.3847/2041-8213/acac92}

\bibitem[{O.~V. {Egorov} {et~al.}(2025){Egorov}, {Leroy}, {Sandstrom}, {Kreckel}, {Baron}, {Belfiore}, {Chown}, {Sutter}, {Boquien}, {Canal i Saguer}, {Congiu}, {Dale}, {Egorova}, {Huber}, {Li}, {Williams}, {Chastenet}, {Chiang}, {Gerasimov}, {Hassani}, {Kim}, {Koziol}, {Lee}, {McClain}, {Delgado}, {Pan}, {Pathak}, {Rosolowsky}, {Sarbadhicary}, {Schinnerer}, {Thilker}, {Ubeda}, \& {Weinbeck}}]{egorov25}
{Egorov}, O.~V., {Leroy}, A.~K., {Sandstrom}, K., {et~al.} 2025, \bibinfo{title}{{Polycyclic aromatic hydrocarbon destruction in star-forming regions across 42 nearby galaxies},} \aap, 703, A103, \dodoi{10.1051/0004-6361/202556427}

\bibitem[{C.~W. {Engelbracht} {et~al.}(2005){Engelbracht}, {Gordon}, {Rieke}, {Werner}, {Dale}, \& {Latter}}]{engel05}
{Engelbracht}, C.~W., {Gordon}, K.~D., {Rieke}, G.~H., {et~al.} 2005, \bibinfo{title}{{Metallicity Effects on Mid-Infrared Colors and the 8 {\ensuremath{\mu}}m PAH Emission in Galaxies},} \apjl, 628, L29, \dodoi{10.1086/432613}

\bibitem[{F. {Galliano} {et~al.}(2018){Galliano}, {Galametz}, \& {Jones}}]{galliano18}
{Galliano}, F., {Galametz}, M., \& {Jones}, A.~P. 2018, \bibinfo{title}{{The Interstellar Dust Properties of Nearby Galaxies},} \araa, 56, 673, \dodoi{10.1146/annurev-astro-081817-051900}

\bibitem[{Y. {Gao} {et~al.}(2022){Gao}, {Tan}, {Gao}, {Fang}, {Chown}, {Jiao}, \& {Luo}}]{gao22}
{Gao}, Y., {Tan}, Q.-H., {Gao}, Y., {et~al.} 2022, \bibinfo{title}{{The Correlation between WISE 12 {\ensuremath{\mu}}m Emission and Molecular Gas Tracers on Subkiloparsec Scales in Nearby Star-forming Galaxies},} \apj, 940, 133, \dodoi{10.3847/1538-4357/ac9af1}

\bibitem[{Y. {Gao} {et~al.}(2019){Gao}, {Xiao}, {Li}, {Jiang}, {Tan}, {Gao}, {Wilson}, {Bureau}, {Saintonge}, {S{\'a}nchez-Gallego}, {Brown}, {Clark}, {Hwang}, {Lamperti}, {Lin}, {Liu}, {Lu}, {Pan}, {Sun}, \& {Williams}}]{gao19}
{Gao}, Y., {Xiao}, T., {Li}, C., {et~al.} 2019, \bibinfo{title}{{Estimating the Molecular Gas Mass of Low-redshift Galaxies from a Combination of Mid-infrared Luminosity and Optical Properties},} \apj, 887, 172, \dodoi{10.3847/1538-4357/ab557c}

\bibitem[{K.~D. {Gordon} {et~al.}(2015){Gordon}, {Chen}, {Anderson}, {Azzollini}, {Bergeron}, {Bouchet}, {Bouwman}, {Cracraft}, {Fischer}, {Friedman}, {Garc{\'\i}a-Mar{\'\i}n}, {Glasse}, {Glauser}, {Goodson}, {Greene}, {Hines}, {Khorrami}, {Lahuis}, {Lajoie}, {Meixner}, {Morrison}, {O'Sullivan}, {Pontoppidan}, {Regan}, {Ressler}, {Rieke}, {Scheithauer}, {Walker}, \& {Wright}}]{gordon15}
{Gordon}, K.~D., {Chen}, C.~H., {Anderson}, R.~E., {et~al.} 2015, \bibinfo{title}{{The Mid-Infrared Instrument for the James Webb Space Telescope, X: Operations and Data Reduction},} \pasp, 127, 696, \dodoi{10.1086/682260}

\bibitem[{B. {Gregg} {et~al.}(2024){Gregg}, {Calzetti}, {Adamo}, {Bajaj}, {Ryon}, {Linden}, {Correnti}, {Cignoni}, {Messa}, {Sabbi}, {Gallagher}, {Grasha}, {Pedrini}, {Gutermuth}, {Melinder}, {Kotulla}, {P{\'e}rez}, {Krumholz}, {Bik}, {{\"O}stlin}, {Johnson}, {Bortolini}, {Smith}, {Tosi}, {Maji}, \& {Faustino Vieira}}]{gregg2024}
{Gregg}, B., {Calzetti}, D., {Adamo}, A., {et~al.} 2024, \bibinfo{title}{{Feedback in Emerging Extragalactic Star Clusters, FEAST: The Relation between 3.3 {\ensuremath{\mu}}m Polycyclic Aromatic Hydrocarbon Emission and Star Formation Rate Traced by Ionized Gas in NGC 628},} \apj, 971, 115, \dodoi{10.3847/1538-4357/ad54b4}

\bibitem[{B. {Gregg} {et~al.}(2025){Gregg}, {Calzetti}, {Adamo}, {Pedrini}, {Linden}, {Bajaj}, {Ryon}, {Bik}, {Bortolini}, {Correnti}, {Draine}, {Elmegreen}, {Faustino Vieira}, {Gallagher}, {Grasha}, {Johnson}, {Lai}, {Messa}, {{\"O}stlin}, {Smith}, \& {Tosi}}]{gregg25}
{Gregg}, B., {Calzetti}, D., {Adamo}, A., {et~al.} 2025, \bibinfo{title}{{The Calibration of Short Wavelength Polycyclic Aromatic Hydrocarbon Emission as Star Formation Rate Indicators with JWST},} arXiv e-prints, arXiv:2511.06481, \dodoi{10.48550/arXiv.2511.06481}

\bibitem[{I.~A. {Grenier} {et~al.}(2005){Grenier}, {Casandjian}, \& {Terrier}}]{grenier05}
{Grenier}, I.~A., {Casandjian}, J.-M., \& {Terrier}, R. 2005, \bibinfo{title}{{Unveiling Extensive Clouds of Dark Gas in the Solar Neighborhood},} Science, 307, 1292, \dodoi{10.1126/science.1106924}

\bibitem[{B. {Groves} {et~al.}(2012){Groves}, {Krause}, {Sandstrom}, {Schmiedeke}, {Leroy}, {Linz}, {Kapala}, {Rix}, {Schinnerer}, {Tabatabaei}, {Walter}, \& {da Cunha}}]{groves12}
{Groves}, B., {Krause}, O., {Sandstrom}, K., {et~al.} 2012, \bibinfo{title}{{The heating of dust by old stellar populations in the bulge of M31},} \mnras, 426, 892, \dodoi{10.1111/j.1365-2966.2012.21696.x}

\bibitem[{B.~A. {Groves} {et~al.}(2015){Groves}, {Schinnerer}, {Leroy}, {Galametz}, {Walter}, {Bolatto}, {Hunt}, {Dale}, {Calzetti}, {Croxall}, \& {Kennicutt}}]{groves15}
{Groves}, B.~A., {Schinnerer}, E., {Leroy}, A., {et~al.} 2015, \bibinfo{title}{{Dust Continuum Emission as a Tracer of Gas Mass in Galaxies},} \apj, 799, 96, \dodoi{10.1088/0004-637X/799/1/96}

\bibitem[{C.-N. {Hao} {et~al.}(2011){Hao}, {Kennicutt}, {Johnson}, {Calzetti}, {Dale}, \& {Moustakas}}]{hao11}
{Hao}, C.-N., {Kennicutt}, R.~C., {Johnson}, B.~D., {et~al.} 2011, \bibinfo{title}{{Dust-corrected Star Formation Rates of Galaxies. II. Combinations of Ultraviolet and Infrared Tracers},} \apj, 741, 124, \dodoi{10.1088/0004-637X/741/2/124}

\bibitem[{G. {Helou} {et~al.}(2004){Helou}, {Roussel}, {Appleton}, {Frayer}, {Stolovy}, {Storrie-Lombardi}, {Hurt}, {Lowrance}, {Makovoz}, {Masci}, {Surace}, {Gordon}, {Alonso-Herrero}, {Engelbracht}, {Misselt}, {Rieke}, {Rieke}, {Willner}, {Pahre}, {Ashby}, {Fazio}, \& {Smith}}]{helou04}
{Helou}, G., {Roussel}, H., {Appleton}, P., {et~al.} 2004, \bibinfo{title}{{The Anatomy of Star Formation in NGC 300},} \apjs, 154, 253, \dodoi{10.1086/422640}

\bibitem[{H. {Hirashita}(2015){Hirashita}}]{hirashita15}
{Hirashita}, H. 2015, \bibinfo{title}{{Two-size approximation: a simple way of treating the evolution of grain size distribution in galaxies},} \mnras, 447, 2937, \dodoi{10.1093/mnras/stu2617}

\bibitem[{L.~K. {Hunt} {et~al.}(2010){Hunt}, {Thuan}, {Izotov}, \& {Sauvage}}]{hunt10}
{Hunt}, L.~K., {Thuan}, T.~X., {Izotov}, Y.~I., \& {Sauvage}, M. 2010, \bibinfo{title}{{The Spitzer View of Low-Metallicity Star Formation. III. Fine-Structure Lines, Aromatic Features, and Molecules},} \apj, 712, 164, \dodoi{10.1088/0004-637X/712/1/164}

\bibitem[{F.~P. {Israel}(1997){Israel}}]{israel97}
{Israel}, F.~P. 1997, \bibinfo{title}{{H\_2 and its relation to CO in the LMC and other magellanic irregular galaxies},} \aap, 328, 471, \dodoi{10.48550/arXiv.astro-ph/9709194}

\bibitem[{B.~C. {Kelly}(2007){Kelly}}]{kelly07}
{Kelly}, B.~C. 2007, \bibinfo{title}{{Some Aspects of Measurement Error in Linear Regression of Astronomical Data},} \apj, 665, 1489, \dodoi{10.1086/519947}

\bibitem[{R.~C. {Kennicutt} \& N.~J. {Evans}(2012){Kennicutt} \& {Evans}}]{kenn12}
{Kennicutt}, R.~C., \& {Evans}, N.~J. 2012, \bibinfo{title}{{Star Formation in the Milky Way and Nearby Galaxies},} \araa, 50, 531, \dodoi{10.1146/annurev-astro-081811-125610}

\bibitem[{R.~C. {Kennicutt} {et~al.}(2003){Kennicutt}, {Armus}, {Bendo}, {Calzetti}, {Dale}, {Draine}, {Engelbracht}, {Gordon}, {Grauer}, {Helou}, {Hollenbach}, {Jarrett}, {Kewley}, {Leitherer}, {Li}, {Malhotra}, {Regan}, {Rieke}, {Rieke}, {Roussel}, {Smith}, {Thornley}, \& {Walter}}]{kenn03}
{Kennicutt}, Jr., R.~C., {Armus}, L., {Bendo}, G., {et~al.} 2003, \bibinfo{title}{{SINGS: The SIRTF Nearby Galaxies Survey},} \pasp, 115, 928, \dodoi{10.1086/376941}

\bibitem[{R.~C. {Kennicutt} {et~al.}(2009){Kennicutt}, {Hao}, {Calzetti}, {Moustakas}, {Dale}, {Bendo}, {Engelbracht}, {Johnson}, \& {Lee}}]{kenn09}
{Kennicutt}, Jr., R.~C., {Hao}, C.-N., {Calzetti}, D., {et~al.} 2009, \bibinfo{title}{{Dust-corrected Star Formation Rates of Galaxies. I. Combinations of H{\ensuremath{\alpha}} and Infrared Tracers},} \apj, 703, 1672, \dodoi{10.1088/0004-637X/703/2/1672}

\bibitem[{J.~M.~D. {Kruijssen} {et~al.}(2019){Kruijssen}, {Schruba}, {Chevance}, {Longmore}, {Hygate}, {Haydon}, {McLeod}, {Dalcanton}, {Tacconi}, \& {van Dishoeck}}]{kruijssen19}
{Kruijssen}, J.~M.~D., {Schruba}, A., {Chevance}, M., {et~al.} 2019, \bibinfo{title}{{Fast and inefficient star formation due to short-lived molecular clouds and rapid feedback},} \nat, 569, 519, \dodoi{10.1038/s41586-019-1194-3}

\bibitem[{M.~R. {Krumholz} {et~al.}(2009){Krumholz}, {McKee}, \& {Tumlinson}}]{krum09}
{Krumholz}, M.~R., {McKee}, C.~F., \& {Tumlinson}, J. 2009, \bibinfo{title}{{The Atomic-to-Molecular Transition in Galaxies. II: H I and H$_{2}$ Column Densities},} \apj, 693, 216, \dodoi{10.1088/0004-637X/693/1/216}

\bibitem[{V. {Lebouteiller} {et~al.}(2011){Lebouteiller}, {Bernard-Salas}, {Whelan}, {Brandl}, {Galliano}, {Charmandaris}, {Madden}, \& {Kunth}}]{lebou11}
{Lebouteiller}, V., {Bernard-Salas}, J., {Whelan}, D.~G., {et~al.} 2011, \bibinfo{title}{{Influence of the Environment on Polycyclic Aromatic Hydrocarbon Emission in Star-forming Regions},} \apj, 728, 45, \dodoi{10.1088/0004-637X/728/1/45}

\bibitem[{A.~K. {Leroy} {et~al.}(2011){Leroy}, {Bolatto}, {Gordon}, {Sandstrom}, {Gratier}, {Rosolowsky}, {Engelbracht}, {Mizuno}, {Corbelli}, {Fukui}, \& {Kawamura}}]{leroy11}
{Leroy}, A.~K., {Bolatto}, A., {Gordon}, K., {et~al.} 2011, \bibinfo{title}{{The CO-to-H$_{2}$ Conversion Factor from Infrared Dust Emission across the Local Group},} \apj, 737, 12, \dodoi{10.1088/0004-637X/737/1/12}

\bibitem[{A.~K. {Leroy} {et~al.}(2012){Leroy}, {Bigiel}, {de Blok}, {Boissier}, {Bolatto}, {Brinks}, {Madore}, {Munoz-Mateos}, {Murphy}, {Sandstrom}, {Schruba}, \& {Walter}}]{leroy12}
{Leroy}, A.~K., {Bigiel}, F., {de Blok}, W.~J.~G., {et~al.} 2012, \bibinfo{title}{{Estimating the Star Formation Rate at 1 kpc Scales in nearby Galaxies},} \aj, 144, 3, \dodoi{10.1088/0004-6256/144/1/3}

\bibitem[{A.~K. {Leroy} {et~al.}(2021){Leroy}, {Hughes}, {Liu}, {Pety}, {Rosolowsky}, {Saito}, {Schinnerer}, {Schruba}, {Usero}, {Faesi}, {Herrera}, {Chevance}, {Hygate}, {Kepley}, {Koch}, {Querejeta}, {Sliwa}, {Will}, {Wilson}, {Anand}, {Barnes}, {Belfiore}, {Be{\v{s}}li{\'c}}, {Bigiel}, {Blanc}, {Bolatto}, {Boquien}, {Cao}, {Chandar}, {Chastenet}, {Chiang}, {Congiu}, {Dale}, {Deger}, {den Brok}, {Eibensteiner}, {Emsellem}, {Garc{\'\i}a-Rodr{\'\i}guez}, {Glover}, {Grasha}, {Groves}, {Henshaw}, {Jim{\'e}nez Donaire}, {Kim}, {Klessen}, {Kreckel}, {Kruijssen}, {Larson}, {Lee}, {Mayker}, {McElroy}, {Meidt}, {Mok}, {Pan}, {Puschnig}, {Razza}, {S{\'a}nchez-Bl'azquez}, {Sandstrom}, {Santoro}, {Sardone}, {Scheuermann}, {Sun}, {Thilker}, {Turner}, {Ubeda}, {Utomo}, {Watkins}, \& {Williams}}]{phangs_alma_pipeline}
{Leroy}, A.~K., {Hughes}, A., {Liu}, D., {et~al.} 2021, \bibinfo{title}{{PHANGS-ALMA Data Processing and Pipeline},} \apjs, 255, 19, \dodoi{10.3847/1538-4365/abec80}

\bibitem[{A.~K. {Leroy} {et~al.}(2023{\natexlab{a}}){Leroy}, {Bolatto}, {Sandstrom}, {Rosolowsky}, {Barnes}, {Bigiel}, {Boquien}, {den Brok}, {Cao}, {Chastenet}, {Chevance}, {Chiang}, {Chown}, {Colombo}, {Ellison}, {Emsellem}, {Grasha}, {Henshaw}, {Hughes}, {Klessen}, {Koch}, {Kim}, {Kreckel}, {Kruijssen}, {Larson}, {Lee}, {Levy}, {Lin}, {Liu}, {Meidt}, {Pety}, {Querejeta}, {Rubio}, {Saito}, {Salim}, {Schinnerer}, {Sormani}, {Sun}, {Thilker}, {Usero}, {Vogel}, {Watkins}, {Whitcomb}, {Williams}, \& {Wilson}}]{leroy23a}
{Leroy}, A.~K., {Bolatto}, A.~D., {Sandstrom}, K., {et~al.} 2023{\natexlab{a}}, \bibinfo{title}{{PHANGS-JWST First Results: A Global and Moderately Resolved View of Mid-infrared and CO Line Emission from Galaxies at the Start of the JWST Era},} \apjl, 944, L10, \dodoi{10.3847/2041-8213/acab01}

\bibitem[{A.~K. {Leroy} {et~al.}(2023{\natexlab{b}}){Leroy}, {Sandstrom}, {Rosolowsky}, {Belfiore}, {Bolatto}, {Cao}, {Koch}, {Schinnerer}, {Barnes}, {Be{\v{s}}li{\'c}}, {Bigiel}, {Blanc}, {Chastenet}, {Chen}, {Chevance}, {Chown}, {Congiu}, {Dale}, {Egorov}, {Emsellem}, {Eibensteiner}, {Faesi}, {Glover}, {Grasha}, {Groves}, {Hassani}, {Henshaw}, {Hughes}, {Jim{\'e}nez-Donaire}, {Kim}, {Klessen}, {Kreckel}, {Kruijssen}, {Larson}, {Lee}, {Levy}, {Liu}, {Lopez}, {Meidt}, {Murphy}, {Neumann}, {Pessa}, {Pety}, {Saito}, {Sardone}, {Sun}, {Thilker}, {Usero}, {Watkins}, {Whitcomb}, \& {Williams}}]{leroy23b}
{Leroy}, A.~K., {Sandstrom}, K., {Rosolowsky}, E., {et~al.} 2023{\natexlab{b}}, \bibinfo{title}{{PHANGS-JWST First Results: Mid-infrared Emission Traces Both Gas Column Density and Heating at 100 pc Scales},} \apjl, 944, L9, \dodoi{10.3847/2041-8213/acaf85}

\bibitem[{A. {Li} \& B.~T. {Draine}(2001){Li} \& {Draine}}]{li01}
{Li}, A., \& {Draine}, B.~T. 2001, \bibinfo{title}{{Infrared Emission from Interstellar Dust. II. The Diffuse Interstellar Medium},} \apj, 554, 778, \dodoi{10.1086/323147}

\bibitem[{S. {Lopez} {et~al.}(2025){Lopez}, {Ring}, {Leroy}, {Cronin}, {Bolatto}, {Lopez}, {Villanueva}, {Fisher}, {Thompson}, {Armus}, {Boeker}, {Boogaard}, {Boyer}, {Chown}, {Dale}, {Donaghue}, {Emig}, {Glover}, {Herrera-Camus}, {Klessen}, {Lai}, {Lenkic}, {Levy}, {Meier}, {Mills}, {Ott}, {Skillman}, {Smith}, {Tarantino}, {Veilleux}, {Walter}, \& {van der Werf}}]{sebastian25}
{Lopez}, S., {Ring}, C., {Leroy}, A.~K., {et~al.} 2025, \bibinfo{title}{{JWST Observations of Starbursts: PAHs Closely Trace the Cool Phase of M82's Galactic Wind},} arXiv e-prints, arXiv:2510.01314, \dodoi{10.48550/arXiv.2510.01314}

\bibitem[{S.~C. {Madden} {et~al.}(2006){Madden}, {Galliano}, {Jones}, \& {Sauvage}}]{madden06}
{Madden}, S.~C., {Galliano}, F., {Jones}, A.~P., \& {Sauvage}, M. 2006, \bibinfo{title}{{ISM properties in low-metallicity environments},} \aap, 446, 877, \dodoi{10.1051/0004-6361:20053890}

\bibitem[{D. {Pathak} {et~al.}(2024){Pathak}, {Leroy}, {Thompson}, {Lopez}, {Belfiore}, {Boquien}, {Dale}, {Glover}, {Klessen}, {Koch}, {Rosolowsky}, {Sandstrom}, {Schinnerer}, {Smith}, {Sun}, {Sutter}, {Williams}, {Bigiel}, {Cao}, {Chastenet}, {Chevance}, {Chown}, {Emsellem}, {Faesi}, {Larson}, {Lee}, {Meidt}, {Ostriker}, {Ramambason}, {Sarbadhicary}, \& {Thilker}}]{pathak24}
{Pathak}, D., {Leroy}, A.~K., {Thompson}, T.~A., {et~al.} 2024, \bibinfo{title}{{A Two-Component Probability Distribution Function Describes the Mid-IR Emission from the Disks of Star-forming Galaxies},} \aj, 167, 39, \dodoi{10.3847/1538-3881/ad110d}

\bibitem[{A. {Pedrini} {et~al.}(2024){Pedrini}, {Adamo}, {Calzetti}, {Bik}, {Gregg}, {Linden}, {Bajaj}, {Ryon}, {Ali}, {Bortolini}, {Correnti}, {Elmegreen}, {Elmegreen}, {Gallagher}, {Grasha}, {Gutermuth}, {Johnson}, {Melinder}, {Messa}, {{\"O}stlin}, {Sabbi}, {Smith}, {Tosi}, \& {Faustino Vieira}}]{pedrini24}
{Pedrini}, A., {Adamo}, A., {Calzetti}, D., {et~al.} 2024, \bibinfo{title}{{FEAST: Feedback in Emerging extragAlactic Star ClusTers: JWST Spots Polycyclic Aromatic Hydrocarbon Destruction in NGC 628 during the Emerging Phase of Star Formation},} \apj, 971, 32, \dodoi{10.3847/1538-4357/ad534d}

\bibitem[{A. {Pedrini} {et~al.}(2025){Pedrini}, {Adamo}, {Bik}, {Calzetti}, {Linden}, {Gregg}, {Bajaj}, {Ryon}, {Buckner}, {Bortolini}, {Cignoni}, {Correnti}, {Duarte-Cabral}, {Elmegreen}, {Faustino Vieira}, {Gallagher}, {Grasha}, {Johnson}, {Krumholz}, {Lapeer}, {Lai}, {Messa}, {{\"O}stlin}, {Roos}, {Smith}, \& {Tosi}}]{pedrini25}
{Pedrini}, A., {Adamo}, A., {Bik}, A., {et~al.} 2025, \bibinfo{title}{{The Near Infrared Spectral Energy Distribution of Young Star Clusters in the FEAST Galaxies: Missing Ingredients at 1─5 {\ensuremath{\mu}}m},} \apj, 992, 96, \dodoi{10.3847/1538-4357/ae0182}

\bibitem[{E. {Peeters} {et~al.}(2004){Peeters}, {Spoon}, \& {Tielens}}]{peeters04}
{Peeters}, E., {Spoon}, H.~W.~W., \& {Tielens}, A.~G.~G.~M. 2004, \bibinfo{title}{{Polycyclic Aromatic Hydrocarbons as a Tracer of Star Formation?},} \apj, 613, 986, \dodoi{10.1086/423237}

\bibitem[{J. {Pety} {et~al.}(2013){Pety}, {Schinnerer}, {Leroy}, {Hughes}, {Meidt}, {Colombo}, {Dumas}, {Garc{\'\i}a-Burillo}, {Schuster}, {Kramer}, {Dobbs}, \& {Thompson}}]{pety13}
{Pety}, J., {Schinnerer}, E., {Leroy}, A.~K., {et~al.} 2013, \bibinfo{title}{{The Plateau de Bure + 30 m Arcsecond Whirlpool Survey Reveals a Thick Disk of Diffuse Molecular Gas in the M51 Galaxy},} \apj, 779, 43, \dodoi{10.1088/0004-637X/779/1/43}

\bibitem[{J.~L. {Pineda} {et~al.}(2013){Pineda}, {Langer}, {Velusamy}, \& {Goldsmith}}]{pineda13}
{Pineda}, J.~L., {Langer}, W.~D., {Velusamy}, T., \& {Goldsmith}, P.~F. 2013, \bibinfo{title}{{A Herschel [C ii] Galactic plane survey. I. The global distribution of ISM gas components},} \aap, 554, A103, \dodoi{10.1051/0004-6361/201321188}

\bibitem[{J.~L. {Puget} {et~al.}(1985){Puget}, {Leger}, \& {Boulanger}}]{puget85}
{Puget}, J.~L., {Leger}, A., \& {Boulanger}, F. 1985, \bibinfo{title}{{Contribution of large polycyclic aromatic molecules to the infrared emission of the interstellar medium.},} \aap, 142, L19

\bibitem[{M.~W. {Regan} {et~al.}(2006){Regan}, {Thornley}, {Vogel}, {Sheth}, {Draine}, {Hollenbach}, {Meyer}, {Dale}, {Engelbracht}, {Kennicutt}, {Armus}, {Buckalew}, {Calzetti}, {Gordon}, {Helou}, {Leitherer}, {Malhotra}, {Murphy}, {Rieke}, {Rieke}, \& {Smith}}]{regan06}
{Regan}, M.~W., {Thornley}, M.~D., {Vogel}, S.~N., {et~al.} 2006, \bibinfo{title}{{The Radial Distribution of the Interstellar Medium in Disk Galaxies: Evidence for Secular Evolution},} \apj, 652, 1112, \dodoi{10.1086/505382}

\bibitem[{M. {Rela{\~n}o} \& R.~C. {Kennicutt}(2009){Rela{\~n}o} \& {Kennicutt}}]{relano09}
{Rela{\~n}o}, M., \& {Kennicutt}, Jr., R.~C. 2009, \bibinfo{title}{{Star Formation in Luminous H II Regions in M33},} \apj, 699, 1125, \dodoi{10.1088/0004-637X/699/2/1125}

\bibitem[{G.~H. {Rieke} {et~al.}(2009){Rieke}, {Alonso-Herrero}, {Weiner}, {P{\'e}rez-Gonz{\'a}lez}, {Blaylock}, {Donley}, \& {Marcillac}}]{rieke09}
{Rieke}, G.~H., {Alonso-Herrero}, A., {Weiner}, B.~J., {et~al.} 2009, \bibinfo{title}{{Determining Star Formation Rates for Infrared Galaxies},} \apj, 692, 556, \dodoi{10.1088/0004-637X/692/1/556}

\bibitem[{G.~H. {Rieke} {et~al.}(2015){Rieke}, {Wright}, {B{\"o}ker}, {Bouwman}, {Colina}, {Glasse}, {Gordon}, {Greene}, {G{\"u}del}, {Henning}, {Justtanont}, {Lagage}, {Meixner}, {N{\o}rgaard-Nielsen}, {Ray}, {Ressler}, {van Dishoeck}, \& {Waelkens}}]{reike15}
{Rieke}, G.~H., {Wright}, G.~S., {B{\"o}ker}, T., {et~al.} 2015, \bibinfo{title}{{The Mid-Infrared Instrument for the James Webb Space Telescope, I: Introduction},} \pasp, 127, 584, \dodoi{10.1086/682252}

\bibitem[{K.~M. {Sandstrom} {et~al.}(2010){Sandstrom}, {Bolatto}, {Draine}, {Bot}, \& {Stanimirovi{\'c}}}]{sandstrom10}
{Sandstrom}, K.~M., {Bolatto}, A.~D., {Draine}, B.~T., {Bot}, C., \& {Stanimirovi{\'c}}, S. 2010, \bibinfo{title}{{The Spitzer Survey of the Small Magellanic Cloud (S$^{3}$MC): Insights into the Life Cycle of Polycyclic Aromatic Hydrocarbons},} \apj, 715, 701, \dodoi{10.1088/0004-637X/715/2/701}

\bibitem[{K.~M. {Sandstrom} {et~al.}(2013){Sandstrom}, {Leroy}, {Walter}, {Bolatto}, {Croxall}, {Draine}, {Wilson}, {Wolfire}, {Calzetti}, {Kennicutt}, {Aniano}, {Donovan Meyer}, {Usero}, {Bigiel}, {Brinks}, {de Blok}, {Crocker}, {Dale}, {Engelbracht}, {Galametz}, {Groves}, {Hunt}, {Koda}, {Kreckel}, {Linz}, {Meidt}, {Pellegrini}, {Rix}, {Roussel}, {Schinnerer}, {Schruba}, {Schuster}, {Skibba}, {van der Laan}, {Appleton}, {Armus}, {Brandl}, {Gordon}, {Hinz}, {Krause}, {Montiel}, {Sauvage}, {Schmiedeke}, {Smith}, \& {Vigroux}}]{sandstrom13}
{Sandstrom}, K.~M., {Leroy}, A.~K., {Walter}, F., {et~al.} 2013, \bibinfo{title}{{The CO-to-H$_{2}$ Conversion Factor and Dust-to-gas Ratio on Kiloparsec Scales in Nearby Galaxies},} \apj, 777, 5, \dodoi{10.1088/0004-637X/777/1/5}

\bibitem[{K.~M. {Sandstrom} {et~al.}(2023){Sandstrom}, {Koch}, {Leroy}, {Rosolowsky}, {Emsellem}, {Smith}, {Egorov}, {Williams}, {Larson}, {Lee}, {Schinnerer}, {Thilker}, {Barnes}, {Belfiore}, {Bigiel}, {Blanc}, {Bolatto}, {Boquien}, {Cao}, {Chastenet}, {Chevance}, {Chiang}, {Dale}, {Faesi}, {Glover}, {Grasha}, {Groves}, {Hassani}, {Henshaw}, {Hughes}, {Kim}, {Klessen}, {Kreckel}, {Kruijssen}, {Lopez}, {Liu}, {Meidt}, {Murphy}, {Pan}, {Querejeta}, {Saito}, {Sardone}, {Sormani}, {Sutter}, {Usero}, \& {Watkins}}]{sandstrom23}
{Sandstrom}, K.~M., {Koch}, E.~W., {Leroy}, A.~K., {et~al.} 2023, \bibinfo{title}{{PHANGS-JWST First Results: Tracing the Diffuse Interstellar Medium with JWST Imaging of Polycyclic Aromatic Hydrocarbon Emission in Nearby Galaxies},} \apjl, 944, L8, \dodoi{10.3847/2041-8213/aca972}

\bibitem[{E. {Schinnerer} {et~al.}(2013){Schinnerer}, {Meidt}, {Pety}, {Hughes}, {Colombo}, {Garc{\'\i}a-Burillo}, {Schuster}, {Dumas}, {Dobbs}, {Leroy}, {Kramer}, {Thompson}, \& {Regan}}]{schin13}
{Schinnerer}, E., {Meidt}, S.~E., {Pety}, J., {et~al.} 2013, \bibinfo{title}{{The PdBI Arcsecond Whirlpool Survey (PAWS). I. A Cloud-scale/Multi-wavelength View of the Interstellar Medium in a Grand-design Spiral Galaxy},} \apj, 779, 42, \dodoi{10.1088/0004-637X/779/1/42}

\bibitem[{H.~V. {Shipley} {et~al.}(2016){Shipley}, {Papovich}, {Rieke}, {Brown}, \& {Moustakas}}]{shipley16}
{Shipley}, H.~V., {Papovich}, C., {Rieke}, G.~H., {Brown}, M. J.~I., \& {Moustakas}, J. 2016, \bibinfo{title}{{A New Star Formation Rate Calibration from Polycyclic Aromatic Hydrocarbon Emission Features and Application to High-redshift Galaxies},} \apj, 818, 60, \dodoi{10.3847/0004-637X/818/1/60}

\bibitem[{I. {Shivaei} \& L.~A. {Boogaard}(2024){Shivaei} \& {Boogaard}}]{shivaei24}
{Shivaei}, I., \& {Boogaard}, L.~A. 2024, \bibinfo{title}{{The tight correlation between PAH and CO emission from z {\ensuremath{\sim}} 0 to 4},} \aap, 691, L2, \dodoi{10.1051/0004-6361/202451826}

\bibitem[{J.~D.~T. {Smith} {et~al.}(2007){Smith}, {Draine}, {Dale}, {Moustakas}, {Kennicutt}, {Helou}, {Armus}, {Roussel}, {Sheth}, {Bendo}, {Buckalew}, {Calzetti}, {Engelbracht}, {Gordon}, {Hollenbach}, {Li}, {Malhotra}, {Murphy}, \& {Walter}}]{smith07}
{Smith}, J.~D.~T., {Draine}, B.~T., {Dale}, D.~A., {et~al.} 2007, \bibinfo{title}{{The Mid-Infrared Spectrum of Star-forming Galaxies: Global Properties of Polycyclic Aromatic Hydrocarbon Emission},} \apj, 656, 770, \dodoi{10.1086/510549}

\bibitem[{J. {Sutter} {et~al.}(2024){Sutter}, {Sandstrom}, {Chastenet}, {Leroy}, {Koch}, {Williams}, {Chown}, {Belfiore}, {Bigiel}, {Boquien}, {Cao}, {Chevance}, {Dale}, {Egorov}, {Glover}, {Groves}, {Klessen}, {Kreckel}, {Larson}, {Oakes}, {Pathak}, {Ramambason}, {Rosolowsky}, \& {Watkins}}]{sutter24}
{Sutter}, J., {Sandstrom}, K., {Chastenet}, J., {et~al.} 2024, \bibinfo{title}{{The Fraction of Dust Mass in the Form of Polycyclic Aromatic Hydrocarbons on 10{\textendash}50 pc Scales in Nearby Galaxies},} \apj, 971, 178, \dodoi{10.3847/1538-4357/ad54bd}

\bibitem[{E.~J. {Tarantino} {et~al.}(2025){Tarantino}, {Roman-Duval}, {Sandstrom}, {Smith}, {Whitcomb}, {Draine}, {Boyer}, {Chastenet}, {Chown}, {Clark}, {Gordon}, {Hensley}, {Lai}, {Lindberg}, {McQuinn}, {Newman}, {Telford}, {Van De Putte}, \& {Williams}}]{tarantino26}
{Tarantino}, E.~J., {Roman-Duval}, J., {Sandstrom}, K.~M., {et~al.} 2025, \bibinfo{title}{{JWST Captures Growth of Aromatic Hydrocarbon Dust Particles in the Extremely Metal-poor Galaxy Sextans A},} arXiv e-prints, arXiv:2512.04060, \dodoi{10.48550/arXiv.2512.04060}

\bibitem[{D.~A. {Thilker} {et~al.}(2000){Thilker}, {Braun}, \& {Walterbos}}]{hiiphot}
{Thilker}, D.~A., {Braun}, R., \& {Walterbos}, R. A.~M. 2000, \bibinfo{title}{{HIIPHOT: Automated Photometry of H II Regions Applied to M51},} \aj, 120, 3070, \dodoi{10.1086/316852}

\bibitem[{A.~G.~G.~M. {Tielens}(2008){Tielens}}]{tielens08}
{Tielens}, A.~G.~G.~M. 2008, \bibinfo{title}{{Interstellar polycyclic aromatic hydrocarbon molecules.},} \araa, 46, 289, \dodoi{10.1146/annurev.astro.46.060407.145211}

\bibitem[{V. {Villanueva} {et~al.}(2025){Villanueva}, {Bolatto}, {Herrera-Camus}, {Leroy}, {Fisher}, {Levy}, {B{\"o}ker}, {Boogaard}, {Cronin}, {Dale}, {Emig}, {De Looze}, {Donnelly}, {Lai}, {Lenkic}, {Lopez}, {Lopez}, {Meier}, {Ott}, {Relano}, {Smith}, {Tarantino}, {Veilleux}, {Walter}, \& {van der Werf}}]{villa25}
{Villanueva}, V., {Bolatto}, A.~D., {Herrera-Camus}, R., {et~al.} 2025, \bibinfo{title}{{JWST Observations of Starbursts: Relations between PAH features and CO clouds in the starburst galaxy M 82},} \aap, 695, A202, \dodoi{10.1051/0004-6361/202553891}

\bibitem[{F. {Walter} {et~al.}(2008){Walter}, {Brinks}, {de Blok}, {Bigiel}, {Kennicutt}, {Thornley}, \& {Leroy}}]{walter08}
{Walter}, F., {Brinks}, E., {de Blok}, W.~J.~G., {et~al.} 2008, \bibinfo{title}{{THINGS: The H I Nearby Galaxy Survey},} \aj, 136, 2563, \dodoi{10.1088/0004-6256/136/6/2563}

\bibitem[{R.~A.~M. {Walterbos} \& P.~B.~W. {Schwering}(1987){Walterbos} \& {Schwering}}]{walterbos87}
{Walterbos}, R.~A.~M., \& {Schwering}, P.~B.~W. 1987, \bibinfo{title}{{Infrared emission from interstellar dust in the Andromeda Galaxy.},} \aap, 180, 27

\bibitem[{C.~M. {Whitcomb} {et~al.}(2023{\natexlab{a}}){Whitcomb}, {Sandstrom}, {Leroy}, \& {Smith}}]{whitcomb23}
{Whitcomb}, C.~M., {Sandstrom}, K., {Leroy}, A., \& {Smith}, J. D.~T. 2023{\natexlab{a}}, \bibinfo{title}{{Star Formation and Molecular Gas Diagnostics with Mid- and Far-infrared Emission},} \apj, 948, 88, \dodoi{10.3847/1538-4357/acc316}

\bibitem[{C.~M. {Whitcomb} {et~al.}(2023{\natexlab{b}}){Whitcomb}, {Sandstrom}, \& {Smith}}]{whitcomb23b}
{Whitcomb}, C.~M., {Sandstrom}, K., \& {Smith}, J.-D.~T. 2023{\natexlab{b}}, \bibinfo{title}{{JWST-MIRI Synthetic Photometry Composition using 460 Spitzer-IRS Spectra of Nearby Galaxies},} Research Notes of the American Astronomical Society, 7, 38, \dodoi{10.3847/2515-5172/acc073}

\bibitem[{T.~G. {Williams} {et~al.}(2024){Williams}, {Lee}, {Larson}, {Leroy}, {Sandstrom}, {Schinnerer}, {Thilker}, {Belfiore}, {Egorov}, {Rosolowsky}, {Sutter}, {DePasquale}, {Pagan}, {Berger}, {Anand}, {Barnes}, {Bigiel}, {Boquien}, {Cao}, {Chastenet}, {Chevance}, {Chown}, {Dale}, {Deger}, {Eibensteiner}, {Emsellem}, {Faesi}, {Glover}, {Grasha}, {Hannon}, {Hassani}, {Henshaw}, {Jim{\'e}nez-Donaire}, {Kim}, {Klessen}, {Koch}, {Li}, {Liu}, {Meidt}, {M{\'e}ndez-Delgado}, {Murphy}, {Neumann}, {Neumann}, {Neumayer}, {Oakes}, {Pathak}, {Pety}, {Pinna}, {Querejeta}, {Ramambason}, {Romanelli}, {Sormani}, {Stuber}, {Sun}, {Teng}, {Usero}, {Watkins}, \& {Weinbeck}}]{will24}
{Williams}, T.~G., {Lee}, J.~C., {Larson}, K.~L., {et~al.} 2024, \bibinfo{title}{{PHANGS-JWST: Data-processing Pipeline and First Full Public Data Release},} \apjs, 273, 13, \dodoi{10.3847/1538-4365/ad4be5}

\bibitem[{M.~G. {Wolfire} {et~al.}(2010){Wolfire}, {Hollenbach}, \& {McKee}}]{wolfire10}
{Wolfire}, M.~G., {Hollenbach}, D., \& {McKee}, C.~F. 2010, \bibinfo{title}{{The Dark Molecular Gas},} \apj, 716, 1191, \dodoi{10.1088/0004-637X/716/2/1191}

\bibitem[{L. {Zhang} \& L.~C. {Ho}(2023){Zhang} \& {Ho}}]{zhang23}
{Zhang}, L., \& {Ho}, L.~C. 2023, \bibinfo{title}{{Estimating Molecular Gas Content in Galaxies from Polycyclic Aromatic Hydrocarbon Emission},} \apj, 943, 1, \dodoi{10.3847/1538-4357/aca8f1}

\end{thebibliography}
\bibliographystyle{aasjournalv7}



\end{document}
